%% file: paper.tex
\documentclass[preprint2]{aastex6}

\usepackage{ulem}
\graphicspath{ {./figs/} } 

\usepackage{color}

\usepackage{mathtools}

\usepackage{marginnote}
\newcommand{\rhocutoff}{\rho_\mathrm{cutoff}}

\newcommand{\gcc}{\mathrm{g~cm^{-3} }}

\newcommand{\Ms}{M_\odot}
\newcommand{\mae}{\textsf{Maestro}}
\newcommand{\mesa}{\textsf{MESA}}
\newcommand{\sch}{sub-M$_\mathrm{Ch}$ }

\input maestrosymbols

\newcommand{\rhostar}{\rho^{\ast}}

\newcommand{\Tstar}{T^{\ast}}
\newcommand{\Astar}{A^{\ast}}

\newcommand{\Ubstar}{{\bf{U}}^{\ast}}
\newcommand{\estar}{e^{\ast}}

\newcommand{\pstar}{p^{\ast}}

\newcommand{\rhoprimestar}{\rho^{\prime,\ast}}
\newcommand{\pprimestar}{p^{\prime,\ast}}

\newcommand{\grad}{{\nabla}}
\newcommand{\entropy}{s}

\newcommand{\calH}{{\mathcal{H}}}

\newcommand{\gb}{{\bf g}}

\begin{document}

%==========================================================================
% Title
%==========================================================================
\title{Low Mach Number Modeling of Convection in Helium Shells on 
       Sub-Chandrasekhar White Dwarfs II: \\ Bulk Properties of Simple Models}
%III will be 
\shorttitle{Sub-Chandra. II. Bulk Properties}
\shortauthors{Jacobs et al.}

\author{A.~M.~Jacobs\altaffilmark{1},
        M.~Zingale\altaffilmark{1},
        A.~Nonaka\altaffilmark{2},
        A.~S.~Almgren\altaffilmark{2},
        J.~B.~Bell\altaffilmark{2}}

\altaffiltext{1}{Department of Physics \& Astronomy,
                 Stony Brook University,
		             Stony Brook, NY 11794-3800, USA}

\altaffiltext{2}{Center for Computational Sciences and Engineering,
                 Lawrence Berkeley National Laboratory,
                 Berkeley, CA 94720, USA}

%==========================================================================
% Abstract
%==========================================================================
\begin{abstract}
  The dynamics of helium shell convection driven by nuclear burning
  establish the conditions for runaway in the sub-Chandrasekhar mass,
  double detonation model for Type Ia supernovae, as well as for a
  variety of other explosive phenomena.  We explore these convection
  dynamics for a range of white dwarf core and helium shell masses in
  three dimensions using the low Mach number hydrodynamics code \mae.
  We present calculations of the bulk properties of this evolution,
  including time-series evolution of global diagnostics, lateral
  averages of the 3D state, and the global 3D state.  We find a
  variety of outcomes including quasi-equilibrium, localized runaway,
  and convective runaway.  Our results suggest the double detonation
  progenitor model is promising and that 3D, dynamic convection plays a
  key role. 
\end{abstract}

%==========================================================================
% Introduction
%==========================================================================
\section{Introduction}
Despite being rather inert embers of evolved stars on their own, white dwarfs 
(WDs) manage to be at the center of many of the Universe's most spectacular
explosions through interactions with companion stars.  One of the most recently
proposed manifestations of such an explosion are ``.Ia'' events 
\citep{BildstenEtAl2007}, 
with the decimal point meant to indicate the events are about a tenth the
brightness for a tenth the time of type Ia supernovae (SNe Ia).

The proposed .Ia progenitor consists of a carbon/oxygen (C/O) WD accreting from
a helium-rich companion.  A prominent example of such
a system are AM Canum Venaticorum (AM CVn) binaries \citep{Warner1995, Nelemans2005}.
In their Letter, \cite{BildstenEtAl2007} calculate that under the right
conditions the thermonuclear timescale in an AM CVn's helium envelope can approach the
dynamical timescale, possibly establishing conditions for a detonation which
consumes the envelope but leaves
the WD core intact.  This yields a relatively faint transient a tenth the
brightness of a normal SNe Ia\footnote{As is conventional
   in the literature, we define normal SNe Ia as having lightcurve and spectral
   properties consistent with the dominant population
   (about 70\%, see \cite{LiEtAl2011}) of observed SNe Ia.  Such properties and
   contrasting peculiar events are discussed in 
   \cite{BranchEtAl1993,Phillips1993,HillebrandtEtAl2013}}.
A unique aspect of these calculations is the unprecedentedly
low ignition pressures, which is related to the unprecedentedly low masses of the
helium envelopes considered.  Previous work considering similar systems in the
context of double detonations, in which the helium detonation triggers a
detonation of the WD core, assumed higher shell masses 
\citep{Nomoto1982b, WoosleyEtAl1986, Livne1990, LivneGlasner1990, 
LivneGlasner1991, WoosleyWeaver1994, GarciaSenzEtAl1999}, excepting a data
point in \cite{Nomoto1982a} and artificial detonations in \cite{LivneArnett1995}.

As suggested by the authors of the Letter, many took on the task of a
detailed reexamination of these systems with lower mass helium shells.
A particularly broad and detailed reexamination was carried out by
\cite{WoosleyKasen2011}.  As they demonstrate, sub-Chandrasekhar mass (\sch)
C/O WDs with low-mass helium shells can yield a variety of explosive phenomena,
including helium novae, double detonations, and deflagrations that consumed
the envelope, leaving behind a hot core.  The potential to produce such a
variety of transient events motivates extensive theoretical inquiry, especially
as we approach first light for the Large Synoptic Survey Telescope
\citep{LSST2008}.  A great deal of this inquiry has been carried out, with
tantalizing results.

Much of the focus has been on these systems as double detonation SNe Ia
progenitors.  Detonation of the C/O core appears to be very robustly triggered
by compression waves if detonation occurs in the helium 
shell~\citep{FinkEtAl2007, FinkEtAl2010, WoosleyKasen2011, 
ShenBildsten2014}, even in the case of asynchronous, asymmetric 
ignition points~\citep{MollWoosley2013}.  This makes \sch promising candidates
as SNe Ia progenitors.  Thin helium shells have been
shown to be capable of carrying sustained detonations, and may even
contribute to features found in SNe Ia observations~\citep{TownsleyEtAl2012}.
Synthetic spectra and light curves indicate that if the C/O core detonates 
and dominates over helium shell effects in the observables, many \sch 
progenitor systems are promising candidates for normal SNe Ia \citep{SimEtAl2010,
  WoosleyKasen2011, FinkEtAl2010, KromerEtAl2010}.  In particular,
work to date favors C/O cores that are near 1.0 $\Ms$ and hot if they are to
produce normal SNe Ia.  The
core-only (no He shell) explosions of \cite{SimEtAl2010} agree best
with normal SNe Ia properties for C/O WDs near 1.0 $\Ms$.  The 1D
models and subsequent synthetic observables of \cite{WoosleyKasen2011}
agree best with normal SNe Ia's for their ``hot'' models in which the
core relaxed to a luminosity of $1.0 ~ L_\odot$ as compared to their
``cool'' models, relaxing to $0.01 ~ L_\odot$ before helium accretion
was modeled.

Further, delay time distribution calculations based on binary population
synthesis find distributions and rates for \sch SNe Ia progenitor models
consistent with being at least one plausible dominant channel for reproducing
distributions and rates based on observations \citep{RuiterEtAl2011}.
Similar calculations focusing on a subset of \sch progenitors find they may
be the progenitors of SNe Iax (\cite{WangEtAl2013}, though see also 
\cite{Liu2015d, Liu2015e}).
\cite{GeierEtAl2013} present observational evidence for both a helium-accreting
\sch progenitor system and a high velocity helium-rich star that matches
the expected properties of the unbound companion star following a \sch
SNe Ia.  \cite{BrownEtAl2011} analyze a sample of WD binary systems
including extremely low mass WDs in the context of AM CVn binaries and
sub-luminous SNe Ia.  They calculate merger rates that are comparable
to the observed rates of sub-luminous SNe Ia.  \cite{DroutEtAl2013}
compare observations of SN 2005ek with many possible models, including
\sch systems.  In particular they argue that if SN 2005ek did have a
\sch progenitor, an edge-lit detonation would be the most viable model.
In the edge-lit scenario, the detonation in the helium layer propagates
directly into the core, setting off a carbon detonation at the core/shell
interface.

We caution that theoretical studies (and the study reported here) have limits
and make many assumptions.  The importance of realistic compositions and
convective mixing have been made clear \citep{KromerEtAl2010, ShenMoore2014,
Piro2015}.  In addition we note that it is currently computationally impossible
in any model of the full core/shell system to fully resolve ignition of core
detonation, which occurs on 0.01 -- 1 cm scales for densities $\rho =
10^7$--$10^8~\gcc$.  Instead, such work must report the critical conditions
achieved in a given computational cell or group of cells and argue the
likelihood of them achieving ignition of detonation.  This challenge is in part
addressed in \cite{ShenBildsten2014}, who carry out small-scale, fully resolved
calculations of detonation ignition in regimes relevant to the C/O core of \sch
systems.  They argue that conditions reported in multi-dimensional studies of the
full core/shell system are sufficient for ignition in many cases, though lower
mass (roughly, below 0.8 $\Ms$) or O/Ne cores are less likely to experience
ignition.

A significant uncertainty remains.  Only one-dimensional (1D) models have
demonstrated development of a detonation in these lower mass helium shells.
Multi-dimensional work has focused on assuming ignition of detonation and
exploring the consequences.  In this series of papers, we hope to begin to
fill this gap by modeling the development of ignition and elucidating the 
detailed 3D properties of the system leading up to and at the moment of such
an ignition.  The first paper in the series details our methodology, carries
out numerical experiments, and demonstrates the development of a localized
runaway in a model with a 1.0 $\Ms$ core and 0.05 $\Ms$ shell 
(\cite{ZingaleEtAl2013}, hereafter Paper I).  In this paper we apply our
methodology to a large number of models at higher resolution, carry out a new
numerical experiment, and include many new analyses and diagnostics.  The
purpose of this paper is to
\begin{itemize}
   \item expand our methodology to a much larger suite of models,
   \item explore what broad outcomes and trends we find for simple initial
      models, and to
   \item characterize the bulk properties of these models, including global 3D
         structure, 1D averages of the 3D state, and peak global properties
         such as the properties of the hottest cell in the domain.
\end{itemize}
This broad exploration lays a foundation for more targeted analysis of a smaller
number of more sophisticated models.

We defer a detailed analysis of potential ignition in these simple models,
including the geometry, timing, number, and statistics of localized igniting
volumes, to the next paper in this series.  We also want to be clear that one
cost of doing such a
broad study is that we must use simple models motivated by detailed 1D models.
Our methodology also fundamentally limits us from modeling the development
of a detonation.  Instead, we are modeling the dynamics we expect immediately
prior (a few convective turnover times) to any ignition.  Thus, we model the
development of potential seeds of a detonation or deflagration.  The ultimate
fate of these seeds should be the focus of future work.

To begin, we describe our models and methodology.

%==========================================================================
% Methodology and Models
%==========================================================================
\section{Methodology and Models}\label{sec:meth}
Our simulations are performed using
\mae\footnote{\label{maegit}https://github.com/BoxLib-Codes/MAESTRO}, a
finite-volume, adaptive mesh stellar hydrodynamics code suitable for flows where
the fluid speed is much less than the sound speed\footnote{Exactly
   quantifying how much less is non-trivial. The
   code has been validated against compressible hydrodynamics up to a Mach
   number of 0.2 \citep{ABRZ:I}, and we generally abide by a rule of thumb that
   \mae\ is valid up to peak Mach numbers of about 0.3.}.  \mae\ models the
flow using a low Mach number approximation---sound waves are filtered out of the
system, but compressibility effects due to stratification and local heat release
are retained.  While the code is mature and has been used in state-of-the-art
astrophysical modeling, it is also under active development.
\cite{NonakaEtAl2010} is the most recent and comprehensive in a series of papers
describing the low Mach number equation set and numerical algorithm.  The
simulations reported here make use of an improved low Mach number equation set;
this and the associated algorithmic changes relative to the model used in
\cite{NonakaEtAl2010} are described in Appendix \ref{app:energy}.  The \mae\
source code, including all the code needed to run the models reported here, and
User's Guide are available in the public release.  Please see these for the full
details of the most current form of the algorithm. 

Paper I lays out in detail our numerical strategy for modeling \sch WDs with
helium shells and demonstrates the robustness of results to parameter variation.
Below we review the methodology and describe the configuration of the suite of
models used to broadly sample the explosive regimes of \sch systems.

% EoS and nuclear burning
\subsection{Microphysics}
\label{ssec:micro}
We utilize a general, publicly available stellar equation of
state~\citep{TimmesSwesty2000,Timmes2008}.  Ions, radiation,
degenerate and relativistic electrons, and Coulomb corrections are all
incorporated.

Our nuclear reaction network is quite simple for the sake of computational
efficiency, enabling a broad sampling of parameter space (see \S
\ref{ssec:modset}).  It is important to note that \cite{WoosleyKasen2011} 
emphasize two crucial reactions for exploring \sch systems: 
$^{14}\rm{N}(e^-, \nu)^{14}\rm{C}(\alpha,\gamma)^{18}\rm{O}$ (NCO) and 
$^{12}$C(p, $\gamma$)$^{13}$N($\alpha$,p)$^{16}$O (CagO-bypass).  Additionally, 
\cite{ShenBildsten2009} demonstrate the importance of 
$^{14}\rm{N}(\alpha, \gamma)^{18}\rm{F}$ (NagF). While we agree these reactions
are crucial to understanding \sch models, we can neglect them for the 
purposes of exploring the dominant energetics in the pre-ignition burning.
For more on our rationale for neglecting them and the conditions under which
they can be neglected, see Appendix \ref{app:rxns}.  We employ a simple network 
consisting of the isotopes $^{12}$C, $^{4}$He, and $^{16}$O and the rates for 
3 $^{4}\rm{He} \rightarrow ^{12}$C (triple-alpha) and 
$^{12}$C($\alpha$,$\gamma$)$^{16}$O (CagO).  CagO is included 
because it can allow for the tracing of $^{16}\rm{O}$ production which in turn 
traces the sites of vigorous burning and how polluted the shell becomes with 
burning products.

Our baseline reaction rates come from \cite{CaughlanFowler1988}, with screening as in 
\cite{GraboskeEtAl1973, WeaverEtAl1978, AlastueyJancovici1978, ItohEtAl1979}.  The 
CagO reaction rate is scaled by a factor of 1.7, as 
recommended in \cite{WeaverWoosley1993, Garnett1997}.  Thermodynamic derivatives are held 
constant over a single timestep as described in \cite{AlmgrenEtAl2008}. 

% Initial models
\subsection{Initial Models}
\label{ssec:initmod}
\begin{figure}
   \centering
   \includegraphics[width=0.5\textwidth]{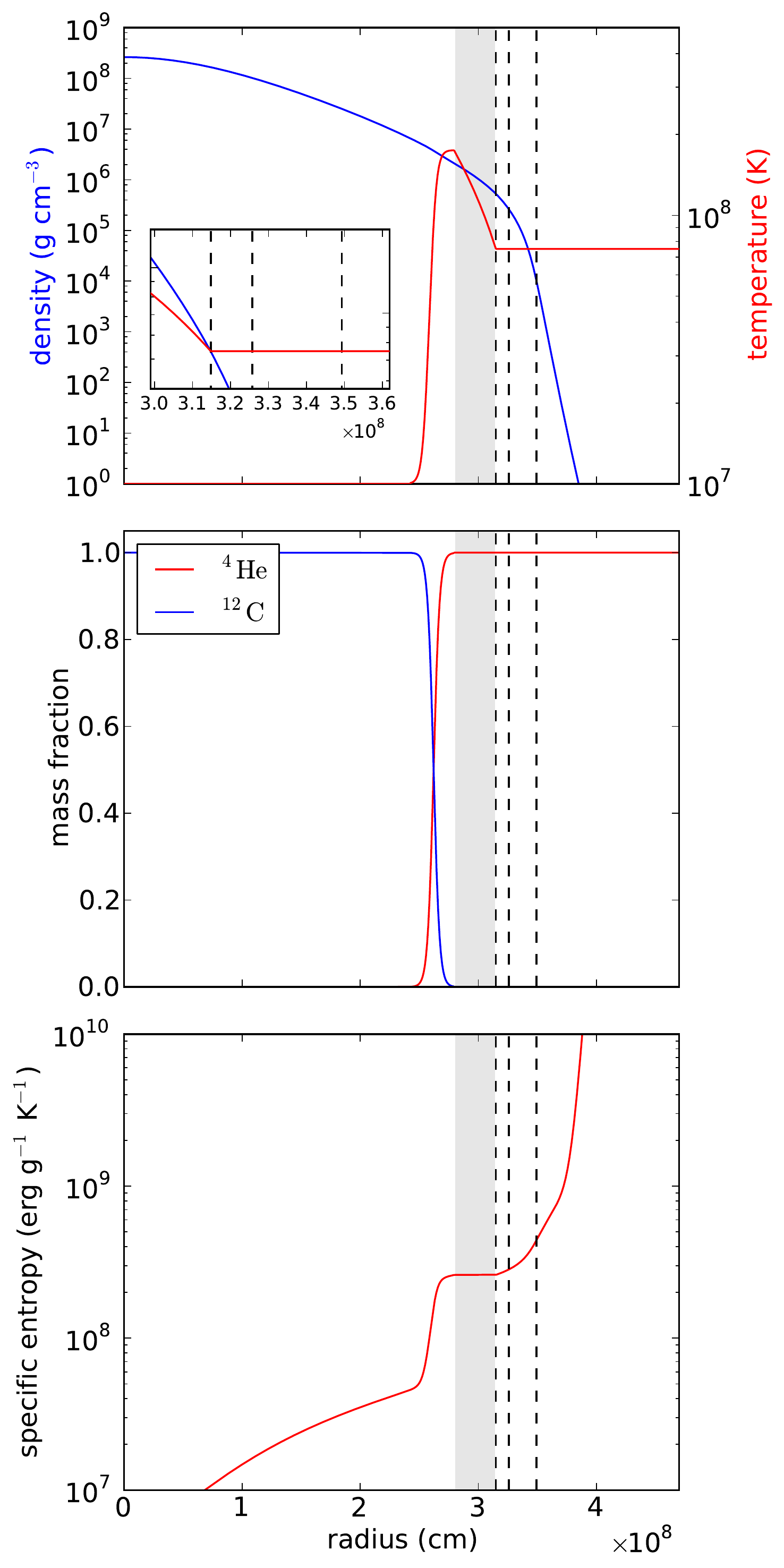}
   
   \caption{\label{fig:initmod} A representative initial model with $M_\mathrm{WD}$ = 1.2 $\Ms$, 
   $M_\mathrm{He}$ = 0.05 $\Ms$, $T_\mathrm{core} = 10^7$ K, $T_\mathrm{base} = 1.75 \times 10^8$ K. The
   shaded region is the convection zone.  The dashed lines from left to right are: the start of the sponge,
   the anelastic cutoff, and the base cutoff density (see \S \ref{ssec:bound}).
   {\bf Top:} Temperature (red) and density (blue) profiles.  The inset zooms in on
   the sponge and cutoff radii.
   {\bf Middle:} Mass fraction profiles of carbon (blue) and helium (red).
   {\bf Bottom:} Specific entropy profile.}
\end{figure}

\mae~evolves both a 1D hydrostatic base state and a 
3D hydrodynamic state.  For spherical
problems, such as the \sch system, this base state is radial.  To set
the initial conditions for our 3D problem we initialize the base state
and map that state onto the 3D grid.  Our \sch initial models are
defined by four parameters: the mass of the WD core, $M_\mathrm{WD}$,
the isentropic helium shell's mass, $M_\mathrm{He}$, the temperature at
the base of the helium shell, $T_\mathrm{base}$, and the core's
isothermal temperature, $T_\mathrm{core}$.  At the interface between
the core and the shell there is a sharp composition and temperature
gradient following the prescription described in \S 2.2 of Paper I.
We generate our own initial models using an iterative scheme that
enforces hydrostatic equilibrium and the values of $T_\mathrm{core}$
and $T_\mathrm{base}$ while converging on the given ($M_\mathrm{WD}$,
$M_\mathrm{He}$). Figure~\ref{fig:initmod} demonstrates a
representative initial model.  

We expand upon Paper I by adding a new parameter test.  Most initial models for
the simulations reported here use the same transition width parameter $\delta$
as in Paper
I: $\delta = 50~\mathrm{km}$.  However, we carried out a supplemental suite of
simulations for model 11030 (see \S \ref{ssec:modset}) in which all parameters
are the same save for a $\delta$ one-fourth, one-half, and twice the original
magnitude of 50 km.  The quadrupled resolution on a side in this paper
allows us to resolve sharper transitions than in Paper I.  Recall that this 
tanh-smoothing is necessary for the problem to be well-posed.  A sharp 
discontinuity in grid-based hydrodynamics makes it impossible to demonstrate
convergence as it offers no resolvable solution to converge to (see Paper I for
convergence tests).  The 50 km value for $\delta$ in the lower
resolution models of Paper I provided roughly 10 cells of
radial resolution over which to resolve the transition.  The lowest $\delta$
examined in this paper, 12.5 km, offers similar resolution of the transition.
In addition, it is not well-known exactly how transitions from core to shell are
realized in nature. Thus, this parameter study has both a numerical and physical
motivation.

The top row of Fig.~\ref{fig:temp_trans} demonstrates the
transition for 11030d0.25 (one-fourth the original $\delta$) and the reference
model 11030.  See \S~\ref{ssec:delta} for discussion of results.

% Lay out structure of grid, resolution, etc
\subsection{Grid Structure}
The 3D grid is Cartesian.  For all models except one (see \S \ref{ssec:modset})
an octant of the \sch WD is modeled, allowing us to capture 3D effects yet
achieve much greater computational efficiency and explore a large number of
models.  The impact of simulating an octant instead of the 
full star is investigated in Paper I. As we discuss in \S \ref{ssec:outcomes},
the higher resolutions and larger model set presented here introduce 
complications at the boundaries for octant runs with localized runaway.

The grid is adaptively refined to focus
resolution and computational power on the regions of greatest
interest.  To study the dynamics of the convection and nuclear burning
in the helium shell, we refine zones in which $\mathrm{X}_\mathrm{He} >
0.01$ at a density greater than $\rhocutoff$ (see \S
\ref{ssec:bound}).  To better resolve the shells, in this
study we further refine cells with temperatures $T > 125$ MK.  We are satisfied
with two levels of refinement for models with a $0.8~\Ms$ core.  The
mass-radius relationship for WDs means these models will have the largest
radius and consequently the thickest shells spatially.  However, for models
having $M_\mathrm{core} \geq 1.0~\Ms$ the spatial extent of the shell is
greatly reduced, which is exacerbated by the fact that more massive cores have
lower mass helium shells.  Thus, for all such models we add an extra level of
refinement for a total of four levels (the base grid, which we label level one,
and three additional, further refined levels).

\begin{figure}
   \centering
   \includegraphics[width=0.5\textwidth]{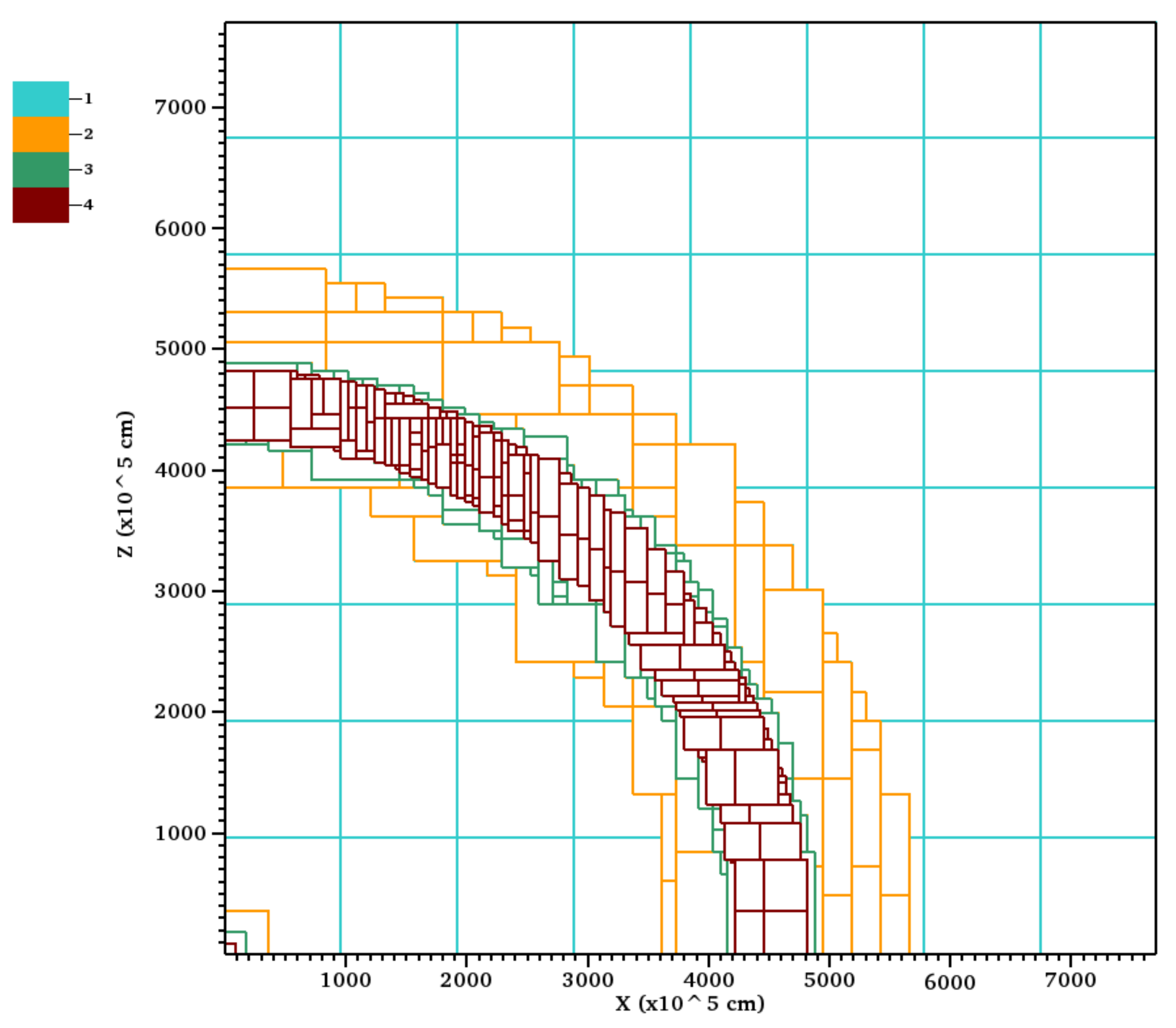}
   
   \caption{\label{fig:grid} A representative slice of the initial grid for $M_\mathrm{WD}$ = 1.0 $\Ms$, 
   $M_\mathrm{He}$ = 0.04 $\Ms$, $T_\mathrm{core} = 10^8$ K, $T_\mathrm{base} =
   1.85 \times 10^8$ K. The different colors indicate grid patches at different levels of
   refinement. Level 1 is the base (coarse) level.} 
\end{figure}  

Figure~\ref{fig:grid} illustrates the grid we have described.  This figure
outlines grid patches for each of the levels in the initial grid for model
10040H (for details about our adaptive mesh algorithm and the definition of grid
patches, see \S 5 of \cite{NonakaEtAl2010}).  At the coarsest
(base) level, all octant runs have a 256$^3$ resolution with a refinement factor
of 2 between levels, leading to subsequent 512$^3$, 1024$^3$, and 2048$^3$
effective resolutions within the refined patches.  We include one full star run
(08130F, see \S \ref{ssec:modset}), which has a 512$^3$ coarse (base)
resolution.  The strong dependence of radius on mass in WDs results in a range
of physical resolutions $\Delta x \approx$ 2.5 - 15.8 km at the finest level.
The 1D base state's resolution is not adaptively refined; instead, it has a
fixed resolution of five times that of the finest level: 5120 cells.  This
factor of five is first used and discussed in \citet{zingale:2009}.

% Describe treatment of boundaries and the regions far from stellar surface
\subsection{Boundaries}
\label{ssec:bound}
The boundary conditions for our simulations are reflecting on the symmetry faces 
of octant domains (lower x, y, and z), and outflow (zero-gradient) on the other faces.
A full star simulation has outflow boundary conditions on all faces of the domain.

As can be seen in Fig.~\ref{fig:grid}, the grid includes a
coarsely-resolved region well outside the convective zone.  This
serves to keep the convective surface insensitive to boundary
conditions.
A steep drop in density occurs at stellar surfaces (as seen in
Fig.~\ref{fig:initmod}).  Without modification, this rapid decline
precipitates a rapid spike in velocity to conserve momentum. The advantages of a
low Mach method become negligible if fluid velocities in any zone
approach the sound speed.  Thus much work has been put into developing
strategies to address steep density gradients at stellar surfaces
without significantly impacting \mae's computational or physical
validity.  The details of these 
treatments can be found in Paper I (\S 2.3), \cite{ZingaleEtAl2011}, 
and \cite{NonakaEtAl2012}.

Briefly, two density cutoffs are implemented: the anelastic cutoff
$\rho_\mathrm{anelastic}$ and the low-density cutoff
$\rho_\mathrm{cutoff}$ (see Fig.~\ref{fig:initmod}).  For zones with densities below
$\rho_\mathrm{anelastic}$, \mae~switches to an anelastic-like velocity
constraint that helps damp velocities (see \cite{AlmgrenEtAl2008}).
Density is held constant once it falls to
$\rho_\mathrm{cutoff}$, halting the steep decline.  To prevent
impacting validity its value is chosen such that the regions with
$\rho \le \rho_\mathrm{cutoff}$ contain an insignificant proportion of
the system's total mass.  These cutoffs are supplemented with a numerical
sponge that damps surface velocities \citep{AlmgrenEtAl2008}.
Cutoff values for all simulations are discussed in \S \ref{ssec:modset}.

This combination of cutoffs, a sponge, and maintaining a buffer zone
in the computational domain between the stellar surface and
the domain's boundaries enables us to study surface convection in detail
over long timescales without surface effects significantly impacting
our results.

% Note: \phn = insert space size of number
%       \phd = insert space size of decimal point
%       \phs = insert space size of minus sign
%       \phm{stuff} = insert space size of ``stuff''
\begin{deluxetable*}{lccccccccccc}
\tabletypesize{\scriptsize}
\tablecolumns{10}
\tablewidth{0pt}
\tablecaption{\label{tab:modset} Model Set}
\tablehead{
   \colhead{label}         & \colhead{($M_\mathrm{core}, M_\mathrm{He}$)} & \colhead{$T_\mathrm{core}$} & \colhead{$T_\mathrm{base}(t=0)$} & \colhead{$\rho_\mathrm{base}$} & \colhead{$x_\mathrm{max}$} & \colhead{$\Delta x_\mathrm{fine}$} & \colhead{$\rho_\mathrm{anelastic}$} & \multicolumn{3}{c}{$t_\mathrm{final}/\langle \tau_\mathrm{conv} \rangle$\tablenotemark{a}} & \colhead{outcome\tablenotemark{b}} \\
   \colhead{}              & \colhead{[$\Ms$]}                            & \colhead{[K]}               & \colhead{[$\times 10^8$ K]}      & \colhead{[$\times 10^5~\gcc$]} & \colhead{[km]}             & \colhead{[km]}                     & \colhead{[$\times 10^5~\gcc$]}      & \colhead{min}     & \colhead{avg}             & \colhead{max}             & \colhead{} 
}
\startdata
12030H                     & (1.2, 0.03)                                  & $10^8$                      & 1.75                             &    10.1                        & \phn5300                   & \phn2.6                            & $1.29$                              & \phn2.3              & \phn8.8                &    12.1                   & l                 \\
12030                      & (1.2, 0.03)                                  & $10^7$                      & 1.75                             &    10.8                        & \phn5100                   & \phn2.5                            & $1.37$                              & \phn1.5              & \phn6.4                & \phn8.6                   & l                 \\
12020H                     & (1.2, 0.02)                                  & $10^8$                      & 1.75                             & \phn6.2                        & \phn5500                   & \phn2.7                            & $0.80$                              & \phn3.1              &    10.3                &    17.1                   & l                 \\
12020                      & (1.2, 0.02)                                  & $10^7$                      & 1.75                             & \phn6.8                        & \phn5300                   & \phn2.6                            & $0.87$                              & \phn3.0              &    12.4                &    19.4                   & l                 \\
11030H                     & (1.1, 0.03)                                  & $10^8$                      & 1.85                             & \phn5.7                        & \phn6600                   & \phn3.2                            & $0.67$                              & \phn2.7              &    10.6                &    17.5                   & l                 \\
11030d0.25                 & (1.1, 0.03)                                  & $10^7$                      & 1.90                             & \phn6.0                        & \phn6400                   & \phn3.1                            & $0.68$                              &    12.1              & \phn4.7                &    35.9                   & l                 \\
11030d0.5                  & (1.1, 0.03)                                  & $10^7$                      & 1.90                             & \phn6.0                        & \phn6400                   & \phn3.1                            & $0.68$                              & \phn9.8              & \phn8.5                &    29.1                   & l                 \\
11030                      & (1.1, 0.03)                                  & $10^7$                      & 1.90                             & \phn6.0                        & \phn6400                   & \phn3.1                            & $0.68$                              & \phn1.9              &    26.3                &    13.6                   & l                 \\
11030d2                    & (1.1, 0.03)                                  & $10^7$                      & 1.90                             & \phn6.0                        & \phn6400                   & \phn3.1                            & $0.68$                              & \phn2.5              &    32.6                & \phn9.4                   & l                 \\
11020H                     & (1.1, 0.02)                                  & $10^8$                      & 1.85                             & \phn3.6                        & \phn6900                   & \phn3.4                            & $0.43$                              & \phn1.5              & \phn3.8                & \phn9.0                   & q                 \\
11020                      & (1.1, 0.02)                                  & $10^7$                      & 1.85                             & \phn3.9                        & \phn6600                   & \phn3.2                            & $0.46$                              & \phn5.2              &    19.5                &    27.2                   & q                 \\
10040H                     & (1.0, 0.04)                                  & $10^8$                      & 1.85                             & \phn5.0                        & \phn7700                   & \phn3.8                            & $0.58$                              & \phn7.1              &    23.4                &    27.6                   & q                 \\
10040                      & (1.0, 0.04)                                  & $10^7$                      & 1.85                             & \phn5.3                        & \phn7400                   & \phn3.6                            & $0.62$                              & \phn4.2              &    14.9                &    18.4                   & q                 \\
10030H                     & (1.0, 0.03)                                  & $10^8$                      & 1.85                             & \phn3.5                        & \phn7900                   & \phn3.9                            & $0.42$                              & \phn1.2              & \phn4.6                & \phn7.8                   & q                 \\
10030                      & (1.0, 0.03)                                  & $10^7$                      & 1.85                             & \phn3.8                        & \phn7600                   & \phn3.7                            & $0.45$                              & \phn1.5              & \phn6.5                & \phn8.3                   & q                 \\
08130H                     & (0.8, 0.13)                                  & $10^8$                      & 1.85                             & \phn9.9                        & \phn8300                   & \phn8.1                            & $1.15$                              & \phn0.6              & \phn3.8                & \phn8.4                   & l                 \\
08130F                     & (0.8, 0.13)                                  & $10^7$                      & 1.85                             &    10.9                        &    16200                   &    15.8                            & $1.26$                              & \phn0.2              & \phn1.3                & \phn6.3                   & l                 \\
08130                      & (0.8, 0.13)                                  & $10^7$                      & 1.85                             &    10.7                        & \phn8100                   & \phn7.9                            & $1.25$                              & \phn0.6              & \phn4.2                &    10.4                   & l                 \\
08120H                     & (0.8, 0.12)                                  & $10^8$                      & 1.85                             & \phn8.8                        & \phn8500                   & \phn8.3                            & $1.03$                              & \phn1.8              & \phn7.7                &    10.8                   & l                 \\
08120                      & (0.8, 0.12)                                  & $10^7$                      & 1.75                             & \phn9.6                        & \phn8100                   & \phn7.9                            & $1.22$                              & \phn8.4              &    25.7                &    27.9                   & l                 \\
08050                      & (0.8, 0.05)                                  & $10^7$                      & 2.50                             & \phn2.6                        &    10100                   & \phn9.9                            & $0.19$                              & \phn0.9              & \phn2.7                & \phn7.1                   & c                 \\
\enddata

\tablenotetext{a}{see \S~\ref{ssec:conv} for how these values are calculated}
\tablenotetext{b}{outcomes are designated as (l) localized runaway, (q) quasi-equilibrium, or (c) for convective runaway.  
   See text for details.}
\end{deluxetable*}  

% Model set
% NOTE: For reasons I can't fathom, if I put the \label under the \subsection as
%    we do in other parts of the paper, the references to it break.  So I'm
%    sticking the label within the \subsection{}
\subsection{Model Set \label{ssec:modset}}
\mae's ability to take large timesteps as well as the nature of \sch
pre-explosive dynamics make a broad sampling of the parameter space in 
3D computationally feasible.  What exactly is the
parameter space of interest?  To determine this we draw on the results
of \cite{BildstenEtAl2007} and the many studies they inspired.

The parameters of greatest interest are the core and helium shell mass
configurations.  The motivating question is how ignition develops and
how it is characterized in minimal helium shell mass systems for a range
of core masses.  Figure 2 of \cite{BildstenEtAl2007} illustrates their
determination of the minimum shell masses for which the nuclear
burning timescale is on the order of the dynamical timescale for
isothermal cores with $T_\mathrm{core} = 3 \times 10^7$ K.  Such a
short nuclear burning timescale suggests the possibility of
thermonuclear runaway even for thin helium shells with $M_\mathrm{He}
\lesssim 0.05 $ -- $ 0.0125 M_\mathrm{core}$ for $1.0$ -- $1.2~\Ms$ cores.
This work is extended and deepened
in subsequent studies, which are largely consistent with the essential
results of the 2007 work \citep{ShenBildsten2009,BrooksEtAl2015}. 
\cite{WoosleyKasen2011} carry out an
extensive set of 1D \sch calculations and generate an analogous figure
(Figure 19).  They include the impact of varying $T_\mathrm{core}$.
For $M_\mathrm{core} = 0.7~\Ms$, runaway can occur with helium shells
having $\sim$ 15\% of the core's mass, perhaps not sufficiently thin for SNe
Ia-like spectra.  As $M_\mathrm{core}$ increases to $1.1~\Ms$, runaway
can be achieved with shells $\sim$ 2.25\% of the core's mass for hotter
cores ($T_\mathrm{core} \sim 7.5 \times 10^7$ K), making SNe Ia-like
spectra more achievable.  The bare (no helium shell) 1D \sch WD
detonation calculations of \cite{SimEtAl2010} suggest systems with
$M_\mathrm{core} \gtrsim 1.0~\Ms$ can yield observables in reasonable
agreement with the range of observed normal SNe Ia while lower
$M_\mathrm{core}$ systems can produce characteristics of observed
sub-luminous SNe Ia. \cite{FinkEtAl2010}'s 2D calculations also find
they can produce many characteristics of the range of observed SNe Ia
and that core detonation is triggered by shell detonations for
($M_\mathrm{core}, M_\mathrm{He}$) = (0.810 -- 1.385, 0.126 -- 0.0035)
$\Ms$.

\begin{figure}
   \centering
   \includegraphics[width=0.5\textwidth]{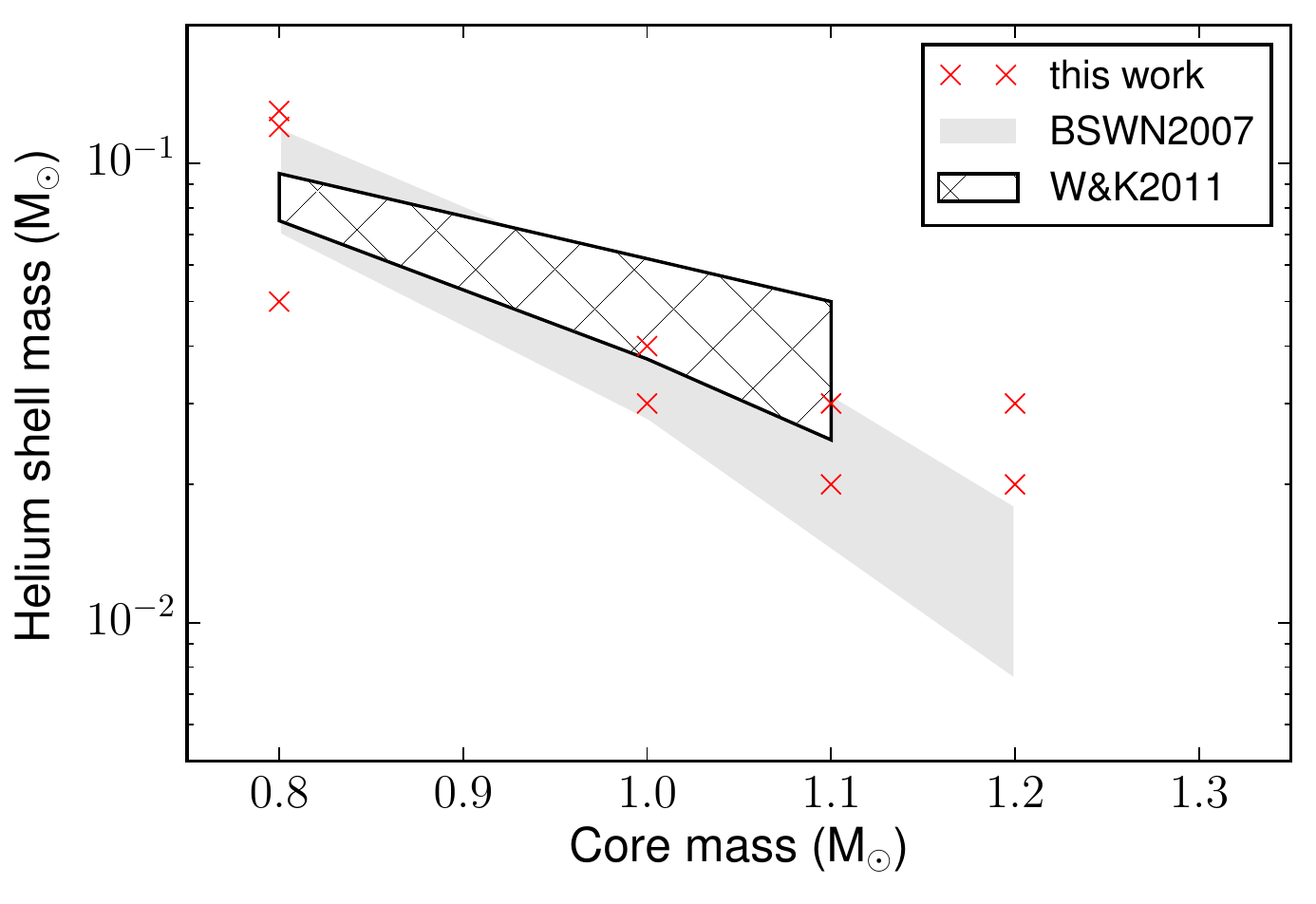}
   
   \caption{\label{fig:minmass} The crosses are the core-shell mass configurations modeled in this paper.  
   For comparison, the shaded region is the range of minimum shell masses capable of initiating ignition
   as given in Fig.~2 of \cite{BildstenEtAl2007} (using their $t_\mathrm{nuc} =
   t_\mathrm{dyn}, 10 t_\mathrm{dyn}$ lines).
   The hatched region is the range of minimum shell masses that yield either a deflagration or detonation
   as given in Fig.~19 of \cite{WoosleyKasen2011}.  The lower bound is for their ``hot'' models, while the
   upper is for ``cold'' models.} 
\end{figure}

Given these studies and the uncertainties involved we investigate
systems with $M_\mathrm{core} = 0.8, 1.0, 1.1, 1.2~\Ms$ and
a range of shell masses including $M_\mathrm{He} = 0.02 - 0.13~\Ms$.  
Our mass configurations are summarized in Fig.~\ref{fig:minmass} and
compared with the minimum shell masses estimated by others.  In choosing shell
masses we had to balance a desire to
model low-mass
shells near the lower limit of models that run away in 1D with a need
for the simulations to be computationally feasible.  Lower mass cores
can take many convective turnover times to reach runaway for minimum mass
helium shells whereas minimum mass shells can be difficult to resolve for
higher mass cores.  As a result, we have the points marked in
Fig.~\ref{fig:minmass} that track near the 1D lower limit but to
varying extents.  

Due to the importance
of $T_\mathrm{core}$ demonstrated in \cite{WoosleyKasen2011}, we
include models with $T_\mathrm{core} = 10^7, 10^8$~K.  Finally, we vary
$T_\mathrm{base}$ from 175 MK to 250 MK.  These interface temperatures are
intended to be roughly what we would expect a few to several convective
turnover times before runaway, based on both our own numerical experiments
and the 1D literature.  Table \ref{tab:modset}
lists the details of our model set.

While our models are motivated by the literature they are
not necessarily likely to be realized in nature and are not the result
of detailed stellar evolution calculations.  Our focus is on broadly
sampling the parameter space, characterizing the relationships between
parameters and possible explosive outcomes, and quantifying the
salient trends that emerge.  This will guide future work studying
particularly interesting parameter configurations using more realistic
initial models and detailed nucleosynthesis.

%==========================================================================
% Simulation Results
%==========================================================================
\section{Results}
\subsection{Outcomes}
\label{ssec:outcomes}
Our broad sampling of parameter space explores several different model
configurations.  We find a range of outcomes for these models:
localized runaway, quasi-equilibrium, and convective runaway.  The last
column of Table \ref{tab:modset} denotes the ultimate outcome of each
model.

\begin{figure}
   \centering
   \includegraphics[width=0.5\textwidth]{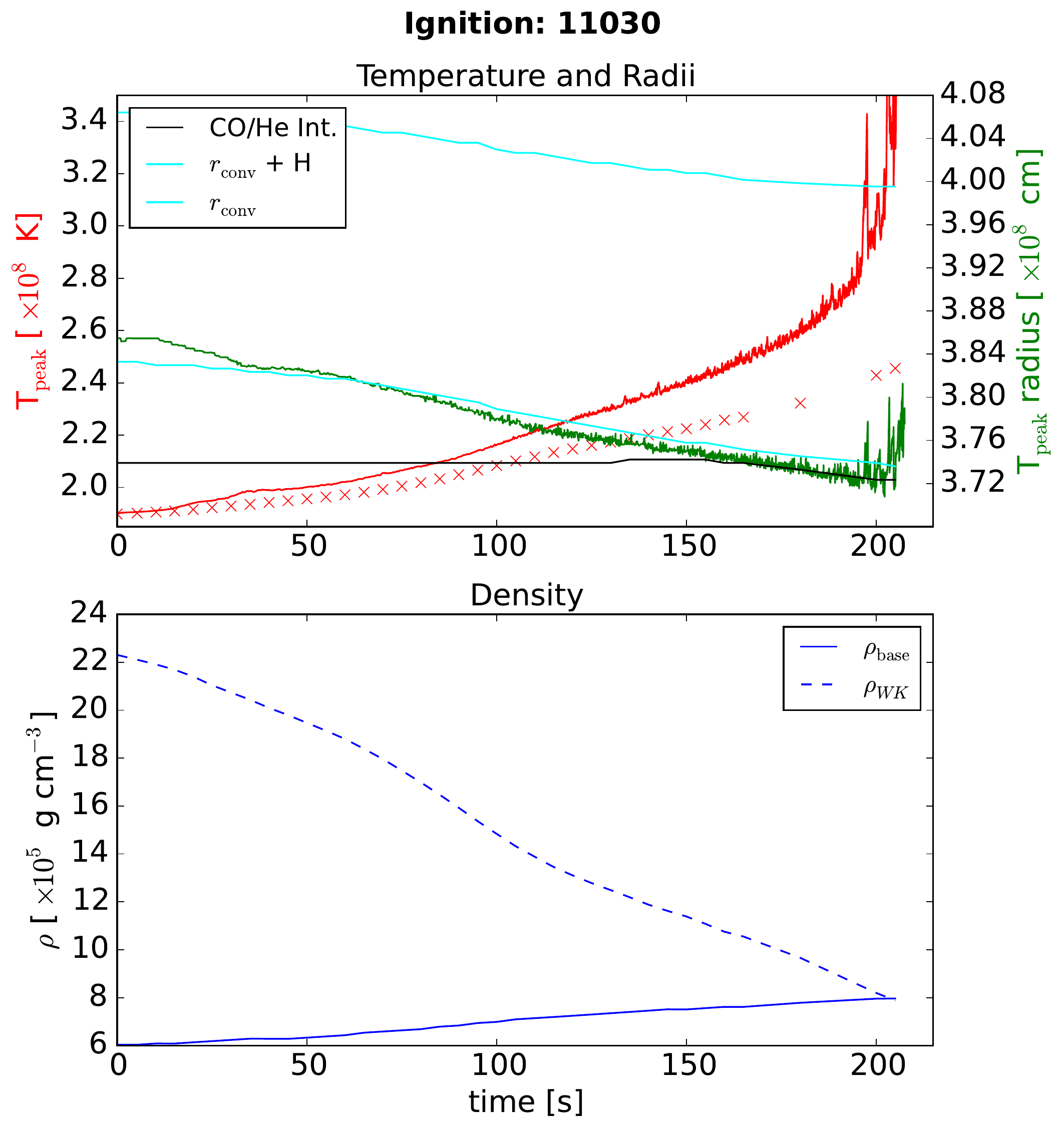}
   
   \caption{\label{fig:outcomes-ig} Temperature, radius, and density over time for model 11030.  
   {\bf Top:} Temperature is plotted in red.  The solid line is the temperature of the hottest cell in
   the entire computational domain.  The x's trace the peak laterally averaged temperature. The green, cyan,
   and black all plot radii.  Green is the radius of the hottest cell in the domain.  Black is a trace of the
   core/shell interface based on the radius at which the average $X_\mathrm{He}$ composition is 0.9. The cyan
   plots the base of the convective region and one pressure scale height above this base.  The inset
   is a 2D temperature slice centered on the site of runaway, demonstrating its
   localized nature. The runaway happens at a boundary, hence half the inset being
   white (no temperature data outside the boundary).
   {\bf Bottom:} The solid line plots the laterally averaged density at the radius of peak average temperature.
   The dashed line is the critical density above which ignition is expected according to Eq.~\ref{eq:critrho}.}
\end{figure}

Localized runaways represent possible seeds of deflagration or
detonation in the helium shell.  All runs in this category have
localized volumes of fluid that experience rapid temperature runaway to about 1
GK.  They also have peak Mach numbers less than 0.3 before runaway, and
less than 0.2 for the majority of the simulated time.
Fig.~\ref{fig:outcomes-ig} plots some of the key properties of interest over
time for an igniting run (model 11030).  This plot is representative of the
general behavior of runs experiencing localized runaway.  The plot demonstrates that the
temperature of the hottest cell in the domain\footnote{We track the cell with
   the largest temperature in the entire domain, but we caution this is not a
   Lagrangian measure---it is simply the hottest cell without regard to the cell's
   mass density} initially follows a trend similar to that of the laterally
averaged peak temperature.  As the model approaches runaway the hottest cell
increasingly deviates from the background conditions.  We also find that the
convectively unstable region moves deeper into the star.  This suggests that
vigorous burning and the convection it drives can result in significant changes
in the density and composition of burning sites (discussed more in \S
\ref{ssec:conv}). 

As one might expect, the radius of the hottest cell moves radially
inward along with the base of the convectively unstable region.
However, at the end we see that this radius moves outward in many models.  If
convection is able to transport an ignition seed to a significant
height above the core/shell interface then it makes ``edge-lit''
double detonation models workable.  In the edge-lit scenario, carbon
detonation of the core is triggered by the propagation of the helium
detonation wave at the surface of the core.  This model is generally
disfavored because it has been shown that it requires the initial
helium detonation to go off at a substantial height above the
core/shell interface \citep{GarciaSenzEtAl1999, WoosleyKasen2011}.  
If convection is more effective than expected at transporting the 
detonation seed then the edge-lit scenario needs to be considered 
more seriously.

Unfortunately, many of the localized runaway events in our models
happen near the boundary of the octant being simulated.  We stress
that we have carried out a full star run for the localized runaway
model 08130 (and also in paper~I for a 10050 model) and still find 
localized runaway as well as a radius significantly above the interface.  
We have also carried out simulations in which the 
temperature of cells is limited to be below 3.5 MK and see that many 
localized runaways occur far from the boundary, though the initial 
runaway happens preferentially at the boundary.  So while we are confident 
the localized runaway is not a boundary effect, the elevated radius of 
the hottest cell in fig.~\ref{fig:outcomes-ig} cannot be ruled out as 
a boundary effect in all runs.
This is an important issue that will be resolved in the next paper in
this series, which focuses on the timing, thermodynamics, and geometry
of ignition in this suite of simple models.

\begin{figure}
   \centering
   \includegraphics[width=0.5\textwidth]{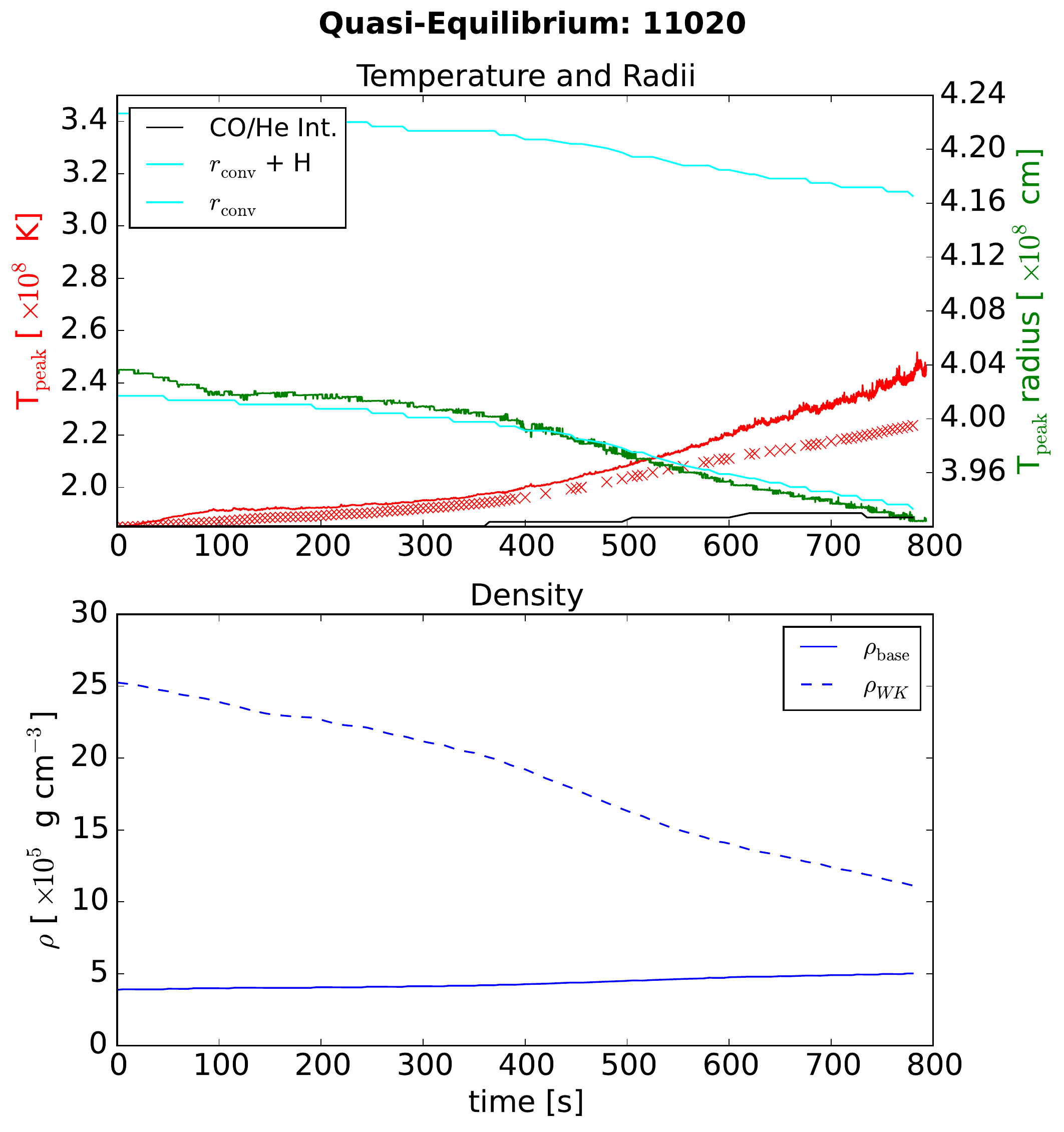}
   
   \caption{\label{fig:outcomes-qe} The same plot as in fig.~\ref{fig:outcomes-ig} for model 11020, 
   demonstrating quasi-equilibrium.}
\end{figure}

In contrast, quasi-equilibrium simulations balance nuclear burning
with convective cooling for many convective turnover times (at least an average
of 3.8 or more turnovers, see Table \ref{tab:modset} for a range of turnover
estimates and \S \ref{ssec:conv} for how we calculate these).
Fig.~\ref{fig:outcomes-qe} demonstrates this case for a model we ran for a
particularly long time.  We cannot say
these runs have reached an equilibrium between burning and convective
cooling because neither peak nor average base temperatures plateau,
nor do we model any energy sinks (cooling).  If we had the
computational resources to run these simulations indefinitely it may
be the case they would experience a runaway.  What we demonstrate
instead is that these models are stable against immediate runaway, 
i.e.~runaway within a few convective turnover times.

\begin{figure}
   \centering
   \includegraphics[width=0.5\textwidth]{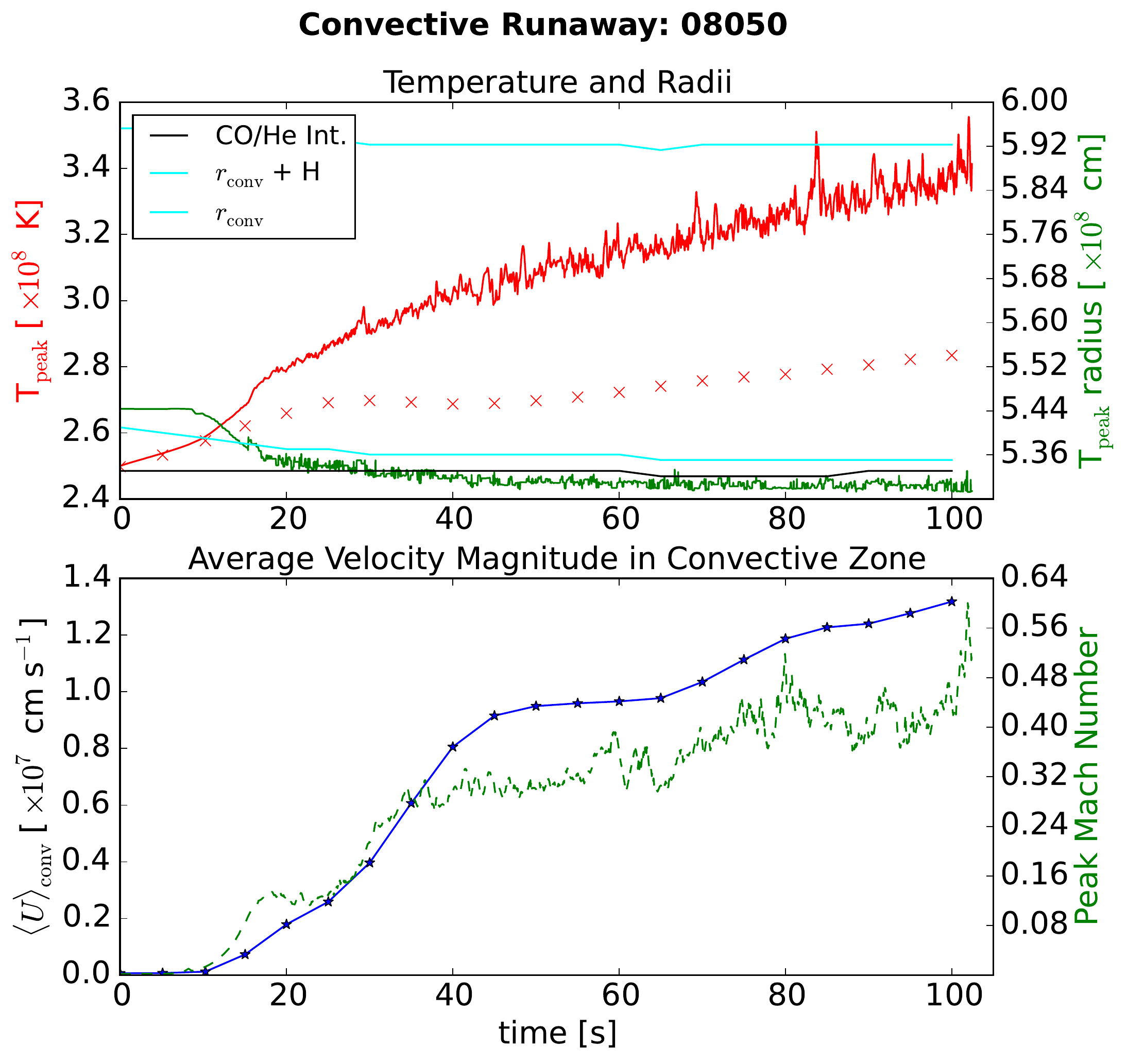}
   
   \caption{\label{fig:outcomes-conv} The top plot is the same as in fig.~\ref{fig:outcomes-ig} for model 08050.
   The bottom plot shows the average velocity magnitude in the convective zone as well as the peak Mach number 
   in the domain.}
\end{figure}

We have also included a model similar to \cite{WoosleyKasen2011}'s
model 8HB, which experiences a helium nova (convective runaway).
This model had a higher interface temperature than most runs, 2.5 MK
instead of around 1.9 MK, to facilitate reaching runaway conditions
without expending more computational resources than necessary.  Within
the low-Mach limits of \mae~we also find convective runaway, even with
the elevated interface temperature.  As the base temperature increases
from burning, the turnover rate of the convective shell is able to
increase without plateauing until the Mach number of the fluid gets
too large for us to track.  This suggests such a thin shell is able to
rapidly transport the energy release of nuclear burning.  In contrast to
localized runaway, this is a more global phenomenon and could develop into
something like a helium nova, as argued in \cite{WoosleyKasen2011}.
The time series data for this convective runaway is plotted in
Fig.~\ref{fig:outcomes-conv}.  The existence of two regimes,
convective and localized runaway, suggests researchers should
investigate the transition from one to the other.  The conditions of
this transition point will be important for determining the minimum
helium shell mass capable of achieving localized runaway.  

\subsection{Temperature}
While the mass of the core and helium shell play a primary role in
determining the thermodynamic conditions at the base of the shell,
there are secondary determinants.  Varying evolutionary histories can
result in accreting C/O WD primaries of varying temperatures.  A
history of helium flashes may heat the WD surface.  This enables
systems with similar mass configurations to have noticeable
differences in burning conditions at the core/shell interface.

Our parameterization of the initial model allows us to vary the
initial temperature of the actively burning base of the helium shell.
However, in our attempts at varying the base temperature we find only
a relatively narrow range of options can be feasibly explored with our
current methods.  Low initial temperatures will either not be able to
initiate sufficiently vigorous burning to allow a study of ignition or
will establish a trend of growing average base temperature that will
steadily build until either ignition or quasi-equilibrium is achieved.
However, the computational resources required to reach a dynamically
interesting stage of burning with a low initial base temperature can
be substantial.  Too high a base temperature will either lead to
unphysical ignition by not allowing time for convection to be
established or will push the convective velocity beyond \mae's ability
to model.  Thus, for a given model we choose an initial base
temperature low enough to allow for convection to be established and
for several convective turnover times of evolution, and high enough to
reach a scientifically interesting stage of burning while using
feasible amounts of computational resources.

The influence of the isothermal core's temperature is easier to
investigate with our methods.  \cite{WoosleyKasen2011} find that their 
synthetic spectra and light curves come closest to resembling observations of 
type Ia's for their ``hot''
models in which the accreting C/O WD relaxes into thermodynamic
equilibrium with a luminosity of $1 ~ \mathrm{L}_\odot$ before
accretion is modeled.  The hotter core enables runaway in thinner
shells than the colder core.  This motivates our exploration of our
own ``hot'' and ``cold'' ($10^8$ and $10^7 ~ \mathrm{K}$) models (hot
models are indicated with an 'H' in their label in
Tab.~\ref{tab:modset}).

\begin{figure}
   \centering
   \includegraphics[width=0.5\textwidth]{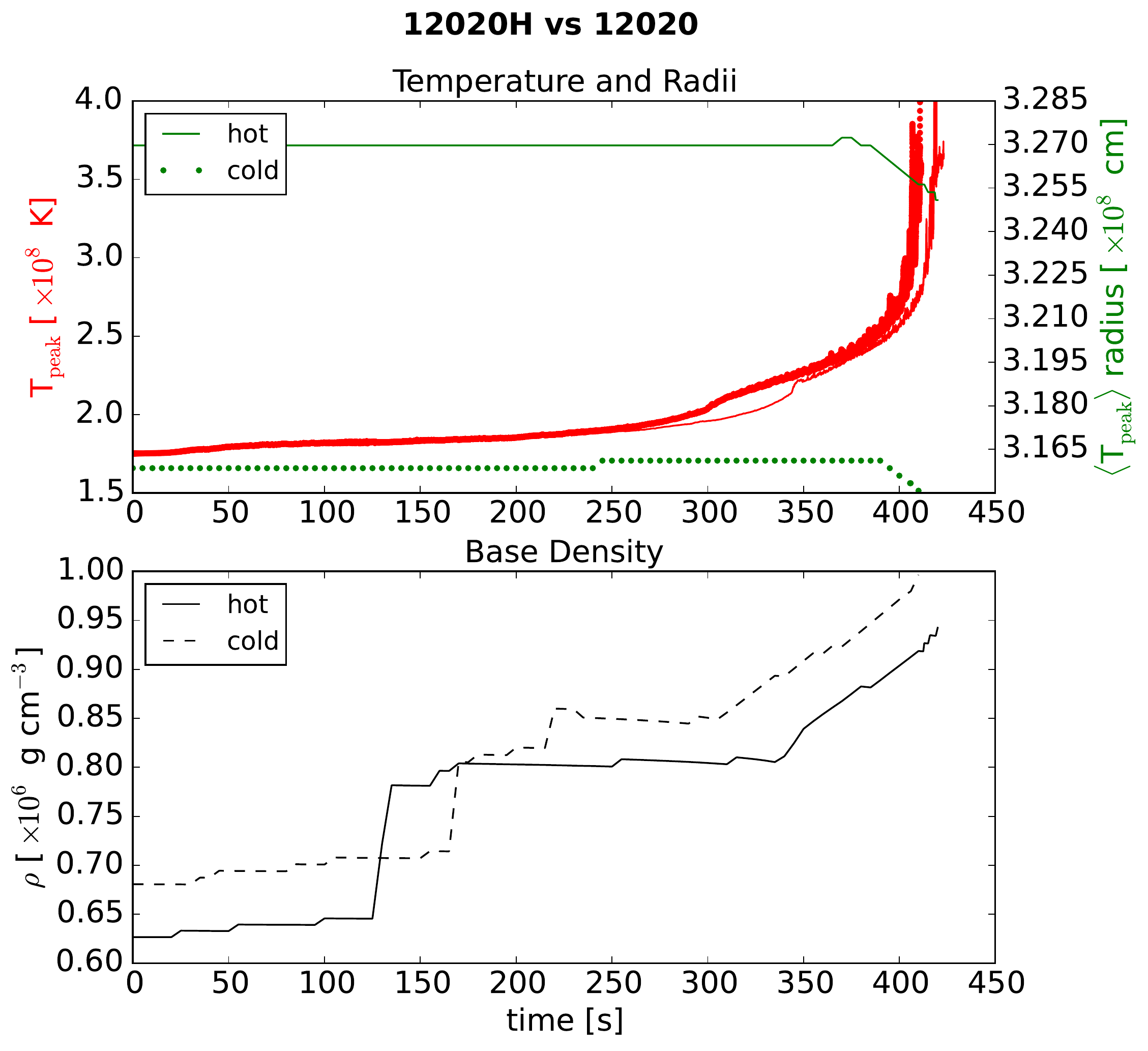}
   
   \caption{\label{fig:hvc} Comparison of several properties for a hot ($10^8$ K
   isothermal core, solid lines) and cold
   ($10^7$ K, dotted and dashed lines) model with a 1.2 $\Ms$ core and 0.02 $\Ms$ shell.
   {\bf Top:} In red we plot the global peak temperature.  In green we plot the
   radius of the temperature peak from a lateral average of the 3D data.
   {\bf Bottom:} In black we plot the laterally averaged density at the radius
   plotted in green in the top panel.}
\end{figure}

Our results are consistent with that of \cite{WoosleyKasen2011}.  Hot
cores allow for initiation of localized runaway at lower densities for
a given core/shell mass configuration.  This is largely due to an
expanded core radius in hot runs, and thus a lower density at the
core/shell interface where the burning occurs.  The lower density
favors higher temperatures at runaway as well.  In Fig.~\ref{fig:hvc}
we compare a hot and cold run to demonstrate these phenomena.

\subsection{Localized Runaway}
We have demonstrated that many of our models achieve localized runaway through 
bulk diagnostics and time-series
data, comparing them to 1D results.  This answers a major question we are
exploring: is ignition found in 1D codes consistent with 3D models?  We argue
the localized runaway we find is consistent, though do caution that localized
runaway should not be thought of as ignition.  The localized runaway reported
here may ignite deflagrations or detonations, but there is insufficient evidence
and analysis in this paper to make a definitive determination.  The next paper
in this series will focus in part on determining the likelihood of such
ignition.  For now, we move to characterizing the localized runaway found in our
models.

The 1D studies we have discussed necessarily model ignition as
simultaneous across a spherical shell. 
\cite{FinkEtAl2007}
contribute 2D simulations including a variety of
detonation seed geometries, following up later in \cite{FinkEtAl2010}
with 2D simulations seeding a single detonation in a larger range of core/shell
masses including thin shells. \cite{MollWoosley2013} have contributed 2- and 3D
studies in which detonation is seeded at multiple points with
variations in geometry as well as timing.  A key conclusion of 
these multi-D
investigations is that detonation in the helium shell very robustly
triggers detonation of the C/O core via radially propagating
compression waves generated by the helium detonation's shock front
traversing the shell\footnote{It is important to note the carbon
  detonation in these studies relies upon assuming certain conditions
  being achieved in a given computational cell will lead to
  detonation.  Current studies of the full core and shell system do
  not have the resolution to model the initiation of carbon detonation
  fully self-consistently. See \cite{ShenBildsten2014} for detailed
  carbon detonation calculations.}.
  
Assumptions about how many detonations to seed, where to seed them,
and their timing impact the ultimate outcome of the double detonation.
Single-point helium detonations lead to viewing-angle dependences not
expected in normal SNe Ia, though the dependence becomes weaker for
lighter helium shells \citep{KromerEtAl2010}.  If detonated at a great
enough altitude above the core/shell interface, helium detonations can
directly ignite carbon burning instead of detonating indirectly with
converging shock fronts \citep{GarciaSenzEtAl1999, WoosleyKasen2011, MollWoosley2013}.
The size and shape of the detonation can also impact how easily core
detonation can be triggered, especially for the thinner shells
considered to be the most promising candidates for modeling normal SNe Ia
\citep{MollWoosley2013}.

\begin{figure*}
   \centering
   \includegraphics[width=0.95\textwidth]{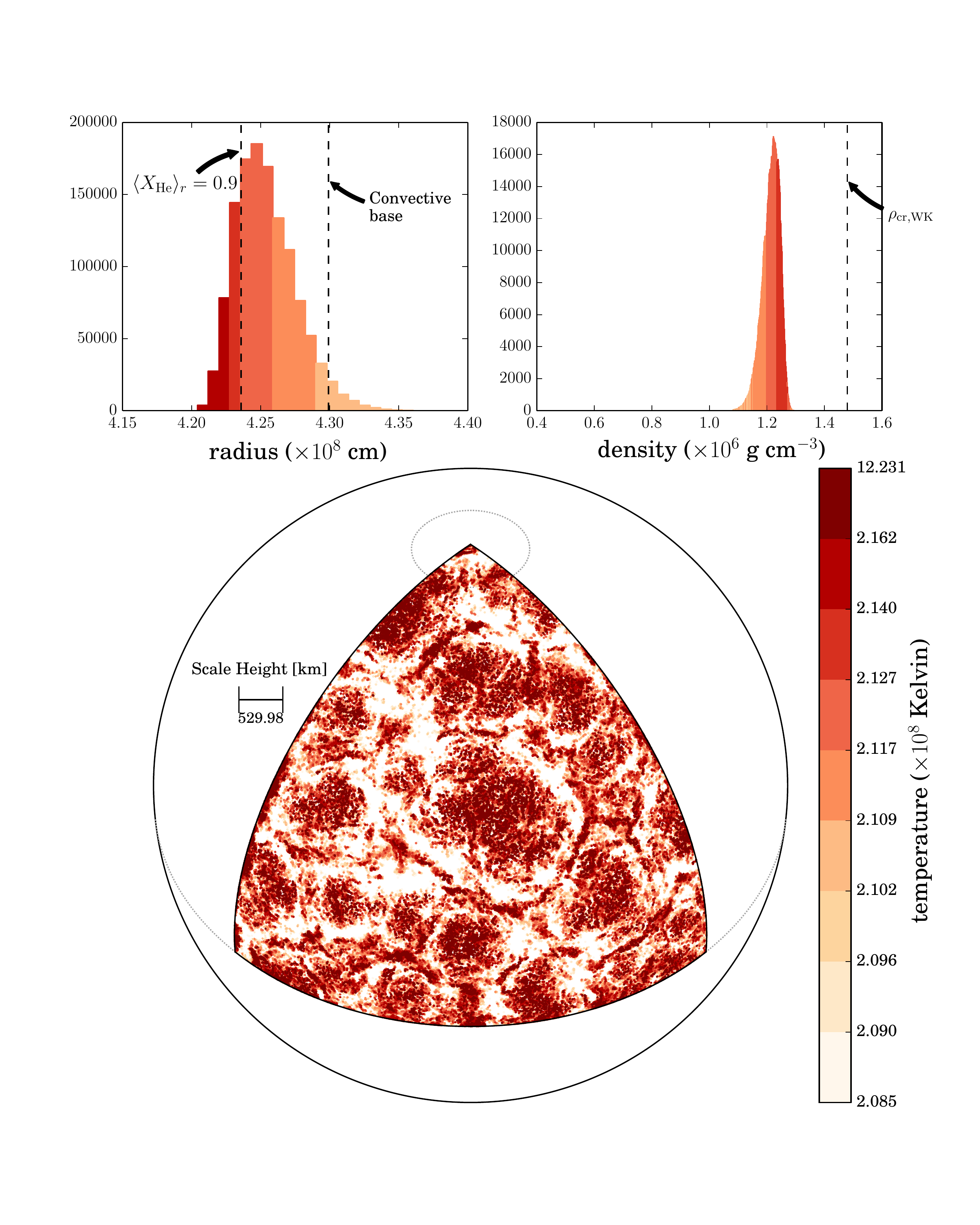}
   
   \caption{\label{fig:hs} Hotspot properties for model 08130 at
   $t=130.0$s.  {\bf Upper left:} histogram of the radii of the
   hotspots, with bin sizes chosen to match the finest level of 3D resolution.  Each bin is color-coded to
   indicate the \emph{maximum} temperature in the bin.  The location where helium's laterally averaged
   mass fraction is 0.9 and the base of the convective envelope are indicated.  {\bf Upper right:} histogram of the
   densities of the hotspots.  Again, bins are color-coded based on \emph{maximum} temperature.  
   The value of Eqn.~\ref{eq:critrho} is indicated.  {\bf Bottom:} projection of the hotspots' angular location
   onto a sphere corresponding to the average radius of the hotspots.  Hotspot pixel sizes correspond to the finest 
   level's physical resolution,
   and where hotspots overlap preference is given to the one with the highest temperature.  The extents of
   the temperature color bar's bins are set such that each bin contains an equal number of hotspots.} 
\end{figure*}  

In light of this, what do our models suggest?  To assess the conditions that
foster ignition we track the hottest
0.005 \% of cells in the computational domain immediately prior to
runaway.  Fig.~\ref{fig:hs} plots histograms of the radii and
densities of these cells in addition to a spherical projection of
their angular locations for model 08130 (see Table~\ref{tab:modset}).
The spherical projection illustrates the two
regions in which volumes of hot fluid develop: at the base of
convective inflows and at the intersection of outflows from
neighboring convective cells (see \S \ref{ssec:conv}).  This is determined by 
contrasting projections like that in Fig.~\ref{fig:hs} with renderings of
convective outflow like that in Fig.~\ref{fig:3dconv}.  The density
histogram includes a reference for a critical runaway density given
by Eq.~8 in \cite{WoosleyKasen2011}:

\begin{equation}
   \label{eq:critrho}
   \rho_\mathrm{cr,WK} = \left( 1.68 \times 10^{-4} ~ \exp(20/T_8) \right)^{1/2.3},
\end{equation}
where $T_8 = T/10^8$ is the temperature in units of $10^8$ K.  This is
a rough critical density above which violent, hydrodynamic runaway is
expected and below which convection is expected to be efficient enough
to transport any energy generated by nuclear burning.
For the purposes of this paper, we
denote it as a critical density above which we expect localized
runaway to be possible.

This density is based on 1D models and thus is only fairly compared to
lateral averages of quantities in our 3D simulations.  Note that
Fig.~\ref{fig:hs} demonstrates localized runaway is achieved even
though the hottest cells have densities significantly below this
critical density.  This is not surprising as the cells trace a 3D
model.  When we consider laterally averaged quantities,
Eq.~\ref{eq:critrho} is often an excellent predictor of ignition.
Ignition is achieved almost exactly as the average density at the peak
burning region surpasses the (temperature-dependent) critical density
in Fig.~\ref{fig:outcomes-ig}, whereas the same average density is
well below the critical value in the quasi-equilibrium case of
Fig.~\ref{fig:outcomes-qe}.

This predictor does not work as well for our models with 0.8 $\Ms$
core masses.  These models achieve localized runaway despite being
quite a bit below the critical density.  This could either be an
effect of our models capturing the 3D dynamics and thus of
scientific interest, or it could be a result of our models being toward
the upper limit of the minimum shell mass the critical density is
calculated for in our 0.8 $\Ms$ models.
In addition, the critical density is only a rough estimate based on
the outcomes of several 1D models.  Still, for cores $\ge 1.0~\Ms$ it
is quite accurate for our models.  We will further investigate this
critical density and determine a 3D version in the next paper of this
series.

Table~\ref{tab:igcon} lays out key properties at the time of runaway
for all models that experienced localized runaway.  The table includes an
estimate of the number of convective turnover times modeled before ignition as
well as the lateral averages of temperature and density at the radius of peak
burning ($r_\mathrm{peak}$).  Note that these are values based on the 3D output
with a timestamp nearest that of the peak temperature.  This is why the peak
times tend to end on round numbers---our 3D state is output at most every tenth
of a second.

\begin{deluxetable}{ccccccc}
\tabletypesize{\scriptsize}
\tablecolumns{6}
\tablewidth{0pt}
\tablecaption{\label{tab:igcon} Ignition Conditions}
\tablehead{
   \colhead{model}      & \colhead{$t_\mathrm{peak}$} & \colhead{$t_\mathrm{peak}/\langle \tau_\mathrm{conv} \rangle$}  & \colhead{$\langle r_\mathrm{base} \rangle$}  & \colhead{$\langle T_\mathrm{base} \rangle$}  & \colhead{$\langle \rho_\mathrm{base} \rangle$}  \\
\colhead{}              & \colhead{[$s$]}             & \colhead{}                                                      & \colhead{[km]}               & \colhead{[$\times 10^8~K$]}                  & \colhead{[$\times 10^5~\gcc$]}                  \\ \hline\noalign{\smallskip}
}                                                                                                                                 
\startdata                                                                                                                        
%                       & t_peak                      & t_peak/t_conv                                                   & <r_base>                       & <T_peak>                                     & <\rho_peak>                                    
 12030H                 & 118.2                       &  8.4                                                            & 3096.4                       &  2.180                                       & 13.877                                          \\
 12030                  & 120.4                       &  6.4                                                            & 3009.4                       &  2.178                                       & 14.541                                          \\
 12020H                 & 420.0                       & 10.3                                                            & 3261.6                       &  2.480                                       &  9.424                                          \\
 12020                  & 410.0                       & 12.4                                                            & 3163.7                       &  2.410                                       &  9.963                                          \\
 11030H                 & 319.5                       & 10.6                                                            & 3839.8                       &  2.462                                       &  7.729                                          \\
 11030                  & 205.0                       &  8.5                                                            & 3739.1                       &  2.456                                       &  7.968                                          \\
 08130H                 & 161.7                       &  3.7                                                           & 4372.9                       &  2.139                                       & 11.464                                          \\
 08130F                 & 137.4                       &  1.2                                                           & 4251.7                       &  2.096                                       & 12.218                                          \\
 08130                  & 130.6                       &  3.3                                                            & 4259.6                       &  2.069                                       & 12.106                                          \\
 08120H                 & 240.0                       &  7.7                                                            & 4470.0                       &  2.181                                       & 10.365                                          \\
 08120                  & 710.0                       & 25.7                                                            & 4338.7                       &  2.041                                       & 11.221                                          \\
\enddata
\end{deluxetable}  

\subsection{Convection}
\label{ssec:conv}
The ignition characterized in the previous section is fostered by the complex interplay between
nuclear burning and convective dynamics.  A detailed understanding of convective evolution is crucial,
as illustrated by the conflicting results of \cite{FinkEtAl2010} and \cite{WoosleyKasen2011}.

\cite{FinkEtAl2010} get their initial 1D models from
\cite{BildstenEtAl2007}, who assume a fully convective shell all the
way up to the point of ignition.  \cite{WoosleyKasen2011} employ a
time-dependent convective model based on mixing-length theory,
allowing for convection to ``freeze out.''  If the runaway timescale
for a volume of fluid at the base of a convective zone, in the absence
of cooling, becomes smaller than the convective turnover timescale,
convection is no longer able to cool the helium-burning layer with the
same efficiency.  It starts to ``freeze out.''  A fully convective
shell requires a larger temperature at its base to achieve runaway
than with one in which freeze-out is allowed.  The temperature
differences translate into entropy differences.  In turn, the lower
entropy of the cooler base makes possible larger base densities in the
\cite{WoosleyKasen2011} models than those of \cite{FinkEtAl2010}, even
when they are modeling similar core and shell masses.  These density
discrepancies lead to significant discrepancies in explosive burning
and the products yielded.

\begin{figure*}
   \centering
   \includegraphics[width=\textwidth]{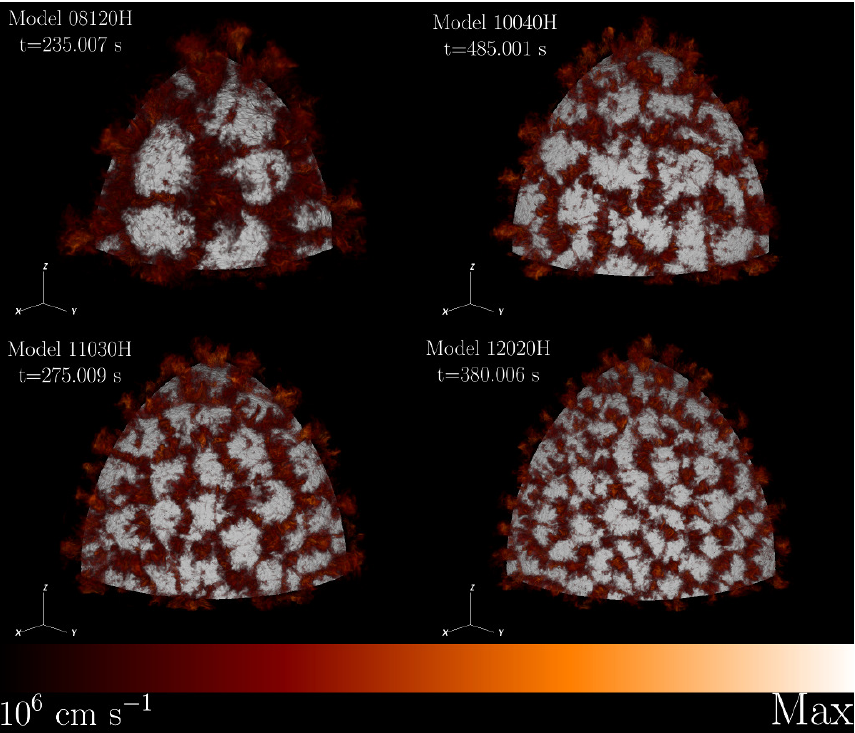}
   
   \caption{\label{fig:3dconv} Volume rendering of radial velocities for some representative models. 
   Maximum velocities are 2.8, 1.3, 2.0, and 1.7 $\times 10^7 ~ \mathrm{cm ~ s}^{-1}$ for models
   08120H, 10040H, 11030H, and 12020H, respectively.}  
\end{figure*}

Our analysis of convective dynamics begins by establishing the broad
context.  Fig.~\ref{fig:3dconv} plots volume renderings of radial
velocity for several models.  These act as a proxy for convective
plumes.  Higher core masses result in more compact systems with
smaller pressure scale heights.  Thus we see the typical size of a
convective cell grows inversely proportionally to core mass.  The
hotspots plotted in Fig.~\ref{fig:hs} occur at the base of convective
plumes of cool in-falling fluid and at regions in which outflows of
adjacent convective cells collide.  The in-falling matter increases
the density, setting off more vigorous nuclear burning.  A competition
between the nuclear burning rate of a volume of fluid and the rate at
which it can be transported and cooled by convective flow is
established.

In our models we see convective dynamics have two key impacts.  First,
convective overshoot serves to push the region of active burning
deeper into the star, thus increasing the ambient density of nuclear
burning sites and altering the ambient composition.  Second,
convection's ability to respond to increasing nuclear energy
generation breaks down for sufficiently energetic burning.  This is
the freeze-out discussed in \cite{WoosleyKasen2011}.

\begin{figure}
   \centering
   \includegraphics[width=0.5\textwidth]{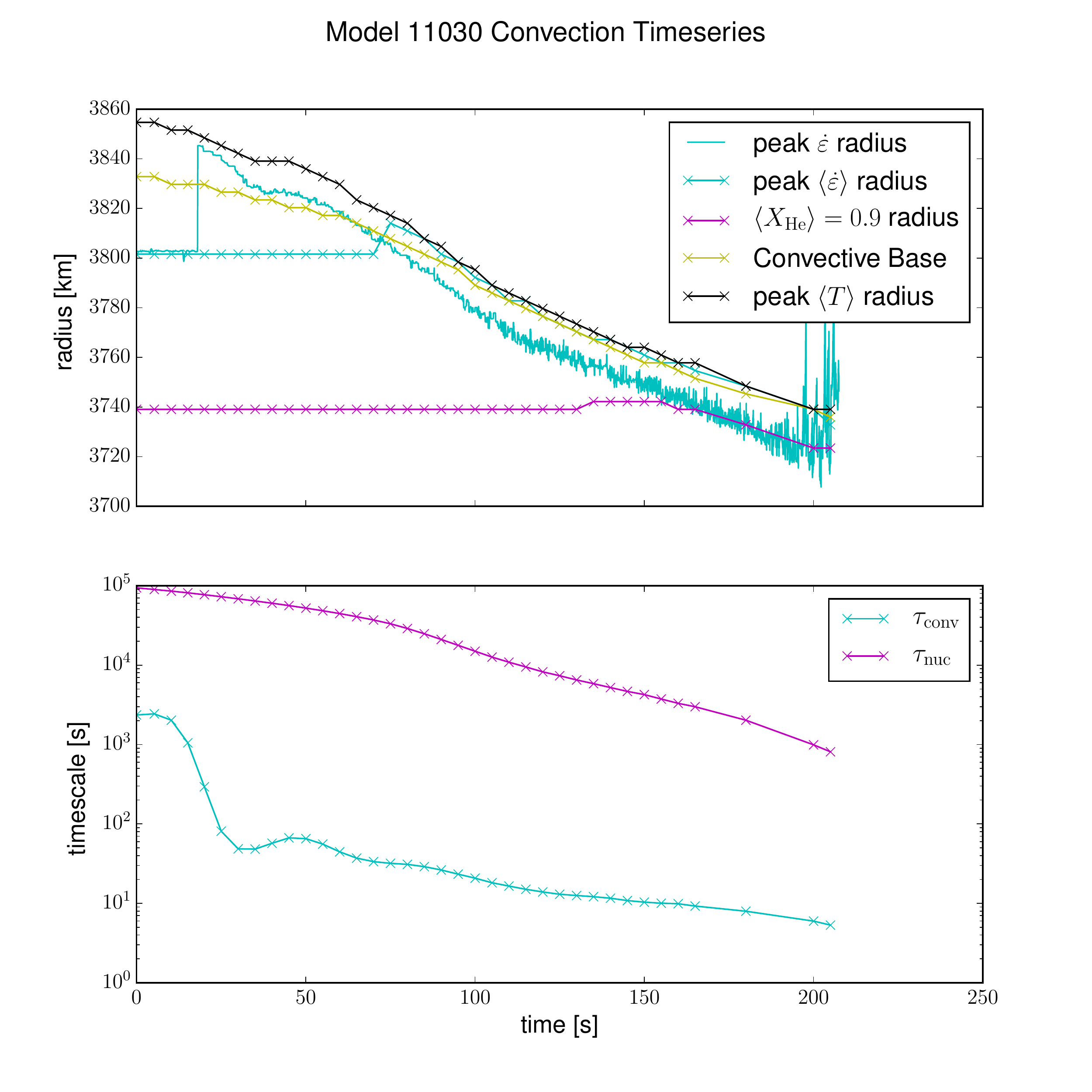}
   
   \caption{\label{fig:ct} {\bf Top:} Here we plot: the radius of the
   cell with the largest nuclear energy generation in the entire domain,
   $r(\dot{\epsilon}_\mathrm{peak})$, in cyan; the radius of this same quantity for
   the laterally averaged data, also in cyan with ``x'' markers indicating the
   timestamp the data was calculated for; the radius at which the
   lateral average of helium's mass fraction $\langle X_\mathrm{He} \rangle$ is 0.9, in magenta;
   the radius of the convective base
   as determined by the radius at which the entropy flattens out (see
   Fig.~\ref{fig:initmod}), in yellow; 
   and the radius at which the lateral average of temperature $\langle T \rangle$ 
   peaks, in black. {\bf Bottom:} A comparison
   of a nuclear burning timescale (magenta) with a convective turnover timescale (cyan).  See 
   text for details.} 
\end{figure}

Convective overshoot is demonstrated in the top plot of Fig.~\ref{fig:ct} for
model 11030.   The first process to happen in a model is for convection to be
established in response to nuclear burning, the bulk temperature profile, and
our initial velocity perturbations (see Fig.~\ref{fig:initmod}).  This is seen
in the plot of the convective timescale at the bottom of Fig.~\ref{fig:ct}.
Initially it experiences a steep drop off until stabilizing as convective cells
are established.  After this we see that the location of the convective base
steadily moves radially inward. In
response, the radius of peak burning moves deeper into the star resulting in
increased ambient density as well as changes in ambient composition.  We find in
most models that experience localized runaway, peak burning radii tend to
stabilize near the location where composition is 90\% helium (shell material),
10\% carbon (core material).  In addition, models in
quasi-equilibrium do not have such substantial radial migration of the
convective base.

Freeze-out is demonstrated in the bottom of Fig.~\ref{fig:ct}.  Here we plot two
different timescales.  To calculate a minimum nuclear burning timescale we
invert the burning rate, $dX_i/dt$, for the most rapidly burning species $i$
(for these models, helium).  This is done for all radius bins $r$ in our lateral
averaging.  The minimum value of this radial slice is used as our nuclear
timescale $\tau_\mathrm{nuc}$.  In sum, 
\begin{equation} 
   \tau_\mathrm{nuc} =
   \min_{r} \left( \min_{i} \left\langle \frac{dX_i}{dt} \right\rangle^{-1}
   \right)_r.  
\end{equation} 

The other timescale is a conservative estimate of the convective
turnover time.  Again using laterally averaged data, we invert the velocity
magnitude $\langle |\mathbf{U}| \rangle$ and integrate over the convective
region, 
\begin{equation} 
   \tau_\mathrm{conv} = \int_{r_b}^{r_t} \left\langle
   \left| \mathbf{U}(r) \right|\right\rangle^{-1} dr.  
\end{equation}
$r_b$ is the smallest radius at which the radial entropy profile satisfies
$ds/dr = 0$ and $r_t$ is largest radius satisfying this condition.  The region
is shaded in Fig.~\ref{fig:initmod}.  This average value is reported in Table
\ref{tab:modset}.  In addition, Table \ref{tab:modset} includes a calculation
of the minimum and maximum estimates for turnover times by simply dividing the
lengthscale $r_t - r_b$ by the maximum and minimum velocities in the convecting
region, respectively.  We want to stress these are conservative estimates.  Not
all plumes that form will extend the full lengthscale, and there may be plumes
with faster fluid than in other plumes.

In the timescale plot of Fig.~\ref{fig:ct} we see that as model 11030 approaches
runaway the nuclear burning timescale drops more rapidly than the convective
turnover timescale.  We note that the focus is not on the magnitudes of the
timescales but their changes.  The highly nonlinear nature of nuclear burning
make comparisons of the rate's magnitude at any given time with other timescales
uninformative.  The changes however suggest the nuclear burning rate is
shrinking faster than the convective turnover rate as runaway is approached,
suggesting a degree of freeze-out.

In contrast, quasi-equilibrium models and the convective runaway model have
relatively flat peak burning radii. Their convection is able to respond to
nuclear burning before any localized runaway can develop.

\subsection{Varying $\delta$ and Comparison with a 1D Model}
\label{ssec:delta}

\begin{figure*}
   \gridline{
      \fig{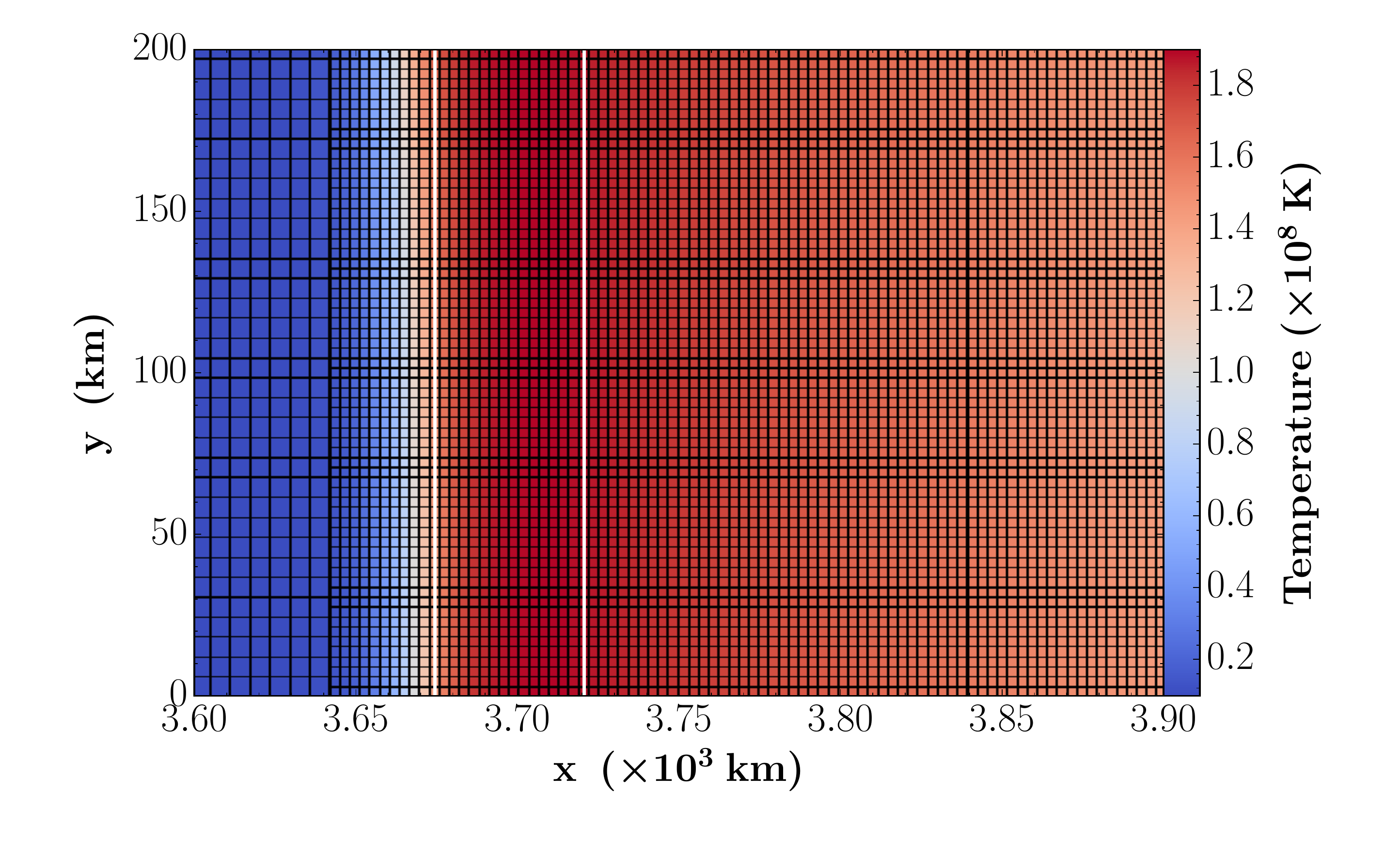}{0.5\textwidth}{(a) 11030d0.25 $t=0$ s}
      \fig{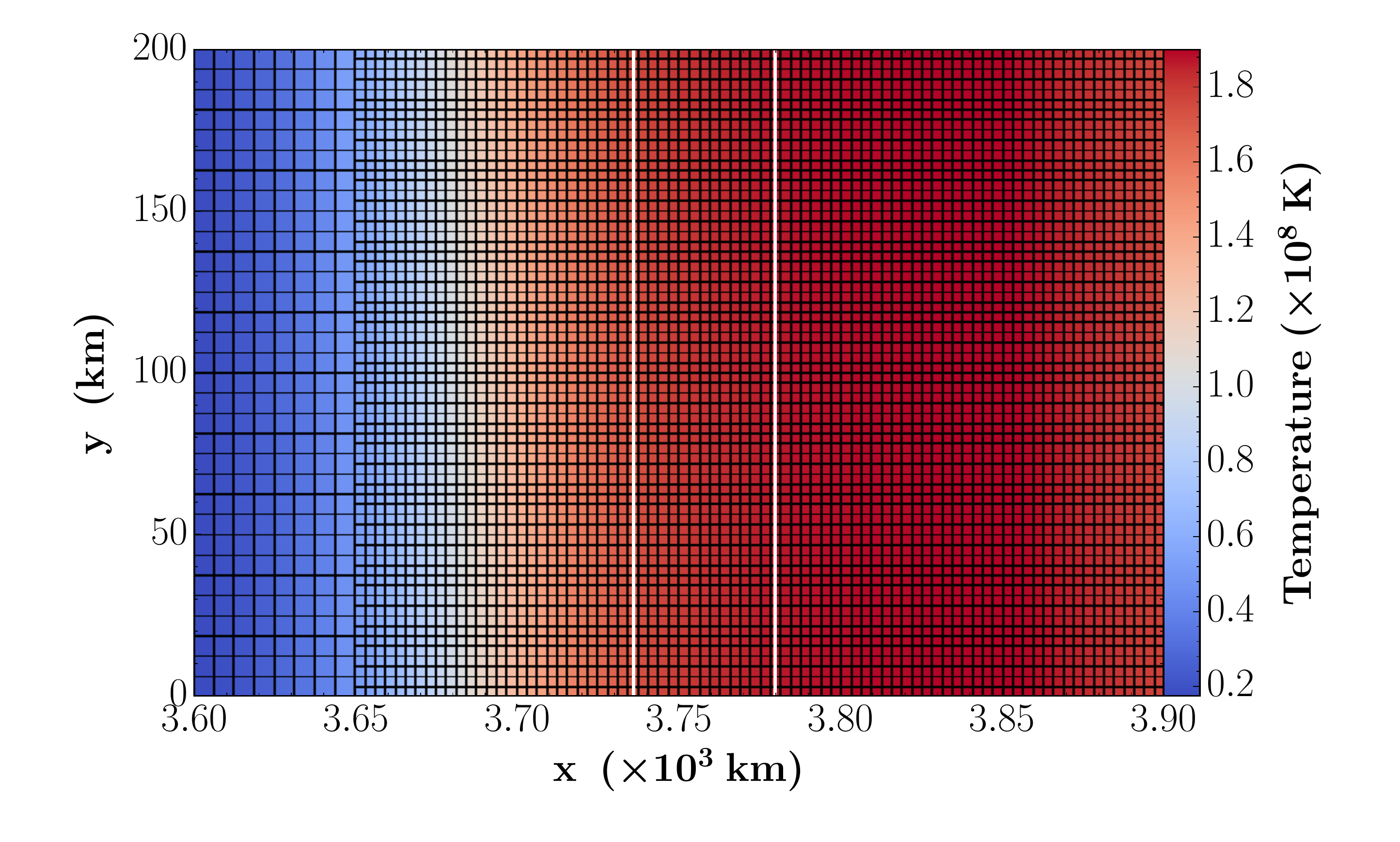}{0.5\textwidth}{(b) 11030 $t=0$ s}
   }
   \gridline{
      \fig{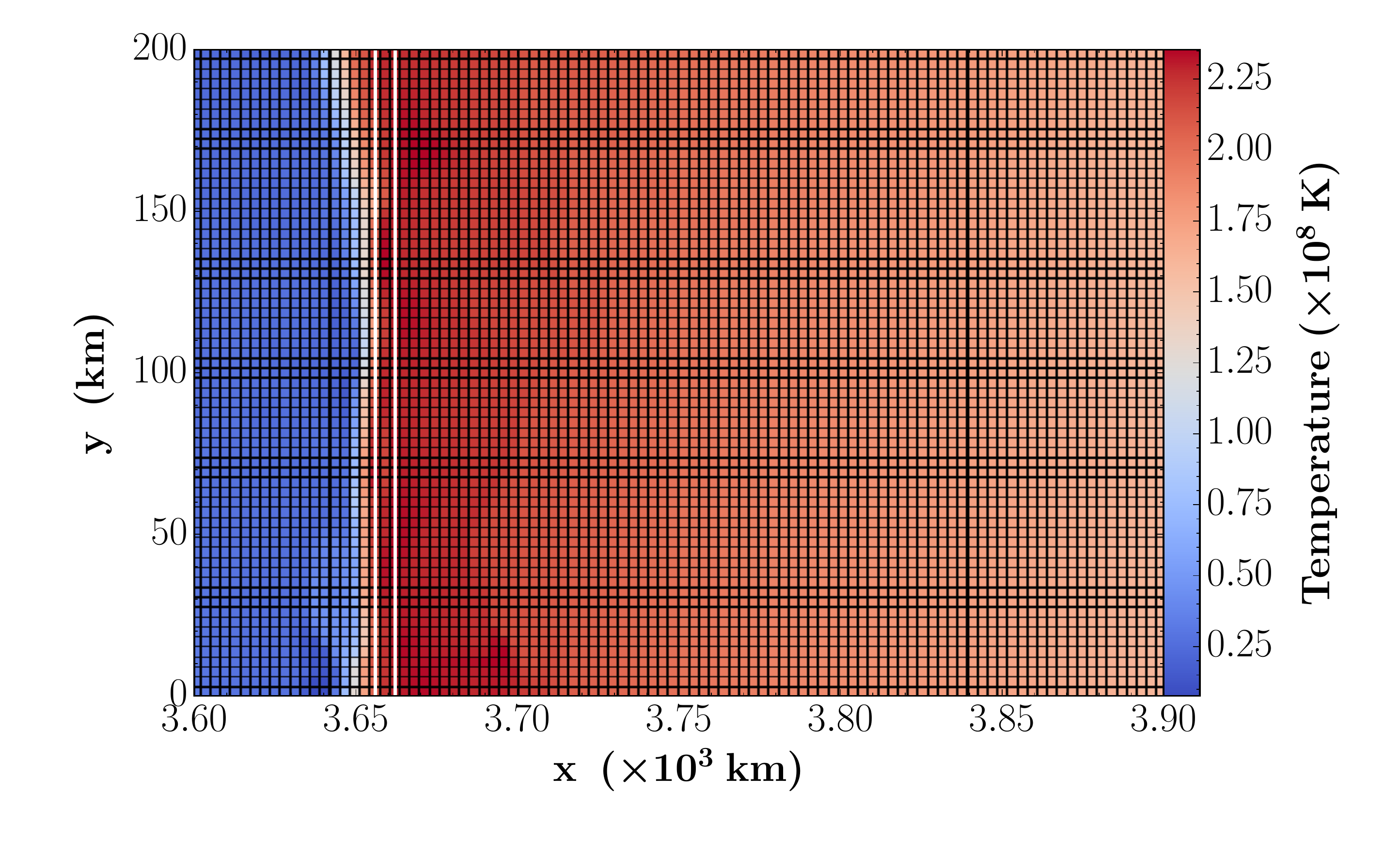}{0.5\textwidth}{(c) 11030d0.25 $t=375$ s}
      \fig{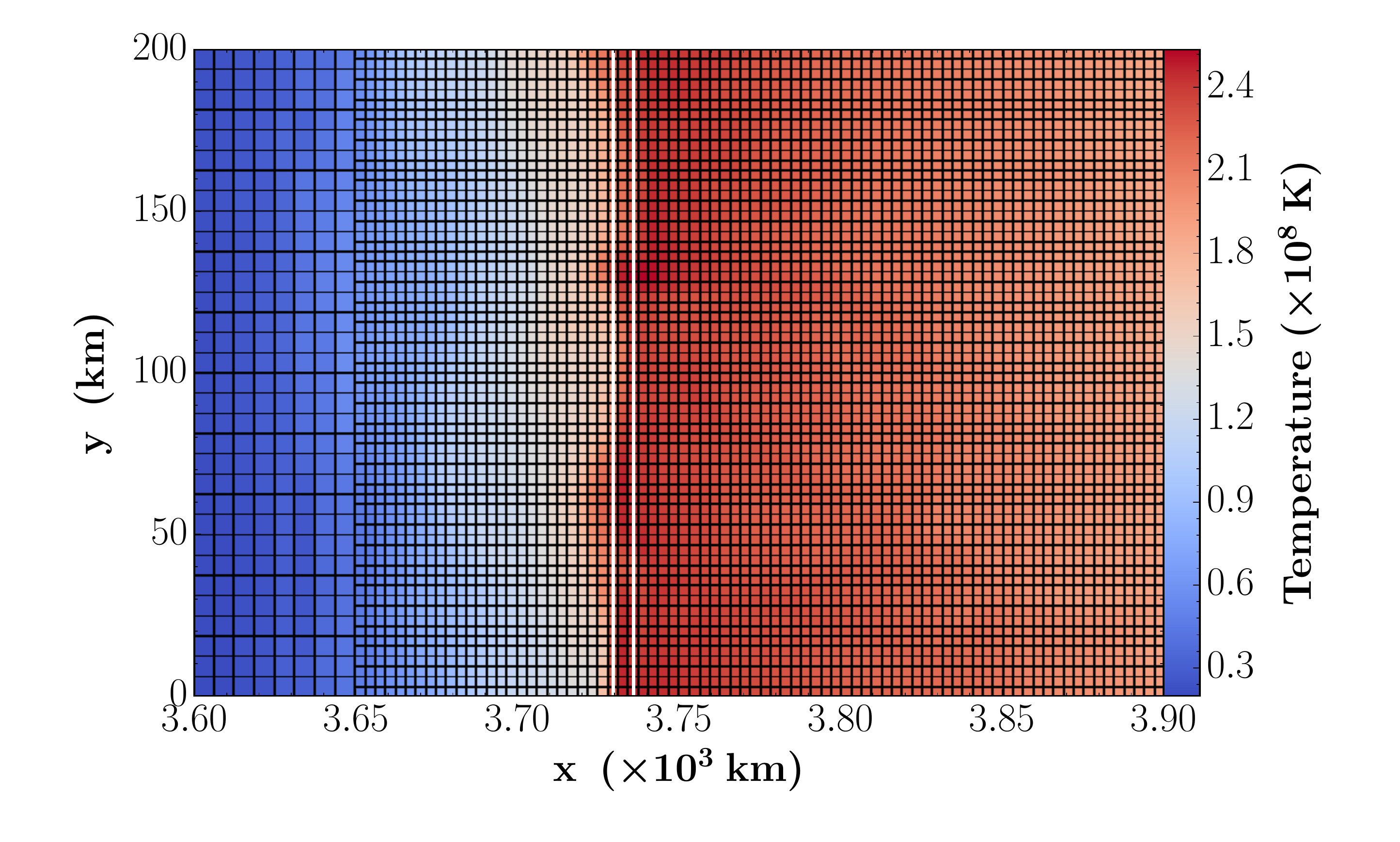}{0.5\textwidth}{(d) 11030 $t=180$ s}
   }
   \caption{\label{fig:temp_trans} Zoomed-in X-Y slices of temperature at the
      base of the convecting helium envelope for models 11030 and 11030d0.25 at
      initial and late times (see subplot labels). Cell edges are overlayed. The
      two white vertical lines mark, left-to-right, the radius at which the lateral
      average of the velocity magnitude is 25\% and 50\% of its peak. }

\end{figure*}

\begin{figure}
   \centering
   \includegraphics[width=0.5\textwidth]{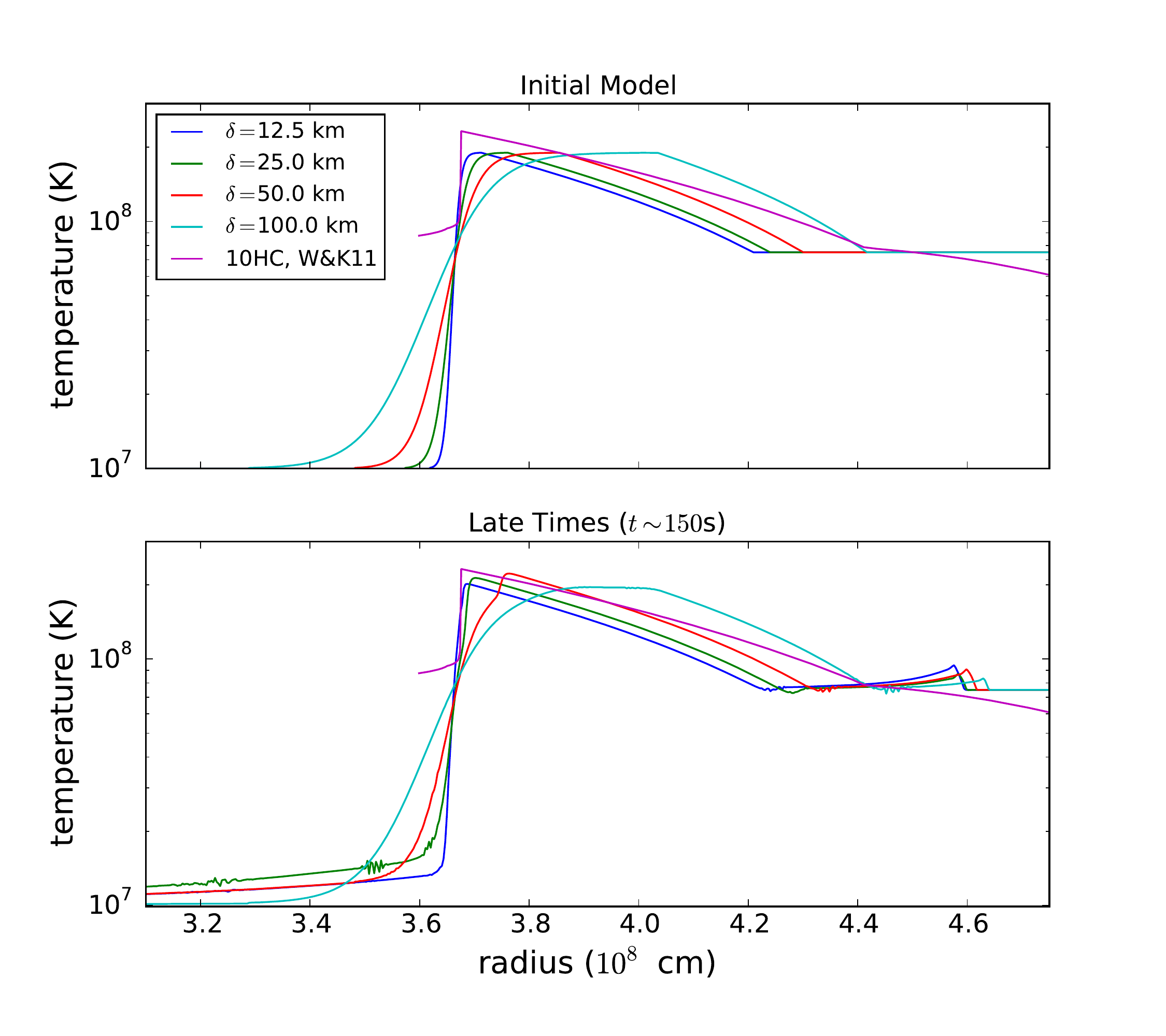}
   
   \caption{\label{fig:deltatemp} Comparison of temperature profiles for varying
   $\delta$.
   {\bf Top:} The temperature profiles of initial models with varying $\delta$ along
   with a realistic 1D reference (model 10HC from \cite{WoosleyKasen2011}, radius
   shifted $-500$ km to facilitate comparison).
   {\bf Bottom:} The same reference model along with profiles for the laterally
   averaged temperature data from models 11030d0.25, 11030d0.5, 11030, 11030d2
   after $\sim 150$s of evolution.}
\end{figure}

As discussed in \S \ref{ssec:initmod}, we have carried out an additional
numerical experiment exploring the impact of varying the $\delta$ parameter that
determines the sharpness of the transition from core to shell (see Paper I for
details).  In Fig.~\ref{fig:deltatemp} we plot the temperature profiles of
models 11030d0.25, 11030d0.5, 11030, and 11030d2, which have identical initial
models except for their $\delta$ parameters: 12.5, 25, 50, and 100 km,
respectively.  In addition, we plot the temperature profile of model 10HC from
\cite{WoosleyKasen2011} at a time near ignition (model data courtesy of Woosley,
private communication).  The radius of 10HC is offset by 500 km to facilitate
comparison of transition widths.  We plot both the initial temperature profile
and those from later times.  Initially, our default $\delta$ value results in a
more gradual transition than in 10HC.  At later times, the temperature profile
develops a more pronounced peak like that of 10HC, though the 
transition continues
to be less sharp than 10HC and leaves a substantial amount of hot fluid
below the temperature peak.  We see that our smaller $\delta$ values, in
particular that of 11030d0.25, are about as sharp as that of 10HC.

One potential worry with the thicker transition is that it allows a thin shell
   of hot fluid (80-90\% of the peak interface temperature) to exist below the region
   initially unstable to convection (see Fig.\ref{fig:initmod}).  If convection is 
   not well-established in these hot
   shells as the model approaches runaway then it becomes hard to determine if
   the runaway is a result of the uncooled hot shell in some initial models
   or the convective dynamics.
   To address this concern, we plot slices of 
   temperature at $t=0$ and late times for 11030d0.25 and 11030 in 
   Fig.~\ref{fig:temp_trans}.  This plot
   includes white vertical lines marking the radius at which the laterally
   averaged velocity magnitude is 25\% and 50\% of its peak value.  This gives
   the reader an idea of how much convective cooling penetrates.  The late-time
   temperature profile develops into a thin, hot layer for both $\delta$ values,
   with similar velocity penetration.  This demonstrates that even with a
   thicker transition the cooling in our models penetrates to the thin shell of
   vigorous burning.  The primary impact of the smaller
   $\delta$ is to shift the radius at which the thin, hot layer develops.  This
   is an expected imprint of the initial model, as the thinner $\delta$ locates
   a more concentrated shell of hot fluid at a lower radius, as seen in
   Fig.~\ref{fig:temp_trans}.  In addition, the thicker $\delta$ results in a
   larger region of intermediate temperature below the thin, hot layer.
   Finally, we note that a history of helium flashes can heat the base of the
   convecting envelope.  Nature may in fact realize configurations in which hot
   fluid exists below the convectively unstable region.

\begin{figure}
   \centering
   \includegraphics[width=0.5\textwidth]{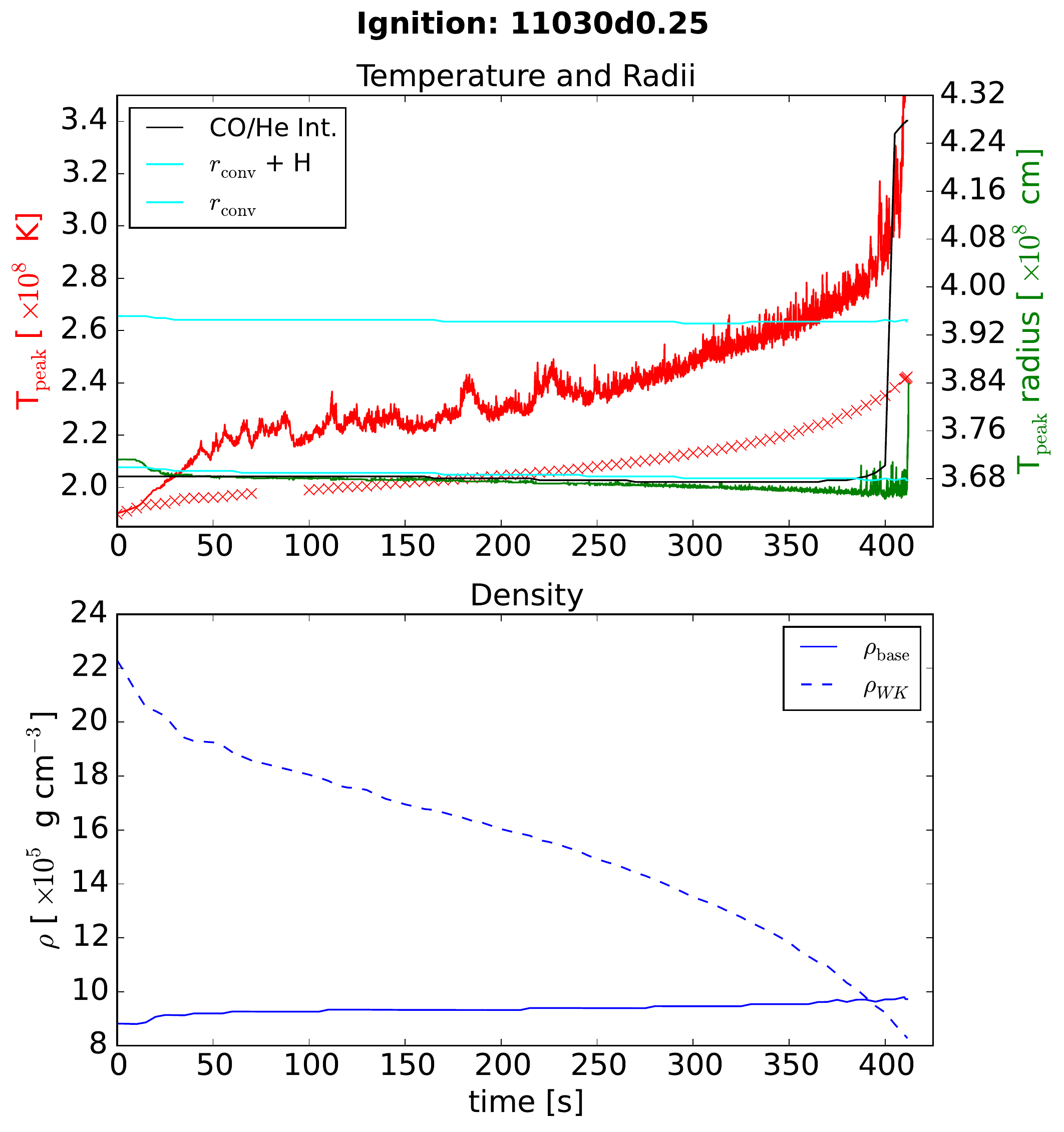}
   
   \caption{\label{fig:deltaig} Same as Fig.~\ref{fig:outcomes-ig} for model 11030d0.25} 
\end{figure}

\begin{figure}
   \centering
   \includegraphics[width=0.5\textwidth]{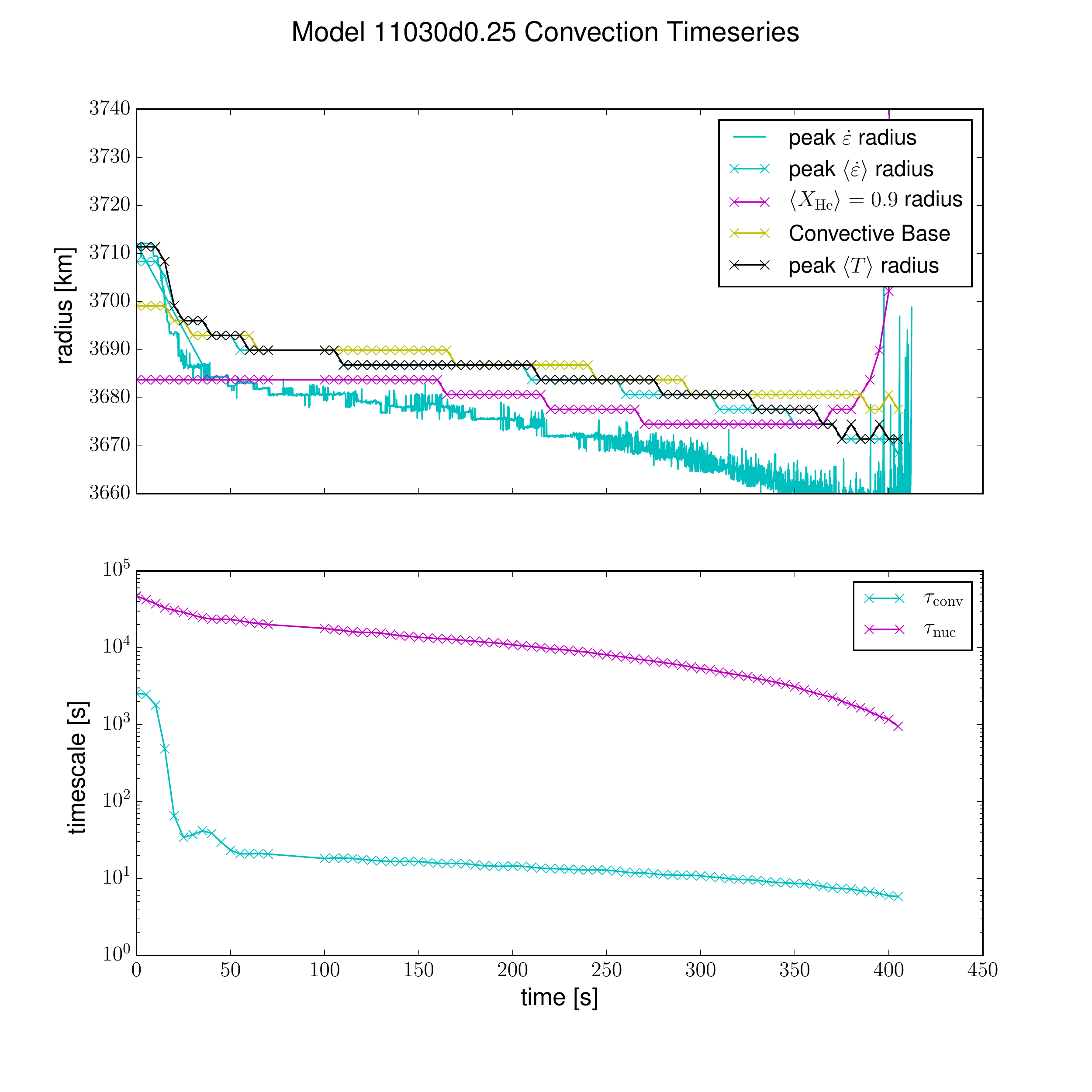}
   
   \caption{\label{fig:deltaconv} Same as Fig.\ref{fig:outcomes-conv} for model 11030d0.25} 
\end{figure}

We find that all of the runs with varied $\delta$'s
also experience localized runaway.  In addition, the convective dynamics
reported in previous sections are
qualitatively insensitive to varied $\delta$.  Figs.~\ref{fig:deltaig} and
\ref{fig:deltaconv} illustrate this using model 11030d0.25.  We also see a new
result: the dramatic rise of the typical radius at which helium's mass fraction
satisfies $X_\mathrm{He}=0.9$.  This suggests substantial mixing as carbon is
dredged up, displacing helium.  This will be explored in more detail in the next
paper in this series.

%==========================================================================
% Conclusions and Discussion
%==========================================================================
\section{Conclusions and Discussion}
We have explored 18 physically motivated, simple models of convective burning in sub-Chandrasekhar 
carbon/oxygen white dwarfs with an accreted, low-mass helium shell, as well as 3
supplemental models exploring a new numerical experiment of the impact of an
initial model parameter.  These systems have been modeled 
in 3D with detailed microphysics using the low-Mach hydrodynamics code \mae.  These are the first
3D simulations of this phase of evolution for such a broad suite of models, and
we have drastically expanded the diagnostics and analysis in comparison to Paper
I.

As we make clear in \S \ref{sec:meth}, our models are simple and limited in the number of species and
reactions we track.  This is sufficient for our focus on exploring a large number of model systems and
capturing the dominant energetics and general dynamical trends.  Our findings here will serve as the 
foundation for future work focused on a smaller number of more detailed models.  From this set of simple 
models we can draw several important insights into potential double detonation SNe Ia or .Ia progenitors 
as well as other runaway events involving sub-Chandrasekhar white dwarfs with envelopes of accreted helium, 
such as possible helium novae.

We find that localized runaway is indeed achievable in 3D.  Whether
this develops into a deflagration or detonation will be the focus of future work.
In particular, it would be valuable to make use of the
\mae-to-\texttt{Castro} mapping developed in \citet{Malone:2014}.  
Like \cite{WoosleyKasen2011}, we find that
hotter cores allow for localized runaway to develop at lower densities
than in models with cooler cores. Thus, hotter cores are able to
achieve runaway with thinner (lower mass) helium shells than cooler
cores of the same mass.

Our results indicate that the complex dynamics of 3D convection should not be
neglected in investigations of double detonation progenitor models.  In cases of
localized runaway, we find evidence of convective overshoot and a steady
freeze-out of convection.  These effects substantially impact the density,
temperature, and composition of the region in which localized runaway is
initiated.

Future papers in this series will characterize the geometry, thermodynamics, statistics, and timing of
ignition in greater detail, carry out more realistic simulations making use of 1D stellar evolution for
initial models and larger reaction networks, and map our low-Mach results into the compressible
code \texttt{Castro} to track the development of deflagration or detonation.

\acknowledgments

We thank Frank Timmes for making his equation of state publicly available, and
we thank Stan Woosley for making available data from his models and helpful
discussions.  We also thank the referee for a detailed reading of our paper and
their constructive commentary.   The work at Stony Brook was supported by
DOE/Office of Nuclear Physics grant DE-FG02-87ER40317 to Stony Brook.  The work
at LBNL was supported by the Applied Mathematics Program of the DOE Office of
Advance Scientific Computing Research under U.S. Department of Energy under
contract No.~DE-AC02-05CH11231.  This research used resources of the National
Energy Research Scientific Computing Center, which is supported by the Office of
Science of the U.S. Department of Energy under Contract No.~DE-AC02-05CH11231.
An award of computer time was provided by the Innovative and Novel Computational
Impact on Theory and Experiment (INCITE) program.  This research used resources
of the Oak Ridge Leadership Computing Facility at the Oak Ridge National
Laboratory, which is supported by the Office of Science of the U.S. Department
of Energy under Contract No.~DE-AC05-00OR22725.  This research is part of the
“Type Ia Supernovae” PRAC allocation of the National Science Foundation (award
number OCI-1036199) and the Blue Waters sustained-petascale computing project,
which is supported by the National Science Foundation (award number OCI
07-25070) and the state of Illinois.  Blue Waters is a joint effort of the
University of Illinois at Urbana-Champagne and its National Center for
Supercomputing Applications.

\software{
   \href{https://github.com/BoxLib-Codes/MAESTRO}{\mae} \citep{NonakaEtAl2010},
   \href{http://mesa.sourceforge.net/}{\textsf{MESA}} \citep{PaxtonEtAl2011, PaxtonEtAl2013},
   \href{https://wci.llnl.gov/simulation/computer-codes/visit}{\textsf{VisIt}},
   \href{http://matplotlib.org/}{\textsf{matplotlib}} \citep{Hunter2007},
   \href{http://ipython.org/}{\textsf{IPython}} \citep{PerezGranger2007}
   \href{http://yt-project.org}{\textsf{yt}} \citep{TurkEtAl2011}
}

\clearpage

\appendix

\section{An Improved Low Mach Number Equation Set} \label{app:energy}

In this Appendix, we discuss the improved low Mach number equation set used for
the simulations presented in this paper, and highlight the differences from the
algorithm presented in \cite{NonakaEtAl2010}.
%All of the simulations presented in this paper use this improved equation set.
We also discuss the effects of this improved model on some standard test
problems.
%and discuss possible impliciations for incorporating non-adiabatic flows and a
%time-varying background state.

Recently, both \citet{KP:2012} and \citet{VLBWZ:2013} introduced a new term in
the vertical momentum equation for low Mach number stratified flows that
enforces conservation of total energy in the low Mach number system in the
absence of external heating or viscous terms.  To compare the new system with
the origianl \mae\ equations, we first define $\rhostar, \Ubstar, \pstar,$ and
$\Tstar,$ as the density, velocity, pressure, and temperature, respectively, in
the low Mach number system,  in keeping with the notation used in
\citet{Durran:1989} and \citet{KP:2012}.  The perturbational quantities,
$\rhoprimestar = \rhostar - \rhozero$ and $\pprimestar = \pstar - p_0,$ are
analogous to those in the compressible system.  The quantities defining the
starred fluid approximate, but are not identical to, the quantities defining the
fully compressible fluid.

The following equations were derived in \citet{ABRZ:I} for low Mach number
stratified flow with a general equation of state: \begin{eqnarray}
   \frac{\partial \rhostar}{\partial t} + \nabla \cdot (\rhostar \Ubstar) &=& 0
   \enskip ,  \nonumber \\ \frac{\partial \Ubstar}{\partial t} + \Ubstar \cdot
   \nabla \Ubstar  + \frac{1}{\rhostar} \grad \pprimestar &=& -
   \frac{\rhoprimestar}{\rhostar} g \er \label{maestro:mom} \enskip ,
\end{eqnarray} with the constraint of the form, \begin{equation} \nabla \cdot
   (\beta_0(r) \Ubstar) = \beta_0 \sigma \mathcal{H} \enskip ,
   \label{LM_constraint} \end{equation} where \[ \beta_0(r) = \beta(0) \exp
   \left( {\int_0^r \frac{dp_0/d r^\prime}{\Gamma_{1_0} p_0(r^\prime)}
\,dr^\prime} \right ) \enskip  \] and $\sigma = p_T / (\rho c_p p_\rho)$
\citep{ABRZ:II}, which in the case of an ideal gas reduces to $1 / (c_p T).$
Also, we define $\Gamma_{1_0}$ as the lateral average of the first adiabatic
exponent, \begin{eqnarray} \Gamma_{1} \equiv \left. {d (\log p)}/{d (\log \rho)}
   \right|_\entropy \label{gamma_def} \enskip .  \end{eqnarray}

\citet{KP:2012} modify the vertical momentum equation by noting that a low Mach
number representation of total energy is conserved if, in moving from the
compressible to the low Mach number system, one substitutes $1 / \rho
\rightarrow 1 / \rhostar - (1 / \rhostar)^2 (\partial \rhostar / \partial p_0)
|_s (\pstar - p_0)$ instead of $1 / \rho \rightarrow 1 / \rhostar$ in the
momentum equation, resulting in \begin{equation} \frac{\partial
   \Ubstar}{\partial t} + \Ubstar \cdot \nabla \Ubstar  + \frac{1}{\rhostar}
   \grad \pprimestar = \dfrac{1}{\rhostar} \left( \left. \dfrac{\partial
   \rhostar}{\partial p_0} \right|_\entropy \pprimestar - \rhoprimestar \right)
   g \er  \label{mom_kp} \; .  \end{equation} \citet{VLBWZ:2013} instead
construct the additional term by deriving the low Mach number momentum equation
from Lagrangian analysis, starting from conservation of total energy.  The
momentum equation given in \citet{VLBWZ:2013} uses the assumption from
\citet{ABRZ:I} that $\Gamma_1$ can be approximated by $\Gamma_{1_0}$, and can be
written in the form \begin{equation} \frac{\partial \Ubstar}{\partial t} +
   \Ubstar \cdot \nabla \Ubstar  + \frac{\beta_0}{\rhostar} \grad \left(
   \frac{\pprimestar}{\beta_0} \right) = -\dfrac{\rhoprimestar}{\rhostar} g \er
   \label{mom_v} \end{equation} We note that (\ref{mom_v}) is analytically
equivalent to (\ref{mom_kp}) when $\Gamma_1 \equiv \Gamma_{1_0}$.  

An essential component of the solution procedure for low Mach number equation
sets is solving a variable coefficient Poisson equation for the perturbational
pressure. 
%(in the form $\pprimestar$ or $\piprimestar$).  
The Poisson equation can be derived by substituting the constraint into the
divergence of the momentum equation, \begin{eqnarray} \nabla \cdot \left
   (\frac{\beta_0}{\rhostar} \nabla \pprimestar\right ) &=& \nabla \cdot
   (\beta_0 \Astar)  - \beta_0 \frac{\partial (\sigma \calH)}{\partial t},
   \label{maestro:poisson} \end{eqnarray} with \[ \Astar = -[\Ubstar \cdot
\nabla \Ubstar] -\frac{(\rhostar - \rhozero)}{\rhostar} \gb \]

Using their formulation, \citet{VLBWZ:2013} show that one can instead solve
\begin{eqnarray} \nabla \cdot \left [\frac{\beta_0^2}{\rhostar} \nabla \left
   (\frac{\pprimestar}{\beta_0} \right )\right ] &=& \nabla \cdot (\beta_0
   \Astar)  - \beta_0 \frac{\partial (\sigma \calH)}{\partial t}
   \label{vlbwz:poisson} \end{eqnarray} for $\pprimestar.$  Analytically, this
looks very similar to the original equations, and numerically, it allows one to
reuse the original solver, simply modifying the coefficients ($\beta_0^2 /
\rhostar$ as opposed to $\beta_0 / \rhostar$) and the interpretation of the
variable being solved for ($\pprimestar/ \beta_0$ as opposed to $\pprimestar$).
While exactly the same solver can be used to solve
Equation~(\ref{vlbwz:poisson}) as to solve Equation~(\ref{maestro:poisson}), we
note that solving for $\pprimestar$ in the new constraint may take more
computational effort.  This arises because in Equation~(\ref{maestro:poisson})
the coefficients of $\pprimestar$ are close to one, since $\beta_0$ is a
density-like variable that is close to $\rhozero$ (and in fact identically
equals $\rhozero$ for an isentropically stratified atmosphere).  The
coefficients in Equation~(\ref{vlbwz:poisson}) are, by contrast, similar to
$\beta_0(r),$ which can have large variation over the scale heights within a
single calculation.  
%In \citet{VLBWZ:2013}, it was shown that the eigenvectors for the gravity waves
%of the modified equation set match those of the compressible equation set, thus
%more correctly modeling the vertical propagation of gravity waves.  In
%addition, in Appendix E, the \mae\ code was used to numerically demonstrate the
%conservation properties of the new approach.

We note also that the simplification allowed by writing the equation for
$\pprimestar$ in the form Equation~(\ref{vlbwz:poisson}) as opposed to more
elaborate formulations used in \citet{KP:2012} occur under the assumptions that
the flow is non-adiabatic and the base state is constant in time.  The
implications of adiabaticity and a time-varying base state are subjects of
future work.

\subsection{Test Problems}

Here we look at the impact of the different formulations of the momentum
equation on two test problems we have used in the development of \mae\ as well
as one of our science cases.  In \citet{VLBWZ:2013} a study of linear gravity
waves using the two different formulations as implemented in the \mae\ code
demonstrated that the eigenfunctions, energy conservation properties and
pseudo-energy conservation were as expected.  We refer the reader interested in
those results to \citet{VLBWZ:2013}.  

%Just an observation: While the analytical formulation of energy conservation
%may ensure proper gravity wave propagation, we don't necessarily numerically
%conserve energy anyway since we don't solve in conservation form

\subsubsection{Reacting Bubble Rise}

\begin{figure}
   \centering
   \includegraphics[width=\textwidth]{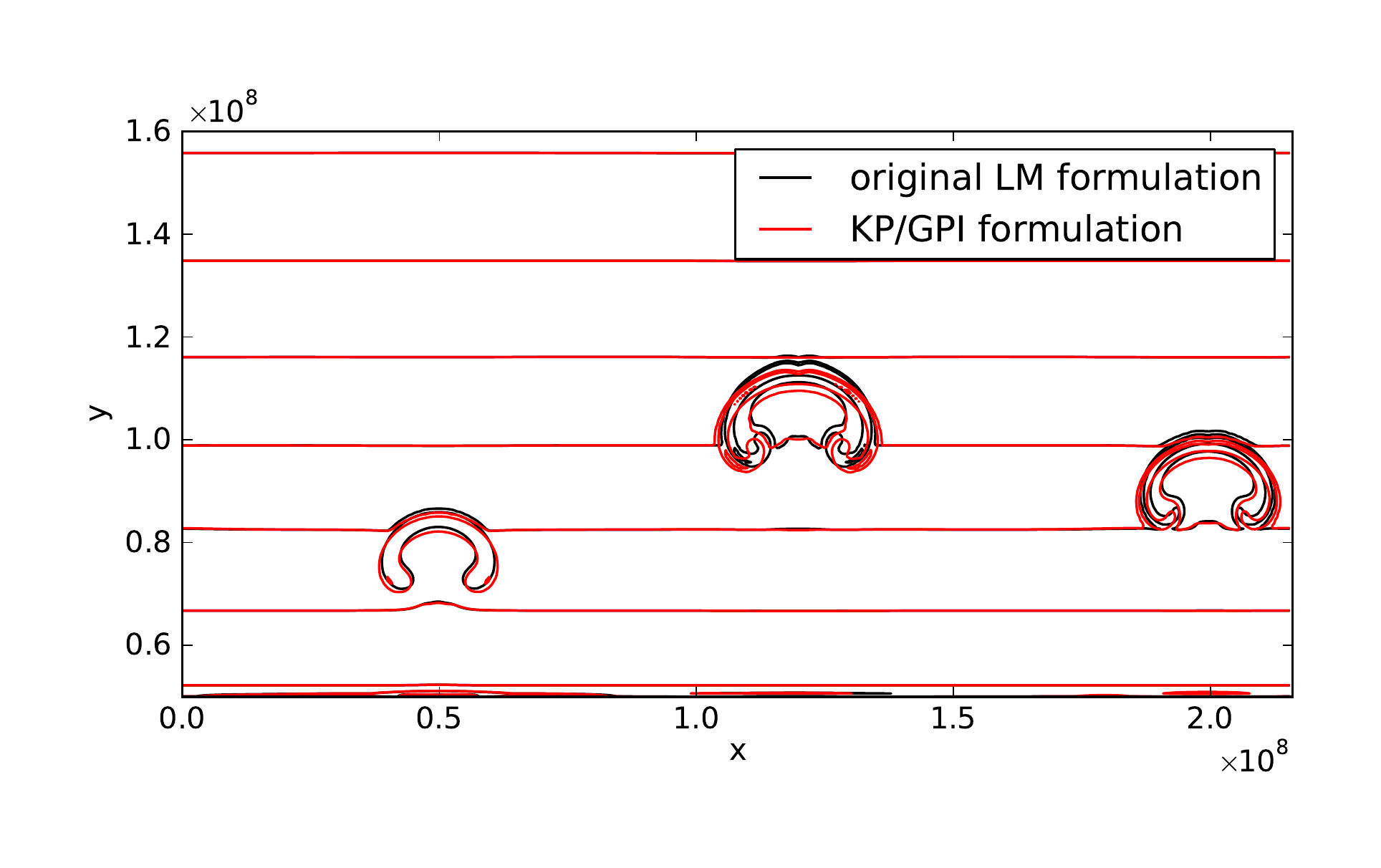}
   
   \caption{\label{fig:test2} Comparison of burning bubbles with the new formulation
   and the original formulation of the \mae\ equation set.}
\end{figure}

In \citet{AlmgrenEtAl2008} we showed an example of three burning bubbles in a
stratified white dwarf atmosphere (modeled in a plane-parallel geometry, for
convenience).  The bubbles are buoyant and rise and roll up as nuclear reactions
release heat in the hotter locations.  Here we revisit that problem with the
modified momentum equation.  Figure~\ref{fig:test2} shows a comparison of the
temperature field ($\Tstar = {\widehat{T}}(p_0, \estar)$).  for the simulation
using the new formulation vs. the original equations.  As we can see, the new
formulation generates slightly wider bubbles.  

This difference in width is consistent with an observation made in
\citet{ABRZ:I}, where we compared low Mach number simulations of non-reacting
buoyant bubbles with compressible simulations.  There we saw variation between
the bubbles as calculated with the different methods (including between the two
compressible formulations), but observed that the bubbles that evolved via the
low Mach number equation set were consistently narrower than the fully
compressible bubbles.  The fact that the new formulation generates slightly
wider bubbles suggests in a general sense that the solution with the new
formulation is closer to the fully compressible solution than that generated
with the original formulation.

This test problem is distributed with the public version of \mae\ as {\tt
reacting\_bubble}.

\subsubsection{Internally Driven Convection}

A simple test of convection driven by an analytic heating source was shown in
\citet{AlmgrenEtAl2008} and again in \citet{NonakaEtAl2010}.  The details of the
setup and heating term are given in Section~4.3 of \citet{AlmgrenEtAl2008}.  The
basic idea of this problem is that there is a convectively unstable region
bounded above and below by stable regions.  An analytic heat source at the base
of the unstable layer drives convection.  The top of the convective region
essentially spans to the top of the atmosphere, up to the point where the
density drops off steeply.  We considered the convectively unstable layer to be
the ``region of interest'' for comparison with results from a compressible code,
and we found good agreement between the original \mae\ equations and the fully
compressible solution.  Here we compare results from simulations using the new
formulation against both the original results and the compressible results.  The
compressible comparison for this paper is done with the \castro\
code~\citep{AlmgrenEtAl2010}.

\begin{figure}
   \centering
   \includegraphics[width=\linewidth]{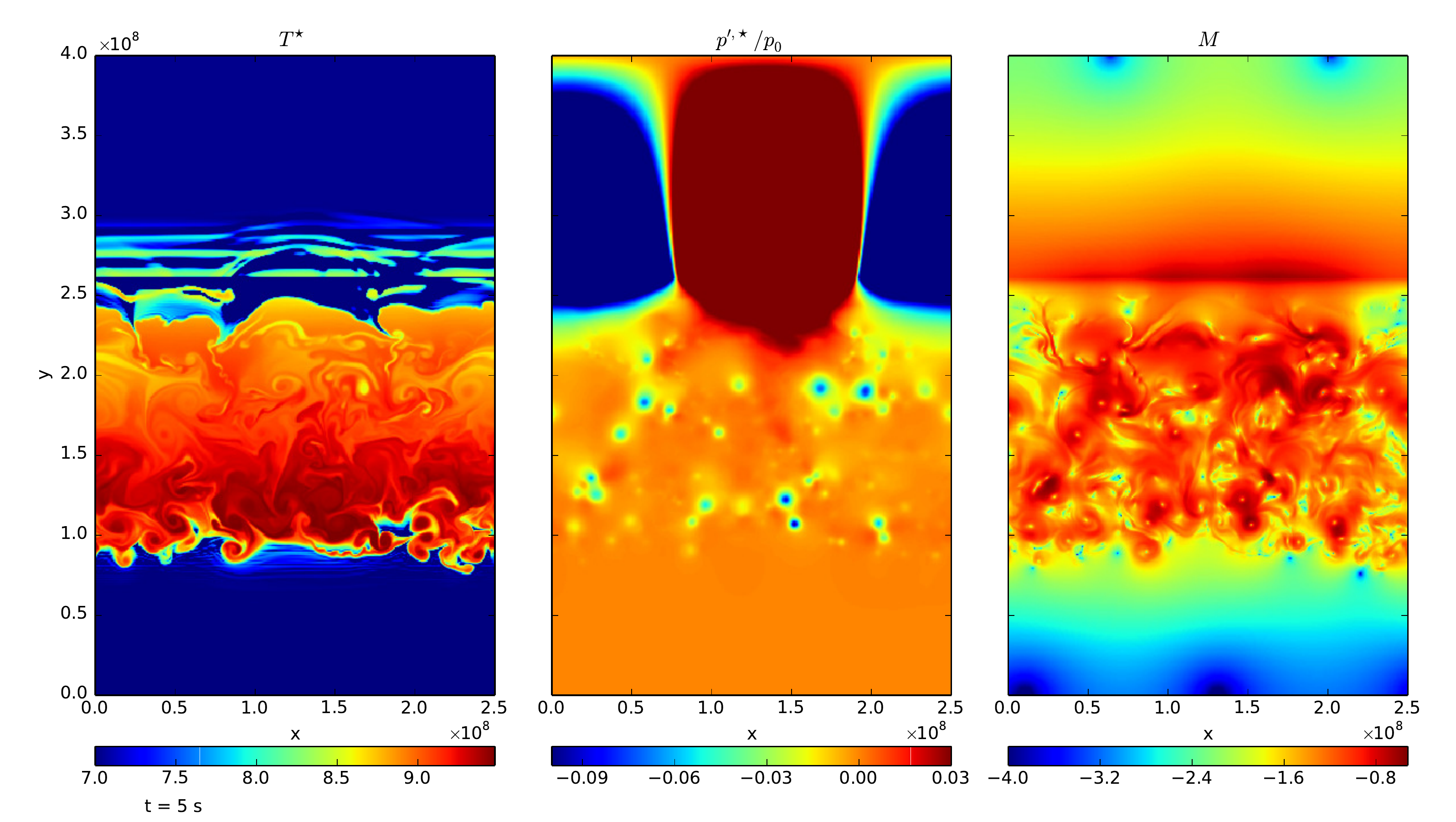} \\
   \includegraphics[width=\linewidth]{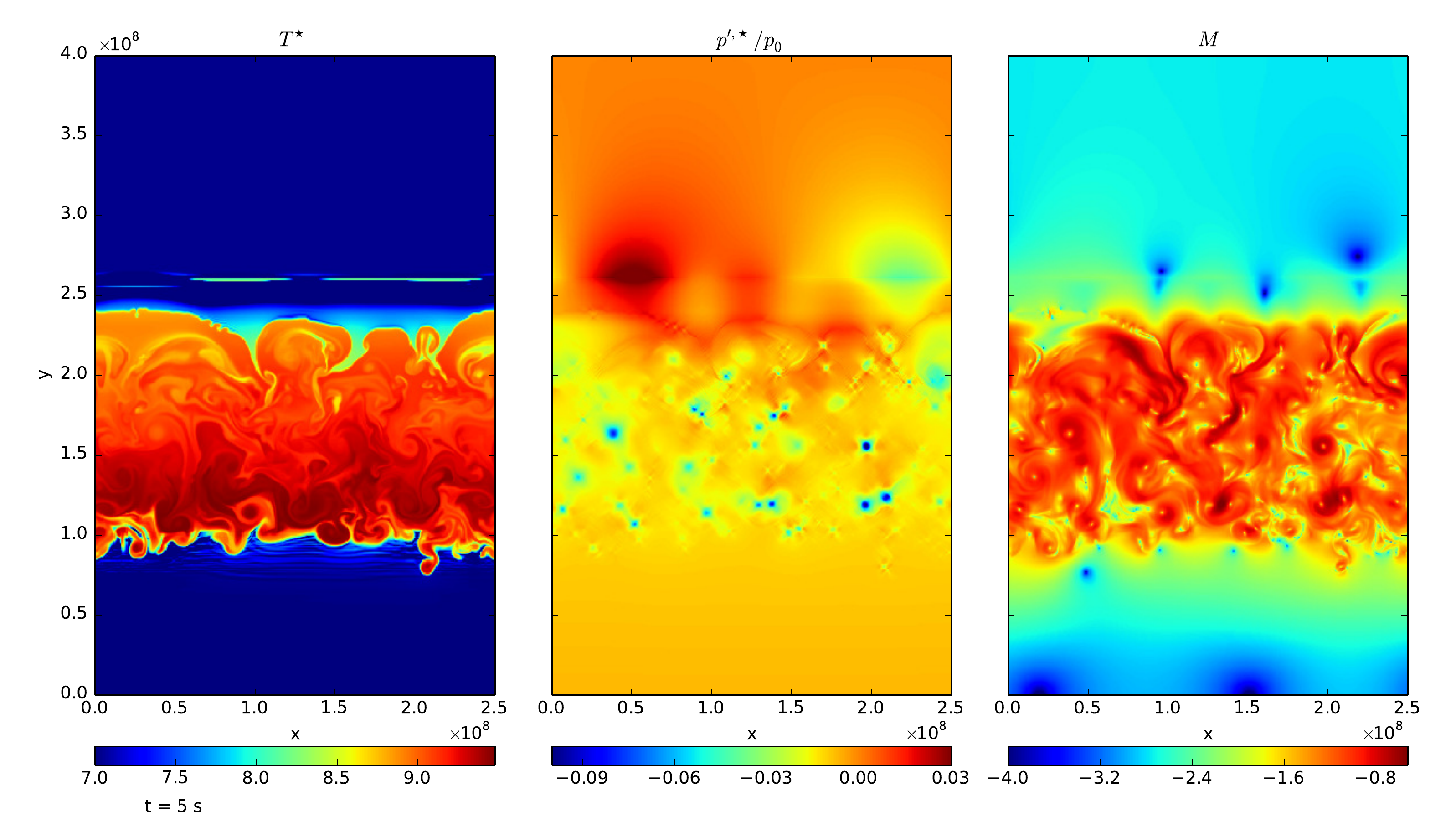} 
   \caption{\label{fig:convect_images} $\Tstar$, $\pprimestar/p_0$, and
     Mach number for the convection test problem using the original (top) and
     new (bottom) algorithms.  Note due to the large difference in
     $\pprimestar/p_0$ between the two runs, the original formulation
     is clipped at the upper value of the plot range.  In reality, $\pprimestar/p_0$
     is two orders-of-magnitude larger in the original formulation than
     in the new energy formulation in the region above the atmosphere.}
\end{figure}

For the calculations presented here, as in the previous two works, a
plane-parallel approximation is used for the stellar atmosphere, and $320\times
512$ cells are used to cover the the domain spanning $2.5\times
10^{8}~\mathrm{cm}$ in the horizontal direction and $4.0\times 10^8~\mathrm{cm}$
in the vertical.  In Figure~\ref{fig:convect_images} we show plots of $\Tstar,$
$\pprimestar/p_0$, and Mach number for the the original and new algorithms.  Two
things are immediately apparent.  First, in the convective region, roughly
between $10^8$~cm and $2.5\times 10^8$~cm, we see good agreement between the
original algorithm and the new version.  The temperature field and Mach number
in that region agree nicely.  However we also note that above the convective
region, the solutions differ.  This is precisely the location of the surface of
the star, so the density drops sharply to the cutoff value here.  We notice that
in the new model, the Mach number is much lower in this surface region and
$p'/p_0$ is also much lower (for the latter, we note that the plots use
different scales to bring out the details in each case).  While these
differences do not appear to influence the behavior of the convective region,
the large $M$ in this surface region does have implications for the timestep.
Overall, the new model seems better behaved in the surface layer.  Looking at
the stable region beneath the convective layer, we do not see much of a
difference between the two models.

\begin{figure}
   \centering
   \includegraphics[width=\textwidth]{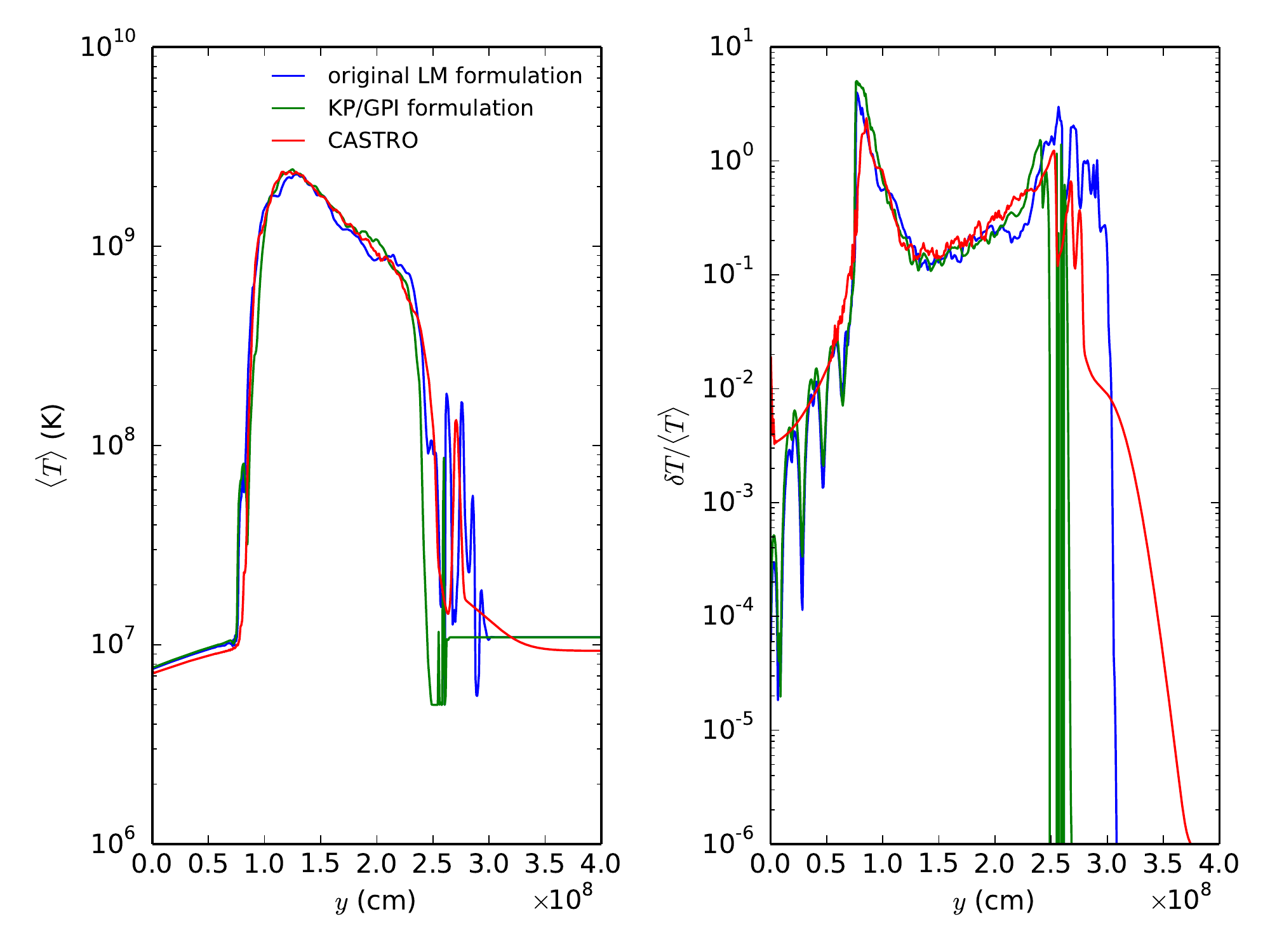}
   \caption{\label{fig:convect} $\langle T \rangle$ and $\delta T/\langle T \rangle$ as a function of height
    for the convection problem, comparing the original \mae\ implementation, the new energy formulation,
    and the compressible solution from \castro.}
\end{figure}

As we did when we previously looked at this test problem, we present two
diagnostics: the lateral average of the temperature, $\Tstar,$ \begin{equation}
   \left \langle T \right \rangle_j = \frac{1}{N_x} \sum_{i=1}^{N_x} T_{i,j}
   \enskip , \end{equation} and the RMS fluctuations: \begin{equation} (\delta
   T)_j = \left [ \frac{1}{N_x} \sum_{i=1}^{N_x} (T_{i,j} - \langle
   T\rangle_j)^2  \right ]^{1/2} \enskip , \end{equation} where $N_x$ is the
number of cells in the lateral direction.  Figure~\ref{fig:convect} shows the
profiles of $\langle T \rangle$ and $\delta T/\langle T \rangle$ as a function
of height.  In the region
of interest, the convective layer, we see that all the solutions agree
strongly.  The differences outside of the convective layer also seem
minimal.  The increase in $T$ that is seen above the convective layer
for the compressible solution is because in the compressible code, the
material above the atmosphere is not in hydrostatic equilibrium (it
has reached our cutoff density) and rains down onto the surface.  This does not
have a large dynamic effect because there is not much mass there.

This test problem is distributed with the public version of \mae\ as {\tt test\_convect}.

\section{Nuclear Reaction Energetics}
\label{app:rxns}

To demonstrate that we utilize the smallest set
of reactions and isotopes capable of yielding the dominant energetics
we calculate the rates for relevant reactions using
\mesa's\footnote{\url{http://mesa.sourceforge.net/}, version 7503} 
flexible and extensive reaction network module 
\citep{PaxtonEtAl2011, PaxtonEtAl2013}. \mesa\ uses multiple sources for 
reaction rates, including \cite{Sakharuk2006, AnguloEtAl1999, CaughlanFowler1988}.
As we do in our own \mae~network, we scale the $ ^{12}$C($\alpha, \gamma$)$^{16}$O
(CagO) reaction by 1.7 \citep{WeaverWoosley1993, Garnett1997}.  The code, data,
and configuration for these \mesa\ calculations are distributed in the
\mae~repository.

\begin{figure}
   \centering
   \includegraphics[width=0.5\textwidth]{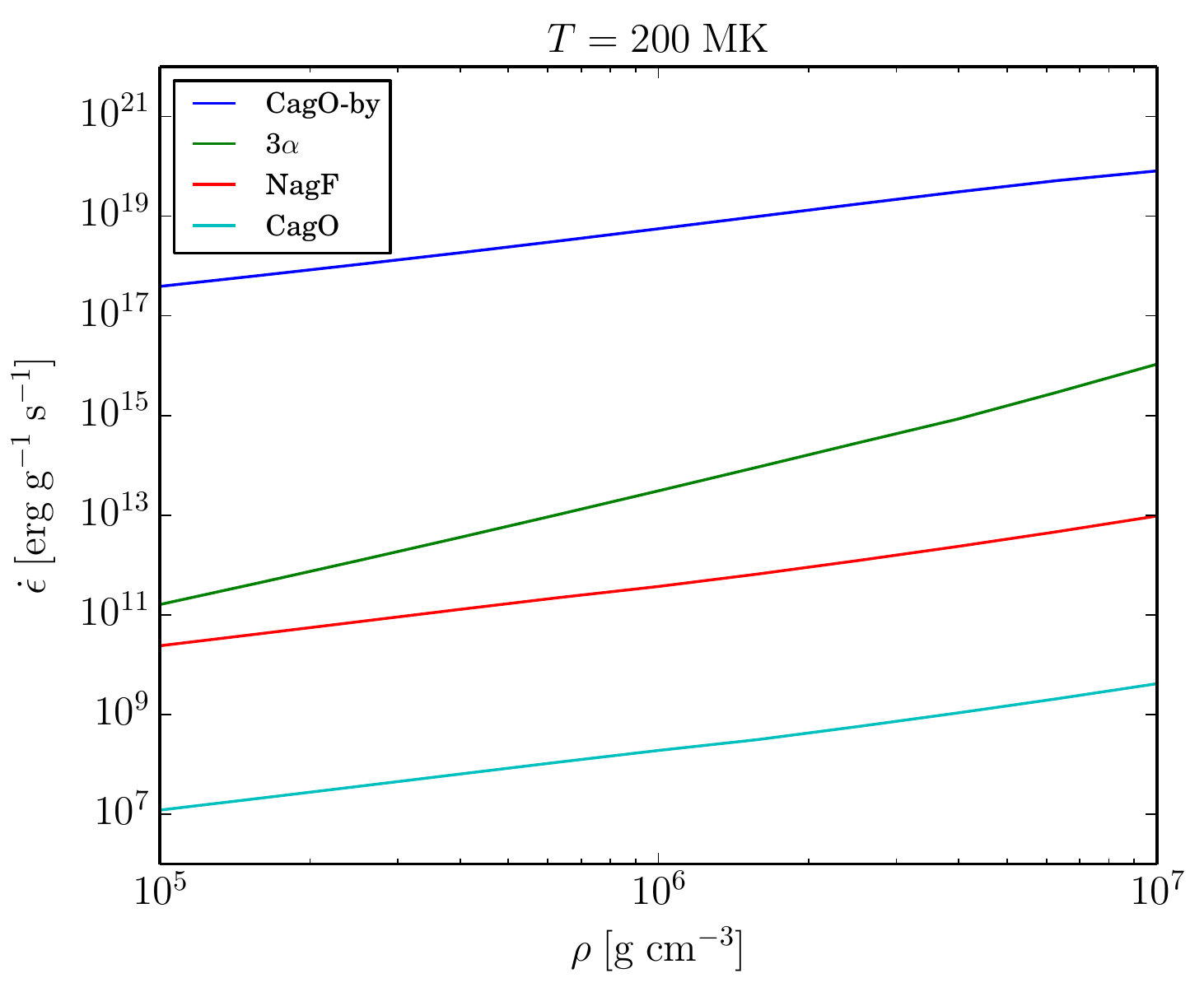}
   \caption{\label{fig:rxns} Energy generation of dominant reactions}
\end{figure}

To explore how the reactions highlighted by others (see \S \ref{ssec:micro}) 
might contribute to the energetics leading into thermonuclear runaway,
our \mesa\ calculations use a composition that allows all the relevant reactions 
to occur and is a reasonable reflection of what we expect in nature.  We  assume a
core of 49.5\% $^{12}\rm{C}$ and $^{16}\rm{O}$, 1\% $^{22}\rm{Ne}$,
and a shell of 99\% $^{4}\rm{He}$, 1\% $^{14}\rm{N}$.  Since the
energy release of $^{14}\rm{N}(e^-, \nu)^{14}\rm{C}(\alpha,\gamma)^{18}\rm{O}$
(NCO) is dominated by the alpha capture on the
intermediate $^{14}\rm{C}$, 10\% of the $^{14}\rm{N}$ is converted
into $^{14}\rm{C}$.  The site of active burning is assumed to be 10\%
core material, 90\% shell material as it is in the simulations being
reported here.  Finally, all mass fractions are reduced by about 5\%
to allow for 5\% of $^{1}\rm{H}$ so that the 
$^{12}$C(p, $\gamma$)$^{13}$N($\alpha$,p)$^{16}$O (CagO-bypass) reaction can
proceed.  The resulting mass fractions are $\mathrm{X}_{^1\mathrm{H}}
= 0.05$, $\mathrm{X}_{^4\mathrm{He}} = 0.84645$,
$\mathrm{X}_{^{12}\mathrm{C}} = 0.047$, $\mathrm{X}_{^{14}\mathrm{N}}
= 0.0077$, $\mathrm{X}_{^{14}\mathrm{C}} = 0.00085$,
$\mathrm{X}_{^{16}\mathrm{O}} = 0.047$, $\mathrm{X}_{^{22}\mathrm{Ne}}
= 0.001$. In the pre-explosive dynamics leading up to thermonuclear
runaway the sites of nuclear burning have $T \approx 200$ MK, $\rho
\approx 10^5$--$10^6~\gcc$ (see \S \ref{ssec:initmod}).  The most
energetic reaction rates for these conditions and the given composition are
plotted in Fig~\ref{fig:rxns}.  

The forward NCO reaction does not become faster than the back-reaction
until about $\rho \approx 4\times 10^6~\gcc$ and even then is only
marginally faster for the densities considered here, so it
is not included in the figure.  A much more important
consequence of a shell with $^{14}\mathrm{N}$ is the quite energetic
NagF reaction which can generate an appreciable fraction of the amount
of energy generated by the triple-alpha reaction.  The triple-alpha reaction 
still dominates and the dynamics we demonstrate here will be similar with or
without the NagF reaction, except that ignition may be achieved more
quickly with it.  For simplicity, we neglect it in this study.  A future study 
building on this work will explore more detailed nucleosynthesis and 
the impact of using a larger reaction network.

The CagO-bypass reaction is incredibly energetic.  We can only justify
neglecting this reaction in our calculations because we are
considering pure helium accretion with no free protons \emph{before}
thermonuclear runaway conditions of $T \approx 1$ GK are achieved.  As
discussed in \cite{ShenBildsten2009}, even in the event of no protons
in the envelope, once conditions of $T \approx 1$ GK are met
alpha-chain reactions of $^{24}\mathrm{Mg}$ and higher can yield
intermediate protons, and the $^{14}\rm{N}(\alpha,
\gamma)^{18}\rm{F}(\alpha, p)^{21}\rm{Ne}$ reaction directly yields
protons (in the pre-explosive regime the $^{18}\rm{F}$ is unable to
alpha-capture at an appreciable rate).  Thus the CagO-bypass reaction
must be included for any studies of the \sch ignition regime.

From these rate calculations we conclude the triple-alpha reaction 
dominates energy generation in the pre-ignition conditions and expected 
composition of C/O WD helium-accretors.  

\clearpage

\bibliographystyle{aasjournal}
\bibliography{paper}

\listofchanges

\end{document}

\appendix

\section{An Improved Low Mach Number Equation Set}
\label{app:energy}

In this Appendix, we discuss the improved low Mach number equation set used
for the simulations presented in this paper, and highlight the 
differences from the algorithm presented in \cite{NonakaEtAl2010}.
%All of the simulations presented in this paper use this improved equation set.
We also discuss the effects of this improved model on some standard test problems.
%and discuss possible impliciations for incorporating non-adiabatic flows and a
%time-varying background state.

Recently, \citet{KP:2012} introduced a new treatment of energy
conservation in a low Mach number model for stratified atmospheric flow
that was later derived in an alternate fashion by \citet{VLBWZ:2013}.
In keeping with the notation used in \citet{Durran:1989} and
\citet{KP:2012}, we define $\rhostar, \Ubstar, \pstar, \Tstar, \Sstar
$ and $\estar$ as the density, velocity pressure, temperature, specific entropy
and specific internal energy, respectively, in the low Mach number system.  The
perturbational quantities, $\rhoprimestar = \rhostar - \rhozero$ and
$\pprimestar = \pstar - p_0,$ are analogous to those in the
compressible system.  The quantities defining the starred fluid
approximate, but are not identical to, the quantities defining the
fully compressible fluid.

The following equations were derived in \citet{ABRZ:I} for low Mach number stratified flow
with a general equation of state:
\begin{eqnarray}
        \frac{\partial \rhostar}{\partial t} + \nabla \cdot (\rhostar \Ubstar)
        &=& 0 \enskip ,  \nonumber \\
        \frac{\partial \Ubstar}{\partial t} + 
        \Ubstar \cdot \nabla \Ubstar  + \frac{1}{\rhostar} \grad \pprimestar
        &=& - \frac{\rhoprimestar}{\rhostar} g \er \label{maestro:mom} \enskip ,
\end{eqnarray}
with the constraint of the form,
\begin{equation}
\nabla \cdot (\beta_0(r) \Ubstar) = \beta_0 \sigma \mathcal{H} \enskip , \label{LM_constraint}
\end{equation}
where 
\[
\beta_0(r) = \beta(0) \exp \left( 
{\int_0^r \frac{dp_0/d r^\prime}{\Gamma_{1_0} p_0(r^\prime)} \,dr^\prime} \right ) \enskip  
\]
and $\sigma = p_T / (\rho c_p p_\rho)$ \citep{ABRZ:II}, which in the case of an ideal gas
reduces to $1 / (c_p T).$  Also, we define $\Gamma_{1_0}$ as the lateral average of the
first adiabatic exponent,
\begin{eqnarray}
\Gamma_{1} \equiv \left. {d (\log p)}/{d (\log \rho)} \right|_\entropy \label{gamma_def} \enskip .
\end{eqnarray}

An essential component of the solution procedure for low Mach number equation sets 
is solving a variable coefficient Poisson equation for the perturbational pressure. 
%(in the form $\pprimestar$ or $\piprimestar$).  
The Poisson equation can be derived
by substituting the constraint into the divergence of the momentum equation,
\begin{eqnarray}
 \nabla \cdot \left (\frac{\beta_0}{\rhostar} 
     \nabla \pprimestar\right ) &=& 
\nabla \cdot (\beta_0 \Astar)  - \beta_0 \frac{\partial (\sigma \calH)}{\partial t},
\label{maestro:poisson}
\end{eqnarray}
with 
\[
\Astar = -[\Ubstar \cdot \nabla \Ubstar] -\frac{(\rhostar - \rhozero)}{\rhostar} \gb 
\]

Both \citet{KP:2012} and \citet{VLBWZ:2013} introduce a new term in the 
vertical momentum equation for low Mach number stratified flows
that enforces conservation of total energy in the absence of external
heating or viscous terms.
\citet{KP:2012} introduce the term by noting that a low Mach number
representation of total energy is conserved if, 
in moving from the compressible to the low Mach number system, one substitutes
$1 / \rho \rightarrow 1 / \rhostar - (1 / \rhostar)^2 (\partial \rhostar / \partial p_0) |_s (\pstar - p_0)$ 
instead of $1 / \rho \rightarrow 1 / \rhostar$ in the momentum equation,
resulting in 
\begin{equation}
        \frac{\partial \Ubstar}{\partial t} + 
        \Ubstar \cdot \nabla \Ubstar  + \frac{1}{\rhostar} \grad \pprimestar
        = \dfrac{1}{\rhostar} \left( \left. \dfrac{\partial \rhostar}{\partial p_0} \right|_\entropy \pprimestar 
           - \rhoprimestar \right) g \er  \label{mom_kp} \; .
\end{equation}
\citet{VLBWZ:2013} construct the additional term 
by deriving the low Mach number momentum equation from Lagrangian analysis, 
starting from conservation of total energy. 
The momentum equation given in \citet{VLBWZ:2013} uses the assumption from \citet{ABRZ:I} that
$\Gamma_1$ can be approximated by $\Gamma_{1_0}$, and can be written in the form
\begin{equation}
        \frac{\partial \Ubstar}{\partial t} + 
        \Ubstar \cdot \nabla \Ubstar  + \frac{\beta_0}{\rhostar} \grad \left( \frac{\pprimestar}{\beta_0} \right)
        = -\dfrac{\rhoprimestar}{\rhostar} g \er  \label{mom_v}
\end{equation}
We note that (\ref{mom_v}) is analytically equivalent to (\ref{mom_kp})
when $\Gamma_1 \equiv \Gamma_{1_0}$.  
%{\color{red} We will discuss the issues associated with this approximation in a later section.
%We will also talk about the effect this form has on the solution procedure.}

%We will refer to the \mae\ equation set as the LM equations and 
%the equation set presented by \citet{KP:2012} as the KP equations;
%\citet{VLBWZ:2013} refer to the modified equations in the form they present
%as the Generalized Pseudo-Incompressible (GPI) equations.  

%% In the original pseudo-incompressible equation set, the divergence constraint has the form
%% \[
%% \nabla \cdot \left(\rhozero \thetazero \Ubstar \right) = \frac{\calH}{c_p \pizero} \enskip .
%% \]
%% If one defines $\Astar$ to be the combined advective and buoyancy terms, i.e. 
%% \[
%% \Astar = -[\Ubstar \cdot \nabla \Ubstar] -\frac{(\thetastar - \thetazero)}{\thetazero} \gb 
%% \]
%% then the resulting Poisson equation for $\piprimestar$ can be written 
%% \[
%% c_p \nabla \cdot \left(\rhozero \thetazero \thetastar \nabla \piprimestar \right) = 
%% \nabla \cdot \left(\rhozero \thetazero \Astar \right) - 
%% \frac{1}{c_p \pizero} \frac{\partial \calH}{\partial t} \enskip .
%% \]
%% Similarly,

{\color{red} (AJN - I say cut out this red text)
One of the issues raised in \citet{KP:2012} is that the addition of
the $p^{\prime,\star}$ term to the momentum equation no longer allows
one to solve an equation of the form Equation~(\ref{maestro:poisson})
for $\pprimestar.$ The Helmholtz-type elliptic equation for
$\pprimestar$ presented in the Appendix of \citet{KP:2012} has the
form
\begin{eqnarray}
 \nabla \cdot \left (\frac{1}{\rhostar} \nabla \pprimestar \right ) + f(\rhostar,p_0) \; \pprimestar 
&=& RHS \enskip, \label{kp:poisson}
\end{eqnarray}
where $f(\rhostar,p_0)$ and $RHS$ are specified in detail in \citet{KP:2012}. 
We note that is no longer a variable coefficient Poisson equation, thus complicating the solution procedure 
for this equation set.  If we allow for $\Gamma_1$ to depart only slightly from ${\Gamma_1}_0$, the
we could use the methodology used to allow for $\Gamma_1$ variations from \citet{AlmgrenEtAl2008} to enforce
this type of constraint.}

Using this formulation, \citet{VLBWZ:2013} show that one can solve
\begin{eqnarray}
 \nabla \cdot \left [\frac{\beta_0^2}{\rhostar} 
     \nabla \left (\frac{\pprimestar}{\beta_0} \right )\right ] &=& 
\nabla \cdot (\beta_0 \Astar)  - \beta_0 \frac{\partial (\sigma \calH)}{\partial t} \label{vlbwz:poisson}
\end{eqnarray}
for $\pprimestar.$  
Analytically, this looks very similar to the original equations, and numerically, 
it allows one to reuse the original solver, simply modifying the coefficients 
($\beta_0^2 / \rhostar$ as opposed to $\beta_0 / \rhostar$) and 
the interpretation of the variable being solved for ($\pprimestar/ \beta_0$ as 
opposed to $\pprimestar$).
While exactly the same solver can be used to solve Equation~(\ref{vlbwz:poisson}) as to solve
Equation~(\ref{maestro:poisson}), we note that solving for $\pprimestar$ in the new constraint
may take more computational effort.  This arises because in Equation~(\ref{maestro:poisson})
the coefficients of $\pprimestar$ are close to one, since $\beta_0$ is a density-like
variable that is close to $\rhozero$ (and in fact identically equals $\rhozero$ for 
an isentropically stratified atmosphere).  The coefficients 
in Equation~(\ref{vlbwz:poisson}) are, by contrast, similar to $\beta_0(r),$ which can have large
variation over the scale heights within a single calculation.  
In \citet{VLBWZ:2013}, it was shown that the eigenvectors for the
gravity waves of the modified equation set match those of the compressible equation set, 
thus more correctly modeling the vertical propagation of gravity waves.  In addition,
in Appendix E, the \mae\ code was 
used to numerically demonstrate the conservation properties of the new approach.

We note also that the simplification allowed by writing the equation for $\pprimestar$ 
in the form Equation~(\ref{vlbwz:poisson}) as opposed to more elaborate formulations
used in \citet{KP:2012}
occur under the assumptions that the flow is non-adiabatic and the base state is constant
in time.  The implications of adiabaticity and a time-varying base state are subjects of future work.

%two assumptions:  first, that
%the constraint can be written as Equation~(\ref{LM_constraint}) rather than in the more general
%form given by the extension to non-adiabatic flow of Equation~(\ref{KP_constraint}),
%and second, that additional terms in $\pprimestar,$ such as those discussed in
%the Appendix of \citet{KP:2012}, do not occur as source terms in the constraint equation.

\subsection{Energy Conservation for Adiabatic Flows}

For the fully compressible form, the evolution of internal energy for flows with no external
heat sources is given by 
\begin{equation*}
 \frac{\partial (\rho e)}{\partial t} + 
 \nabla \cdot \left( \rho \Ub e \right) + p \nabla \cdot \Ub = 0 \enskip ;
\end{equation*}
if we add that to the expression for the evolution of kinetic ($K = 1/2 \; \rho \Ub \cdot \Ub$) 
and gravitational ($\rho \Phi$) energy for $\Phi_t = 0,$
\begin{equation*}
\frac{\partial (K + \rho \Phi)}{\partial t} + 
\nabla \cdot \left [ (K + \rho \Phi + p) \Ub) \right ] - p \nabla \cdot \Ub = 0 \enskip ,
\end{equation*}
then we see that the $p \divu$ terms cancel,  and we obtain
\[
\frac{\partial (\rho E_T)}{\partial t} + 
\nabla \cdot \left [ (\rho E_T + p) \Ub \right ] = 0   \enskip .
\]
where $\rho E_T = \rho e + K + \rho \Phi.$

The original LM equations as used in \mae\ violate energy conservation for adiabatic flows
because the internal energy equation has the form 
\begin{equation}
 \frac{\partial (\rhostar \estar)}{\partial t} + 
 \nabla \cdot \left( \rhostar \Ubstar \estar \right) + p_0 \nabla \cdot \Ubstar = 0 \enskip ,
\label{rhostar_estar}
\end{equation}
but the evolution of kinetic and gravitational energy still contains the full pressure, i.e.
\begin{equation}
\frac{\partial (\Kstar + \rhostar \Phi)}{\partial t} + 
\nabla \cdot \left [ (\Kstar + \rhostar \Phi + \pstar) \Ub) \right ] - \pstar \nabla \cdot \Ubstar = 0 \enskip , \label{Kstar_LM}
\end{equation}
Thus
\begin{equation}
\frac{\partial (\rhostar E^\ast_T)}{\partial t} + 
\nabla \cdot \left[ (\rhostar E^\ast_T + \pstar) \Ubstar \right] = \pprimestar \nabla \cdot \Ubstar
\enskip .  \label{total_Estar}
\end{equation}
where $\rhostar E^\ast_T = \rhostar \estar + \Kstar + \rhostar \Phi.$

The KP and GPI vertical momentum equations carry an additional term of the form \\
$-(\rhozero /\rhostar) (\partial \rhostar/\partial p_0) \pprimestar g \er$
or $\pprimestar \nabla \beta_0 / \beta_0,$ respectively.  \MarginPar{this part is unclear -- MZ}
Using hydrostatic equilibrium to replace $\rhozero g \er$ by $-\nabla p_0,$
this results in an extra term in the equation for the evolution of kinetic plus gravitational energy
of the form $-\pprimestar/(\Gamma_1 p_0)\, \Ubstar \cdot \nabla p_0$
or $\pprimestar (\Ubstar \cdot \nabla \beta_0) / \beta_0,$ respectively.   
The right hand side term in the evolution of the total energy then looks like 
\begin{equation} 
\pprimestar \left [ \nabla \cdot \Ubstar - \left (\dfrac{1}{\Gamma_1 p_0} \right )
               \Ubstar \cdot \nabla p_0 \right ]  \label{cancel_1}
\end{equation} 
in the KP equations, and 
\begin{equation}
\pprimestar  (\nabla \cdot \Ubstar +  \frac{1}{\beta_0} \nabla \beta_0 \cdot \Ubstar)
= 
\frac{1}{\beta_0} \pprimestar \nabla \cdot (\beta_0 \Ubstar)  \label{cancel_2}
\enskip ,
\end{equation} 
in the GPI equations.  

Recalling the divergence constraint as written in Equation~(\ref{LM_constraint}),
we see that the right hand side for the GPI equations is identically zero when ${\mathcal H} =0.$
The KP equations are more general, and assume only that the constraint is written in the form,
\begin{equation}
\nabla \cdot \Ubstar + \dfrac{1}{{\rhostar}} \left. \dfrac{\partial \rhostar}{\partial p_0}\right|_\entropy 
                       \Ubstar \cdot \nabla p_0 = 0 \label{KP_constraint}
\end{equation}
when ${\mathcal H} = 0$.  We note that both the KP and GPI equations
conserve the total energy, $\rhostar E^\ast_T,$ while the loss of
energy conservation in the LM equations is $O(\pstar-p_0) = O(M^2).$
We also note that in the case of low Mach number adiabatic flows for
which stratification effects do not enter the constraint, the LM
equations do conserve total energy.  {\color{red} We remind the
reader that since the `$\star$' state approximates the compressible
state, it is not necessarily the case that $E^\star_T = E_T$.}

\subsection{Extension to Nonadiabatic Motions}

%We now consider the extension to nonadiabatic motions.
The GPI equations were derived with no diffusion or heat sources ({\color{red} although, 
see Lecoanet's new paper});
here we extend the discussion to include energy evolution when external sources are present, such 
as those that result from nuclear reactions in astrophysical modeling or moisture phase
changes in atmospheric modeling.  For convenience we assume 
$\Gamma_{1} = \Gamma_{1_0}(r)$ and recall the divergence constraint in the form 
of Equation~(\ref{LM_constraint}).  

For the fully compressible form, the heating term that appears in the fully compressible energy equation
contributes to the evolution of total energy in exactly the same form as it contributes to the 
evolution of internal energy, and leaves the evolution of kinetic energy unchanged, hence
\[
\frac{\partial (\rho E_T)}{\partial t} + \nabla \cdot \left [ (\rho E_T + p) \Ub  \right ] = \rho \calH \enskip .
\]
However, in the low Mach number framework, the internal energy equation takes the form
\begin{equation}
 \frac{\partial (\rhostar \estar)}{\partial t} + 
 \nabla \cdot \left [ (\rhostar \estar) \Ubstar \right ] + p_0 \nabla \cdot \Ubstar =
 \rhostar \calH \enskip ,
\label{rhostar_estar_H}
\end{equation}
while the evolution of kinetic plus gravitational energy in the LM equation set is
still given by Equation~(\ref{Kstar_LM}).
%\begin{equation}
%\frac{\partial (\Kstar + \rhostar \Phi)}{\partial t} + 
%\nabla \cdot \left [ (\Kstar + \rhostar \Phi + \pstar) \Ubstar  \right ] - 
%   \pstar \nabla \cdot \Ubstar = 0 
%\enskip . 
%\label{Kstar_LM}
%\end{equation}
As we recall from the previous section, the LM equation set violates 
conservation of total energy in the adiabatic context
because the addition of Equations~(\ref{rhostar_estar}) and (\ref{Kstar_LM}) 
leaves an uncanceled $\pprimestar \divustar$ term in Equation~(\ref{total_Estar}).
The GPI equations eliminate this contribution 
to the total energy in the adiabatic case by adding 
$\pprimestar \nabla \beta_0 / \beta_0$ to the LM momentum equation.  
However, in the nonadiabatic case in which $\calH \neq 0,$  
%substituting 
%\[
%\divustar =  -\frac{\Ubstar \cdot \nabla \beta_0}{\beta_0} + \sigma \calH\enskip ,
%\]
%into Equation~(\ref{total_Estar}), we find 
the evolution for total energy in the GPI or KP equation set becomes
\[
\frac{\partial (\rhostar \Estar_T)}{\partial t} + 
\nabla \cdot \left [ (\rhostar \Estar_T + \pstar) \Ubstar  \right ] 
= \rhostar \calH + \pprimestar \sigma \calH \enskip .
\label{total_energy_with_H}
\]
In the KP and GPI systems, as in the original LM system, the contribution of heat sources to the
total energy is not identical to the contribution to the internal energy. 
%The momentum equation itself is unchanged in the presence of heating but the 
%kinetic energy does change with heating in the low Mach number system.
In low Mach number flows without stratification effects in the constraint,  and in
compressible flows, the contribution of heat sources to the internal energy 
is identical to the contribution to the total energy. 

%There are two different ways to pose the question about how external
%heating affects the solution of the compressible vs.\ the low Mach number systems.  
%The difference in how the external heating affects the solution of the 
%compressible vs.\ the low Mach number systems is discussed 
In \citet{KP:2012} the source term is written in terms of $\sdot,$ 
the change in entropy for both the compressible and low Mach
number systems.  \citet{KP:2012} point out that while
$\sdot$ contributes to the internal and total energy of the compressible system 
via a source term of the form
$\rho T \sdot,$ where T is the temperature, it contributes to the internal energy 
(and to the enthalpy)
of the low Mach number system in the form $\rhostar \Tstar \sdotstar,$  
but to the total energy of the low Mach number system in the form 
$\rhostar T^{\ast \ast} \sdotstar,$
where, given $T = \That(p,\entropy)$ in the 
compressible system, $T^\ast = \That(p_0,\Sstar)$ and  
\[
T^{\ast \ast} = \Tstar + 
    \left. \frac{\partial \That}{\partial p_0}\right|_{\entropy^\ast}  \; \pprimestar \enskip .
\]
%Considering for simplicity the case of an ideal gas, 
The difference between the contribution to the internal energy and to the total energy is shown above to be 
$\pprimestar \sigma \calH$; the difference in the KP system can be shown to be the same:
\begin{eqnarray*}
\rhostar (T^{\ast \ast} - T^\ast) \sdotstar &=& 
    \left. \rhostar \frac{\partial \That}{\partial p}\right|_\entropy  \; \pprimestar \sdotstar  \\
   &=&   \pprimestar \left ( \left. \frac{\rhostar}{\Tstar} \frac{\partial \That}{\partial p} \right|_\entropy  \; 
        \right )  \; \Tstar \sdotstar \\
   &=&  \pprimestar \sigma \Tstar \sdotstar \\
   &=&  \pprimestar \sigma \calH \enskip .
\end{eqnarray*} \MarginPar{is it really $T \dot{s} = \calH$---isn't there an advective term?}
where we use the fact that $\sigma$ can be written as
$\sigma = -1 / (\rho T) \left. (\partial \rho / \partial \entropy) \right|_p
= \rho / T \left. (\partial T / \partial p) \right|_\entropy,$
\MarginPar{at the moment, I have this derivation in the Lambda2 paper---should I move it here?}

\subsection{Extension to Time-Varying Reference State}

We recall from \citet{almgren:2000} that large-scale heating must be absorbed into the 
base state pressure rather than the perturbational pressure in order to give the correct
thermodynamic response, as measured by comparison with the compressible adjustment.
%We explore here whether allowing $p_0$ to vary in response to $\calH$ allows a
%form of energy conservation more in line with the compressible expression.

For simplicity we consider here the equations in a plane-parallel geometry.  
When we allow the base state to respond to base state heating by evolving in time,
the constraint takes the form
\[
\divustar =  -\frac{\Ubstar \cdot \nabla \beta_0}{\beta_0} + \sigma \calH
             -\frac{1}{\Gamma p_0} \frac{\partial p_0}{\partial t} \enskip .
\]
Here 
\[
\frac{\partial p_0}{\partial t}  = - w_0 \frac{\partial p_0}{\partial r}  
\]
and
\[
\frac{\partial (\beta_0 w_0)}{\partial r} = \beta_0 \left( (\overline{\sigma \calH}) - 
        \frac{1}{\Gamma p_0} \frac{\partial p_0}{\partial t} \right) \enskip ,
\]
where $(\overline{\sigma \calH})$ is the lateral average of $(\sigma \calH).$
The KP/GPI energy equation takes the form
\begin{eqnarray*}
\frac{\partial \Estar_T}{\partial t} + 
\nabla \cdot \left [ (\Estar_T + \pstar) \Ubstar  \right ] 
&=& \rhostar \calH - \pprime \left( \sigma \calH - \frac{1}{\Gamma p_0} \frac{\partial p_0}{\partial t} \right) \\
&=& \rhostar \calH - 
\pprime \left( \sigma \calH + \frac{1}{\Gamma p_0} w_0 \frac{\partial p_0}{\partial r} \right) 
\enskip .
\end{eqnarray*}
The original LM equations would have the form
\begin{eqnarray*}
\frac{\partial \Estar_T}{\partial t} + 
\nabla \cdot \left [ (\Estar_T + \pstar) \Ubstar  \right ] 
&=& \rhostar \calH - \pprime \left( \sigma \calH - \frac{1}{\Gamma p_0} \frac{\partial p_0}{\partial t} 
-\frac{\Ubstar \cdot \nabla \beta_0}{\beta_0}  \right) \\
&=& \rhostar \calH - \pprime \left( \sigma \calH + \frac{1}{\Gamma p_0} w_0 \frac{\partial p_0}{\partial r} 
-\frac{\widetilde{\Ubstar} \cdot \nabla p_0}{\Gamma p_0}  
-\frac{1}{\Gamma p_0} w_0 \frac{\partial p_0}{\partial r} \right) \\
&=& \rhostar \calH - \pprime \left( \sigma \calH -\frac{1}{\Gamma p_0} \widetilde{\Ubstar} \cdot \nabla p_0 \right)
\enskip .
\end{eqnarray*}
where $\widetilde{\Ubstar} = \Ubstar - w_0 \er.$  \MarginPar{so we don't have a strong statement here?}

It is difficult to assess which of these expressions yields a solution
closer to the compressible solution.  We can observe, however, that in
the limit as the heating becomes uniform, i.e., $(\sigma \calH)
\rightarrow (\overline{\sigma \calH}),$ the difference between the
contributions to the internal energy and to the total energy should
become smaller, because the contribution of $\calH$ to $\pprimestar$
itself goes to zero, which would not be the case if the base state did
not adjust to large-scale heating.

%\noindent{\bf Observations}
%
%\begin{itemize}
%\item In the KP/GPI equations, allowing the base state to adjust to large-scale heating 
%reduces the degree to which the low Mach number and compressible solutions diverge as 
%measured by energy conservation.  If $\calH > 0$ then $w_0 > 0$; given that $\partial p_0 / \partial r < 0,$ 
%then the presence of the $\partial p_0 / \partial r$ term reduces the magnitude of the 
%$(\sigma \calH)$ contribution.  In the original LM equations, this is not the case because we can
%make no assumption about the orientation of $\widetilde{\Ubstar}$ relative to $\nabla p_0.$
%
%\item More importantly, much of the effect on pressure from any large-scale heating is absorbed by $p_0$
%rather than $\pprime,$ reducing the magnitude of $\pprime$ and thus the overall degree 
%to which the low Mach energy diverges from the compressible energy.  
%In the case where $\calH = \calH(r,t)$ is the only dynamic influence, $\pprime$ is identically
%zero and the total energy increases by $(\rhostar \calH)$, analogous to the compressible solution.
%If the base state were not allowed to respond to the heating, then this would not be the case.
%This second effect is independent of the presence of the additional term in the KP/GPI momentum equations.
%\end{itemize}

While energy conservation is a fundamental and desirable feature of a closed fluid system, 
an interesting question remains as to the effect of the additional term 
on the closeness of the low Mach number solution to the compressible solution. 

\subsection{Test Problems}

Here we look at the impact of the different formulations of the momentum equation 
on two test problems we have used in the development of \mae\ as 
well as one of our science cases.  In \citet{VLBWZ:2013} a study of linear gravity
waves using the two different formulations as implemented in the \mae\ code
demonstrated that the eigenfunctions, energy conservation properties and
pseudo-energy conservation were as expected.  We refer the reader
interested in those results to \citet{VLBWZ:2013}.  

%Just an observation:
%While the analytical formulation of energy conservation may ensure proper 
%gravity wave propagation, we don't necessarily numerically
%conserve energy anyway since we don't solve in conservation form

\subsubsection{Reacting Bubble Rise}

In \citet{AlmgrenEtAl2008} we showed an example of three burning bubbles in a
stratified white dwarf atmosphere (modeled in a plane-parallel
geometry, for convenience).  
The bubbles are buoyant and rise and roll up as nuclear reactions 
release heat in the hotter locations.
Here we revisit that problem with the modified momentum equation.  
Figure~\ref{fig:test2} shows a comparison 
of the temperature field ($\Tstar = {\widehat{T}}(p_0, \estar)$).
for the simulation using the KP/GPI formulation vs. the original LM equations.
As we can see, the KP/GPI formulation has slightly wider bubbles.  

This difference in width is consistent with an observation made in 
\citet{ABRZ:I}, where we compared low Mach number simulations of
non-reacting buoyant bubbles with compressible simulations.  
There we saw variation between the bubbles as calculated with the
different methods (including between the two compressible
formulations), but observed that the bubbles that evolved via the low Mach number
equation set were consistently narrower than the fully compressible bubbles.
The fact that the new formulation generates slightly wider bubbles suggests
in a general sense that the solution with the KP/GPI formulation is closer to 
the fully compressible solution than that generated with the original LM formulation.

This test problem is distributed with the public version of \mae\ as {\tt reacting\_bubble}.

\subsubsection{Internally Driven Convection}

A simple test of convection driven by an analytic heating source was
shown in \citet{AlmgrenEtAl2008} and again in \citet{NonakaEtAl2010}.  The
details of the setup and heating term are given in Section~4.3 of
\citet{AlmgrenEtAl2008}.  The basic idea of this problem is that there is a
convectively unstable region bounded above and below by stable
regions.  An analytic heat source at the base of the unstable layer
drives convection.  The top of the convective region essentially spans
to the top of the atmosphere, up to the point where the density drops
off steeply.  We considered the convectively unstable layer to be the
``region of interest'' for comparison with results from a compressible
code, and we found good agreement between the LM equations and the
fully compressible solution.  Here we compare results from simulations
using the KP/GPI formulation against both the original LM results and
the compressible results.  The compressible comparison for this paper
is done with the CASTRO code~\citep{AlmgrenEtAl2010}.

\MarginPar{pictures really don't look the same -- we need to do 
something with these.  Maybe add horiz avg of KE.  Maybe it is 
developing more slowly?}

For the calculations presented here, as in the previous two works, a
plane-parallel approximation is used for the stellar atmosphere, and
$320\times 512$ cells are used to cover the the domain spanning
$2.5\times 10^{8}~\mathrm{cm}$ in the horizontal direction and
$4.0\times 10^8~\mathrm{cm}$ in the vertical.  In
Figure~\ref{fig:convect_images} we show plots of $\Tstar,$
$\pprimestar/p_0$, and Mach number for the the LM and KP/GPI
algorithms.  Two things are immediately apparent.  First, in the
convective region, roughly between $10^8$~cm and $2.5\times 10^8$~cm,
we see good agreement between the original LM algorithm and the new
KP/GPI version.  The temperature field and Mach number in that region
agree nicely.  However we also note that above the convective region,
the solutions differ.  This is precisely the location of the surface
of the star, so the density drops sharply to the cutoff value
here.  We notice that in the KP/GPI model, the Mach number is much
lower in this surface region and $p'/p_0$ is also much lower (for the
latter, we note that the plots use different scales to bring out the
details in each case).  While these differences do not appear to
influence the behavior of the convective region, the large $M$ in this
surface region does have implications for the timestep.  Overall, the
KP/GPI model seems better behaved in the surface layer.  Looking at
the stable region beneath the convective layer, we do not see much of
a difference between the two models.

As we did when we previously looked at this test problem, we present
two diagnostics: the lateral average of the temperature, $\Tstar,$
\begin{equation}
\left \langle T \right \rangle_j = \frac{1}{N_x} \sum_{i=1}^{N_x} T_{i,j} \enskip ,
\end{equation}
and the RMS fluctuations:
\begin{equation}
(\delta T)_j = \left [ \frac{1}{N_x} \sum_{i=1}^{N_x} (T_{i,j} - \langle T\rangle_j)^2  \right ]^{1/2} \enskip ,
\end{equation}
where $N_x$ is the number of cells in the lateral direction.
Figure~\ref{fig:convect} shows the profiles of $\langle T \rangle$ and
$\delta T/\langle T \rangle$ as a function of height.  In the region
of interest, the convective layer, we see that all the solutions agree
strongly.  The differences outside of the convective layer also seem
minimal.  The increase in $T$ that is seen above the convective layer
for the compressible solution is because in the compressible code, the
material above the atmosphere is not in hydrostatic equilibrium (it
has reached our cutoff density) and rains down onto the surface.  This does not
have a large dynamic effect because there is not much mass there.
 
This test problem is distributed with the public version of \mae\ as {\tt test\_convect}.

%% file: maestrosymbols.tex
\newcommand{\castro}{{\sf CASTRO}}

\newcommand{\rhozero}{\rho_0}

\newcommand{\sfrac}[2]{\mathchoice
  {\kern0em\raise.5ex\hbox{\the\scriptfont0 #1}\kern-.15em/
   \kern-.15em\lower.25ex\hbox{\the\scriptfont0 #2}}
  {\kern0em\raise.5ex\hbox{\the\scriptfont0 #1}\kern-.15em/
   \kern-.15em\lower.25ex\hbox{\the\scriptfont0 #2}}
  {\kern0em\raise.5ex\hbox{\the\scriptscriptfont0 #1}\kern-.2em/
   \kern-.15em\lower.25ex\hbox{\the\scriptscriptfont0 #2}}
  {#1\!/#2}}

\newcommand{\er}{\mathbf{e}_r}

%% file: paper.bbl
\begin{thebibliography}{}
\expandafter\ifx\csname natexlab\endcsname\relax\def\natexlab#1{#1}\fi

\bibitem[{{Alastuey} \& {Jancovici}(1978)}]{AlastueyJancovici1978}
{Alastuey}, A., \& {Jancovici}, B. 1978, \apj, 226, 1034

\bibitem[{{Almgren} {et~al.}(2008){Almgren}, {Bell}, {Nonaka}, \&
  {Zingale}}]{AlmgrenEtAl2008}
{Almgren}, A.~S., {Bell}, J.~B., {Nonaka}, A., \& {Zingale}, M. 2008, \apj,
  684, 449

\bibitem[{Almgren {et~al.}(2006{\natexlab{a}})Almgren, Bell, Rendleman, \&
  Zingale}]{ABRZ:I}
Almgren, A.~S., Bell, J.~B., Rendleman, C.~A., \& Zingale, M.
  2006{\natexlab{a}}, \apj, 637, 922

\bibitem[{Almgren {et~al.}(2006{\natexlab{b}})Almgren, Bell, Rendleman, \&
  Zingale}]{ABRZ:II}
---. 2006{\natexlab{b}}, \apj, 649, 927

\bibitem[{{Almgren} {et~al.}(2010){Almgren}, {Beckner}, {Bell}, {Day},
  {Howell}, {Joggerst}, {Lijewski}, {Nonaka}, {Singer}, \&
  {Zingale}}]{AlmgrenEtAl2010}
{Almgren}, A.~S., {Beckner}, V.~E., {Bell}, J.~B., {et~al.} 2010, \apj, 715,
  1221

\bibitem[{Angulo {et~al.}(1999)Angulo, Arnould, Rayet, Descouvemont, Baye,
  Leclercq-Willain, Coc, Barhoumi, Aguer, Rolfs, Kunz, Hammer, Mayer,
  Paradellis, Kossionides, Chronidou, Spyrou, Degl'Innocenti, Fiorentini,
  Ricci, Zavatarelli, Providencia, Wolters, Soares, Grama, Rahighi, Shotter, \&
  {Lamehi Rachti}}]{AnguloEtAl1999}
Angulo, C., Arnould, M., Rayet, M., {et~al.} 1999, Nuclear Physics A, 656, 3

\bibitem[{{Bildsten} {et~al.}(2007){Bildsten}, {Shen}, {Weinberg}, \&
  {Nelemans}}]{BildstenEtAl2007}
{Bildsten}, L., {Shen}, K.~J., {Weinberg}, N.~N., \& {Nelemans}, G. 2007,
  \apjl, 662, L95

\bibitem[{{Branch} {et~al.}(1993){Branch}, {Fisher}, \&
  {Nugent}}]{BranchEtAl1993}
{Branch}, D., {Fisher}, A., \& {Nugent}, P. 1993, \aj, 106, 2383

\bibitem[{Brooks {et~al.}(2015)Brooks, Bildsten, Marchant, \&
  Paxton}]{BrooksEtAl2015}
Brooks, J., Bildsten, L., Marchant, P., \& Paxton, B. 2015, The Astrophysical
  Journal, 807, 74

\bibitem[{{Brown} {et~al.}(2011){Brown}, {Kilic}, {Allende Prieto}, \&
  {Kenyon}}]{BrownEtAl2011}
{Brown}, W.~R., {Kilic}, M., {Allende Prieto}, C., \& {Kenyon}, S.~J. 2011,
  \mnras, 411, L31

\bibitem[{{Caughlan} \& {Fowler}(1988)}]{CaughlanFowler1988}
{Caughlan}, G.~R., \& {Fowler}, W.~A. 1988, Atomic Data and Nuclear Data
  Tables, 40, 283

\bibitem[{{Drout} {et~al.}(2013){Drout}, {Soderberg}, {Mazzali}, {Parrent},
  {Margutti}, {Milisavljevic}, {Sanders}, {Chornock}, {Foley}, {Kirshner},
  {Filippenko}, {Li}, {Brown}, {Cenko}, {Chakraborti}, {Challis}, {Friedman},
  {Ganeshalingam}, {Hicken}, {Jensen}, {Modjaz}, {Perets}, {Silverman}, \&
  {Wong}}]{DroutEtAl2013}
{Drout}, M.~R., {Soderberg}, A.~M., {Mazzali}, P.~A., {et~al.} 2013, \apj, 774,
  58

\bibitem[{Durran(1989)}]{Durran:1989}
Durran, D.~R. 1989, 46, 1453

\bibitem[{{Fink} {et~al.}(2007){Fink}, {Hillebrandt}, \&
  {R{\"o}pke}}]{FinkEtAl2007}
{Fink}, M., {Hillebrandt}, W., \& {R{\"o}pke}, F.~K. 2007, \aap, 476, 1133

\bibitem[{{Fink} {et~al.}(2010){Fink}, {R{\"o}pke}, {Hillebrandt},
  {Seitenzahl}, {Sim}, \& {Kromer}}]{FinkEtAl2010}
{Fink}, M., {R{\"o}pke}, F.~K., {Hillebrandt}, W., {et~al.} 2010, \aap, 514,
  A53

\bibitem[{{Garc{\'{\i}}a-Senz} {et~al.}(1999){Garc{\'{\i}}a-Senz}, {Bravo}, \&
  {Woosley}}]{GarciaSenzEtAl1999}
{Garc{\'{\i}}a-Senz}, D., {Bravo}, E., \& {Woosley}, S.~E. 1999, \aap, 349, 177

\bibitem[{{Garnett}(1997)}]{Garnett1997}
{Garnett}, D.~R. 1997, Nuclear Physics A, 621, 27

\bibitem[{Geier {et~al.}(2013)Geier, Marsh, Wang, Dunlap, Barlow, Schaffenroth,
  Chen, Irrgang, Maxted, Ziegerer, Kupfer, Miszalski, Heber, Han, Shporer,
  Telting, G\"{a}nsicke, \O~stensen, O'Toole, \& Napiwotzki}]{GeierEtAl2013}
Geier, S., Marsh, T.~R., Wang, B., {et~al.} 2013, Astronomy \& Astrophysics,
  554, A54

\bibitem[{{Graboske} {et~al.}(1973){Graboske}, {Dewitt}, {Grossman}, \&
  {Cooper}}]{GraboskeEtAl1973}
{Graboske}, H.~C., {Dewitt}, H.~E., {Grossman}, A.~S., \& {Cooper}, M.~S. 1973,
  \apj, 181, 457

\bibitem[{Hillebrandt {et~al.}(2013)Hillebrandt, Kromer, Röpke, \&
  Ruiter}]{HillebrandtEtAl2013}
Hillebrandt, W., Kromer, M., Röpke, F., \& Ruiter, A. 2013, Frontiers of
  Physics, 8, 116

\bibitem[{Hunter(2007)}]{Hunter2007}
Hunter, J.~D. 2007, Computing In Science \& Engineering, 9, 90

\bibitem[{{Itoh} {et~al.}(1979){Itoh}, {Honda}, \& {Suzuki}}]{ItohEtAl1979}
{Itoh}, F., {Honda}, T., \& {Suzuki}, K. 1979, Journal of the Physical Society
  of Japan, 46, 1201

\bibitem[{{Ivezic} {et~al.}(2008){Ivezic}, {Tyson}, {Abel}, {Acosta},
  {Allsman}, {AlSayyad}, {Anderson}, {Andrew}, {Angel}, {Angeli}, {Ansari},
  {Antilogus}, {Arndt}, {Astier}, {Aubourg}, {Axelrod}, {Bard}, {Barr},
  {Barrau}, {Bartlett}, {Bauman}, {Beaumont}, {Becker}, {Becla}, {Beldica},
  {Bellavia}, {Blanc}, {Blandford}, {Bloom}, {Bogart}, {Borne}, {Bosch},
  {Boutigny}, {Brandt}, {Brown}, {Bullock}, {Burchat}, {Burke}, {Cagnoli},
  {Calabrese}, {Chandrasekharan}, {Chesley}, {Cheu}, {Chiang}, {Claver},
  {Connolly}, {Cook}, {Cooray}, {Covey}, {Cribbs}, {Cui}, {Cutri}, {Daubard},
  {Daues}, {Delgado}, {Digel}, {Doherty}, {Dubois}, {Dubois-Felsmann},
  {Durech}, {Eracleous}, {Ferguson}, {Frank}, {Freemon}, {Gangler}, {Gawiser},
  {Geary}, {Gee}, {Geha}, {Gibson}, {Gilmore}, {Glanzman}, {Goodenow},
  {Gressler}, {Gris}, {Guyonnet}, {Hascall}, {Haupt}, {Hernandez}, {Hogan},
  {Huang}, {Huffer}, {Innes}, {Jacoby}, {Jain}, {Jee}, {Jernigan},
  {Jevremovic}, {Johns}, {Jones}, {Juramy-Gilles}, {Juric}, {Kahn}, {Kalirai},
  {Kallivayalil}, {Kalmbach}, {Kantor}, {Kasliwal}, {Kessler}, {Kirkby},
  {Knox}, {Kotov}, {Krabbendam}, {Krughoff}, {Kubanek}, {Kuczewski},
  {Kulkarni}, {Lambert}, {Le Guillou}, {Levine}, {Liang}, {Lim}, {Lintott},
  {Lupton}, {Mahabal}, {Marshall}, {Marshall}, {May}, {McKercher}, {Migliore},
  {Miller}, {Mills}, {Monet}, {Moniez}, {Neill}, {Nief}, {Nomerotski},
  {Nordby}, {O'Connor}, {Oliver}, {Olivier}, {Olsen}, {Ortiz}, {Owen}, {Pain},
  {Peterson}, {Petry}, {Pierfederici}, {Pietrowicz}, {Pike}, {Pinto}, {Plante},
  {Plate}, {Price}, {Prouza}, {Radeka}, {Rajagopal}, {Rasmussen}, {Regnault},
  {Ridgway}, {Ritz}, {Rosing}, {Roucelle}, {Rumore}, {Russo}, {Saha},
  {Sassolas}, {Schalk}, {Schindler}, {Schneider}, {Schumacher}, {Sebag},
  {Sembroski}, {Seppala}, {Shipsey}, {Silvestri}, {Smith}, {Smith}, {Strauss},
  {Stubbs}, {Sweeney}, {Szalay}, {Takacs}, {Thaler}, {Van Berg}, {Vanden Berk},
  {Vetter}, {Virieux}, {Xin}, {Walkowicz}, {Walter}, {Wang}, {Warner},
  {Willman}, {Wittman}, {Wolff}, {Wood-Vasey}, {Yoachim}, {Zhan}, \& {for the
  LSST Collaboration}}]{LSST2008}
{Ivezic}, Z., {Tyson}, J.~A., {Abel}, B., {et~al.} 2008, ArXiv e-prints,
  arXiv:0805.2366

\bibitem[{Klein \& Pauluis(2012)}]{KP:2012}
Klein, R., \& Pauluis, O. 2012

\bibitem[{{Kromer} {et~al.}(2010){Kromer}, {Sim}, {Fink}, {R{\"o}pke},
  {Seitenzahl}, \& {Hillebrandt}}]{KromerEtAl2010}
{Kromer}, M., {Sim}, S.~A., {Fink}, M., {et~al.} 2010, \apj, 719, 1067

\bibitem[{{Li} {et~al.}(2011){Li}, {Leaman}, {Chornock}, {Filippenko},
  {Poznanski}, {Ganeshalingam}, {Wang}, {Modjaz}, {Jha}, {Foley}, \&
  {Smith}}]{LiEtAl2011}
{Li}, W., {Leaman}, J., {Chornock}, R., {et~al.} 2011, \mnras, 412, 1441

\bibitem[{Liu {et~al.}(2015{\natexlab{a}})Liu, Moriya, Stancliffe, \&
  Wang}]{Liu2015e}
Liu, Z.-W., Moriya, T.~J., Stancliffe, R.~J., \& Wang, B. 2015{\natexlab{a}},
  Astronomy {\&} Astrophysics, 574, A12

\bibitem[{Liu {et~al.}(2015{\natexlab{b}})Liu, Stancliffe, Abate, \&
  Wang}]{Liu2015d}
Liu, Z.-W., Stancliffe, R.~J., Abate, C., \& Wang, B. 2015{\natexlab{b}}, The
  Astrophysical Journal, 808, 138

\bibitem[{{Livne}(1990)}]{Livne1990}
{Livne}, E. 1990, \apjl, 354, L53

\bibitem[{{Livne} \& {Arnett}(1995)}]{LivneArnett1995}
{Livne}, E., \& {Arnett}, D. 1995, \apj, 452, 62

\bibitem[{{Livne} \& {Glasner}(1990)}]{LivneGlasner1990}
{Livne}, E., \& {Glasner}, A.~S. 1990, \apj, 361, 244

\bibitem[{{Livne} \& {Glasner}(1991)}]{LivneGlasner1991}
---. 1991, \apj, 370, 272

\bibitem[{{Malone} {et~al.}(2014){Malone}, {Nonaka}, {Woosley}, {Almgren},
  {Bell}, {Dong}, \& {Zingale}}]{Malone:2014}
{Malone}, C.~M., {Nonaka}, A., {Woosley}, S.~E., {et~al.} 2014, \apj, 782, 11

\bibitem[{Moll \& Woosley(2013)}]{MollWoosley2013}
Moll, R., \& Woosley, S.~E. 2013, The Astrophysical Journal, 774, 137

\bibitem[{Nelemans(2005)}]{Nelemans2005}
Nelemans, Â. 2005, The Astrophysics of Cataclysmic Variables and Related
  Objects, 330

\bibitem[{{Nomoto}(1982{\natexlab{a}})}]{Nomoto1982b}
{Nomoto}, K. 1982{\natexlab{a}}, \apj, 257, 780

\bibitem[{{Nomoto}(1982{\natexlab{b}})}]{Nomoto1982a}
---. 1982{\natexlab{b}}, \apj, 253, 798

\bibitem[{{Nonaka} {et~al.}(2010){Nonaka}, {Almgren}, {Bell}, {Lijewski},
  {Malone}, \& {Zingale}}]{NonakaEtAl2010}
{Nonaka}, A., {Almgren}, A.~S., {Bell}, J.~B., {et~al.} 2010, \apjs, 188, 358

\bibitem[{Nonaka {et~al.}(2012)Nonaka, Aspden, Zingale, Almgren, Bell, \&
  Woosley}]{NonakaEtAl2012}
Nonaka, A., Aspden, A.~J., Zingale, M., {et~al.} 2012, The Astrophysical
  Journal, 745, 73

\bibitem[{{Paxton} {et~al.}(2011){Paxton}, {Bildsten}, {Dotter}, {Herwig},
  {Lesaffre}, \& {Timmes}}]{PaxtonEtAl2011}
{Paxton}, B., {Bildsten}, L., {Dotter}, A., {et~al.} 2011, \apjs, 192, 3

\bibitem[{{Paxton} {et~al.}(2013){Paxton}, {Cantiello}, {Arras}, {Bildsten},
  {Brown}, {Dotter}, {Mankovich}, {Montgomery}, {Stello}, {Timmes}, \&
  {Townsend}}]{PaxtonEtAl2013}
{Paxton}, B., {Cantiello}, M., {Arras}, P., {et~al.} 2013, \apjs, 208, 4

\bibitem[{P\'erez \& Granger(2007)}]{PerezGranger2007}
P\'erez, F., \& Granger, B.~E. 2007, Computing in Science and Engineering, 9,
  21

\bibitem[{{Phillips}(1993)}]{Phillips1993}
{Phillips}, M.~M. 1993, \apjl, 413, L105

\bibitem[{Piro(2015)}]{Piro2015}
Piro, A.~L. 2015, \apj, 801, 137

\bibitem[{{Ruiter} {et~al.}(2011){Ruiter}, {Belczynski}, {Sim}, {Hillebrandt},
  {Fryer}, {Fink}, \& {Kromer}}]{RuiterEtAl2011}
{Ruiter}, A.~J., {Belczynski}, K., {Sim}, S.~A., {et~al.} 2011, \mnras, 417,
  408

\bibitem[{Sakharuk(2006)}]{Sakharuk2006}
Sakharuk, A. 2006, in AIP Conference Proceedings, Vol. 819 (AIP), 118--122

\bibitem[{Shen \& Bildsten(2009)}]{ShenBildsten2009}
Shen, K.~J., \& Bildsten, L. 2009, The Astrophysical Journal, 699, 1365

\bibitem[{{Shen} \& {Bildsten}(2014)}]{ShenBildsten2014}
{Shen}, K.~J., \& {Bildsten}, L. 2014, \apj, 785, 61

\bibitem[{Shen \& Moore(2014)}]{ShenMoore2014}
Shen, K.~J., \& Moore, K. 2014, The Astrophysical Journal, 797, 46

\bibitem[{{Sim} {et~al.}(2010){Sim}, {R{\"o}pke}, {Hillebrandt}, {Kromer},
  {Pakmor}, {Fink}, {Ruiter}, \& {Seitenzahl}}]{SimEtAl2010}
{Sim}, S.~A., {R{\"o}pke}, F.~K., {Hillebrandt}, W., {et~al.} 2010, \apjl, 714,
  L52

\bibitem[{{Timmes}(2008)}]{Timmes2008}
{Timmes}, F.~X. 2008, {Stellar Equations Of State}, ,

\bibitem[{{Timmes} \& {Swesty}(2000)}]{TimmesSwesty2000}
{Timmes}, F.~X., \& {Swesty}, F.~D. 2000, \apjs, 126, 501

\bibitem[{Townsley {et~al.}(2012)Townsley, Moore, \&
  Bildsten}]{TownsleyEtAl2012}
Townsley, D.~M., Moore, K., \& Bildsten, L. 2012, The Astrophysical Journal,
  755, 4

\bibitem[{{Turk} {et~al.}(2011){Turk}, {Smith}, {Oishi}, {Skory}, {Skillman},
  {Abel}, \& {Norman}}]{TurkEtAl2011}
{Turk}, M.~J., {Smith}, B.~D., {Oishi}, J.~S., {et~al.} 2011, \apjs, 192, 9

\bibitem[{Vasil {et~al.}(2013)Vasil, Lecoanet, Brown, Wood, \&
  Zweibel}]{VLBWZ:2013}
Vasil, G.~M., Lecoanet, D., Brown, B.~P., Wood, T.~S., \& Zweibel, E.~G. 2013,
  \apj, 773, 169

\bibitem[{{Wang} {et~al.}(2013){Wang}, {Justham}, \& {Han}}]{WangEtAl2013}
{Wang}, B., {Justham}, S., \& {Han}, Z. 2013, \aap, 559, A94

\bibitem[{Warner(1995)}]{Warner1995}
Warner, B. 1995, Astrophysics and Space Science, 225, 249

\bibitem[{{Weaver} \& {Woosley}(1993)}]{WeaverWoosley1993}
{Weaver}, T.~A., \& {Woosley}, S.~E. 1993, \physrep, 227, 65

\bibitem[{{Weaver} {et~al.}(1978){Weaver}, {Zimmerman}, \&
  {Woosley}}]{WeaverEtAl1978}
{Weaver}, T.~A., {Zimmerman}, G.~B., \& {Woosley}, S.~E. 1978, \apj, 225, 1021

\bibitem[{{Woosley} \& {Kasen}(2011)}]{WoosleyKasen2011}
{Woosley}, S.~E., \& {Kasen}, D. 2011, \apj, 734, 38

\bibitem[{{Woosley} {et~al.}(1986){Woosley}, {Taam}, \&
  {Weaver}}]{WoosleyEtAl1986}
{Woosley}, S.~E., {Taam}, R.~E., \& {Weaver}, T.~A. 1986, \apj, 301, 601

\bibitem[{{Woosley} \& {Weaver}(1994)}]{WoosleyWeaver1994}
{Woosley}, S.~E., \& {Weaver}, T.~A. 1994, \apj, 423, 371

\bibitem[{{Zingale} {et~al.}(2009){Zingale}, {Almgren}, {Bell}, {Nonaka}, \&
  {Woosley}}]{zingale:2009}
{Zingale}, M., {Almgren}, A.~S., {Bell}, J.~B., {Nonaka}, A., \& {Woosley},
  S.~E. 2009, \apj, 704, 196

\bibitem[{{Zingale} {et~al.}(2013){Zingale}, {Nonaka}, {Almgren}, {Bell},
  {Malone}, \& {Orvedahl}}]{ZingaleEtAl2013}
{Zingale}, M., {Nonaka}, A., {Almgren}, A.~S., {et~al.} 2013, \apj, 764, 97

\bibitem[{Zingale {et~al.}(2011)Zingale, Nonaka, Almgren, Bell, Malone, \&
  Woosley}]{ZingaleEtAl2011}
Zingale, M., Nonaka, A., Almgren, A.~S., {et~al.} 2011, The Astrophysical
  Journal, 740, 8

\end{thebibliography}
